\documentclass[12pt,preprint]{aastex61}

\accepted {ApJS}

\def\MO{M_\odot}

\def\LO{L_\odot}

\shorttitle{Chemical survey in the Perseus}
\shortauthors{Higuchi et al.}

\begin{document}
\title{Chemical survey toward young stellar objects in the Perseus molecular cloud complex}

\correspondingauthor{Aya E. Higuchi}
\email{aya.higuchi@riken.jp}

\author[0000-0002-9221-2910]{Aya E. Higuchi}
\affil{RIKEN Cluster for Pioneering Research, 2-1, Hirosawa, Wako-shi, Saitama 351-0198, Japan}

\author{Nami Sakai}
\affiliation{RIKEN Cluster for Pioneering Research, 2-1, Hirosawa, Wako-shi, Saitama 351-0198, Japan}

\author{Yoshimasa Watanabe}
\affiliation{Division of Physics, Faculty of Pure and Applied Sciences, University of Tsukuba, Tsukuba, Ibaraki 305-8571, Japan}

\author{Ana L{\'o}pez-Sepulcre}
\affiliation{Univ.  Grenoble  Alpes, CNRS, Institut de Plan \'etologie et d’Astrophysique de Grenoble (IPAG), 38000 Grenoble, France}
\affiliation{Institut  de  Radioastronomie Millim\'etrique, 
300  rue  de  la Piscine, Domaine  Universitaire  de  Grenoble,  38406  Saint-Martin  d'H \`eres, France}

\author{Kento Yoshida}
\affiliation{RIKEN Cluster for Pioneering Research, 2-1, Hirosawa, Wako-shi, Saitama 351-0198, Japan}
\affiliation{Department of Physics, The University of Tokyo, Hongo, Bunkyo-ku, Tokyo 113-0033, Japan}

\author{Yoko Oya}
\affiliation{Department of Physics, The University of Tokyo, Hongo, Bunkyo-ku, Tokyo 113-0033, Japan}

\author{Muneaki Imai}
\affiliation{Department of Physics, The University of Tokyo, Hongo, Bunkyo-ku, Tokyo 113-0033, Japan}

\author{Yichen Zhang}
\affiliation{RIKEN Cluster for Pioneering Research, 2-1, Hirosawa, Wako-shi, Saitama 351-0198, Japan}

\author{Cecilia Ceccarelli}
\affiliation{Univ. Grenoble Alpes, CNRS, Institut de Plan \'etologie et d’Astrophysique de Grenoble (IPAG), 38000 Grenoble, France}

\author{Bertrand Lefloch}
\affiliation{Univ. Grenoble Alpes, CNRS, Institut de Plan \'etologie et d’Astrophysique de Grenoble (IPAG), 38000 Grenoble, France}

\author{Claudio Codella}
\affiliation{INAF, Osservatorio Astrofisico di Arcetri, Largo E. Fermi 5, 50125, Firenze, Italy}

\author{Rafael Bachiller}
\affiliation{IGN Observatorio Astron\'omico Nacional, Apartado 1143, E-28800 Alcal\'a de Henares, Spain}

\author{Tomoya Hirota}
\affiliation{National Astronomical Observatory of Japan, Osawa, Mitaka, Tokyo 181-8588, Japan}

\author{Takeshi Sakai}
\affiliation{Department of Communication Engineering and Informatics, Graduate School of Informatics 
and Engineering, The University of Electro-Communications, Chofugaoka, Chofu, Tokyo 182-8585, Japan}

\author{Satoshi Yamamoto}
\affiliation{Department of Physics, The University of Tokyo, Hongo, Bunkyo-ku, Tokyo 113-0033, Japan}

\begin{abstract}

Chemical diversity of the gas in low-mass protostellar cores is widely recognized. 
In order to explore its origin, a survey of chemical composition toward 36 Class 0/I protostars 
in the Perseus molecular cloud complex, which are selected in an unbiased way under certain physical conditions,
has been conducted with IRAM~30~m and NRO~45~m telescope.
Multiple lines of C$_{2}$H, c-C$_{3}$H$_{2}$ and CH$_{3}$OH have been observed to characterize the chemical composition averaged over 
a 1000~au scale around the protostar.
The derived beam-averaged column densities show significant chemical diversity among the sources, 
where the column density ratios of C$_{2}$H/CH$_{3}$OH are spread out by 2 orders of magnitude.
From previous studies, the hot corino sources have abundant CH$_{3}$OH but deficient C$_{2}$H, their C$_{2}$H/CH$_{3}$OH
column density ratios being relatively low.
In contrast, the warm-carbon-chain chemistry (WCCC) sources are found to reveal the high C$_{2}$H/CH$_{3}$OH column density ratios. 
We find that the majority of the sources have intermediate characters between these two distinct chemistry types.
A possible trend is seen between the C$_{2}$H/CH$_{3}$OH ratio and the distance of the source from the edge of a molecular cloud.
The sources located near cloud edges or in isolated clouds tend to have a high C$_{2}$H/CH$_{3}$OH ratio.
On the other hand, the sources having a low C$_{2}$H/CH$_{3}$OH ratio tend to be located in inner regions of the molecular cloud complex.
This result gives an important clue to an understanding of the origin of the chemical diversity of protostellar cores in terms of environmental effects.

\end{abstract}



\section{Introduction} \label{intro}

Chemical compositions of protostellar cores are of fundamental importance, because they are related to the initial condition for chemical evolution toward protoplanetary disks.
During the last decade, they have extensively been studied by radioastronomical observations.
Now, it is well known that chemical compositions of low-mass protostellar cores ($r$ $<$ $\sim$ 1000~au) show significant diversity (Sakai $\&$ Yamamoto 2013). 
One distinct case of the diversity is hot corino chemistry.
It is characterized by rich  complex-organic molecules (COMs) such as HCOOCH$_{3}$ and (CH$_{3}$)$_{2}$O and deficient 
carbon-chain molecules such as C$_{2}$H, c-C$_{3}$H$_{2}$ and C$_{4}$H.  
A prototypical hot corino source is IRAS~16293-2422 (e.g., Cazaux et al 2003; Bottinelli et al. 2004). 
Another distinct case is warm-carbon-chain chemistry (WCCC). It is characterized by abundant carbon-chain molecules and deficient COMs.  
A prototypical WCCC source is IRAS 04368+2557 in L1527 (Sakai et al. 2008; 2010a). Such an exclusive chemical feature between 
COMs and carbon-chain molecules stands for a major axis of chemical diversity.
It should be noted that we do not know at this moment whether any other axes exist.
In this paper, we therefore use the word of ``chemical diversity'' to represent the hot corino chemistry vs. WCCC axis.  


Sakai et al. (2009) proposed that one possible origin of the above chemical diversity of low-mass protostellar cores is 
the difference of a duration time of their starless phase after 
the shielding of the interstellar UV radiation.
After the UV shielding, formation of molecules starts both in the gas phase and on dust grains, 
whose time scale is comparable to the dynamical time scale of a parent cloud (i.e., free-fall time).  
A longer duration time of the starless phase tends to result in the hot corino chemistry, while a shorter duration time the WCCC. 
This mechanism can explain the various observational results obtained so far (Sakai $\&$ Yamamoto 2013). 
For instance, lower deuterium fractionation ratios and association of young starless cores near the WCCC source are consistent with this picture 
(Sakai et al. 2010; Sakai $\&$ Yamamoto 2013). 
However, other mechanisms such as shocks (outflows, cloud-cloud/filament-filament collision) and UV radiation from nearby 
OB stars may also contribute to chemical diversity 
(e.g., Buckle \& Fuller 2002; Watanabe et al. 2012; Lindberg $\&$ J{\o}rgensen 2012; Higuchi et al. 2010; 2014; Fukui et al. 2015; Spezzano et al. 2016). 
Hence, the origin of the above chemical diversity is still controversial.


So far, only a few sources have unambiguously been identified as the hot corino sources and WCCC sources each.
The former examples are IRAS~16293-2422, NGC~1333 IRAS~4A, NGC~1333 IRAS~4B, NGC~1333 IRAS~2, Serpens SMM1, Serpens SMM4, and HH212  
(e.g., Cazaux et al 2003; Bottinelli et al. 2004; Sakai et al. 2006;{\"O}berg et al. 2011; Codella et al. 2016). 
The latter examples are L1527, IRAS 15398-3359, and TMC-1A (e.g., Sakai et al. 2008; Sakai et al. 2014b).
Thus, the statistics is very poor. 
To step forward to understanding the origin of the chemical diversity of protostellar cores, 
we need to know what kind of chemistry is a common occurrence.
Recently, Lindberg et al. (2016) and Graninger et al. (2016) reported statistical studies of the CH$_{3}$OH and C$_{4}$H abundances toward
low mass star forming cores. They use CH$_{3}$OH and C$_{4}$H as representative COM and carbon-chain molecule, respectively.
Although these studies provide us with rich information on chemical diversity, different distances to the sources as well as regional difference of physical conditions
(UV radiation field, star-formation activities) may complicate an interpretation of the observed chemical diversity.

A powerful approach to overcome this situation is an unbiased survey of all protostellar cores in a single molecular cloud complex.
Such a study allows us to explore environmental effects on chemical composition of protostellar sources in the molecular cloud complex without any preconception. 
In addition, all the targets are at almost the same distance, and are therefore affected similarly by beam dilution effects. 
This feature makes statistical arguments easier.  
With these in mind, we have conducted the unbiased chemical survey of the Perseus molecular cloud complex in the 3~mm and 1.3~mm bands.
We observed the CH$_{3}$OH lines as a proxy of the COMs, because CH$_{3}$OH is a parent molecule for production of larger COMs.  
We employed the C$_{2}$H and c-C$_{3}$H$_{2}$ lines as a proxy of carbon-chain related molecules, because they are 
the most fundamental carbon-chain molecules giving bright emission. 
By comparing the results for these molecules, we discuss the chemical diversity of protostellar cores in Perseus.

\section{Observations} \label{obs}
\subsection{Observed sources and molecules}

The Perseus molecular cloud complex is one of the most famous and well-studied nearby low-mass star forming regions 
(e.g., Hatchell et al. 2005; J{\o}rgensen et al. 2006). 
The distance from the Sun is reported to be 235~pc -- 238~pc (Hirota et al. 2008; 2011).
In this paper, we employ the distance of 235~pc.
It consists of a few molecular clouds including NGC~1333, L1455, L1448, IC~348, B1 and B5, which show different star-formation activities.
In the whole Perseus molecular cloud extending over a 10~pc scale, about 400 sources are identified as young-stellar-object (YSO) candidates, 
among which more than 50 are thought to be Class 0 or Class I protostars (Hatchell et al. 2005).  
We selected the target sources from the list by Hatchell et al. (2007) under the following criteria: 
(1) The protostellar sources are in the Class 0/I stage. 
(2) The bolometric luminosity is higher than 1~$\LO$ (except for B1-5; 0.7~$\LO$). 
(3) The envelope mass is higher than 1~$\MO$ to ensure association of a substantial amount of molecular gas. 
In total, 36 protostellar sources are in our target-source list (Table \ref{t1}).  
Our sample is unbiased under the above three conditions.
It should be stressed that it is unbiased from a view point of chemical condition.

We observed the CH$_{3}$OH, C$_{2}$H, and c-C$_{3}$H$_{2}$ lines listed in Table \ref{t2}.
CH$_{3}$OH is the most fundamental saturated organic molecule which is abundant in hot corino sources 
(e.g., Maret et al. 2005; Kristensen et al. 2010; Sakai et al. 2012).
On the other hand, C$_{2}$H and c-C$_{3}$H$_{2}$ are basic carbon-chain related molecules which are abundant in WCCC sources 
(e.g., Sakai et al. 2008; Sakai et al. 2009).
Hence, we can characterize the chemical composition of the sources with these species.

\subsection{Observation with Nobeyama 45~m telescope}

Observations of the CH$_{3}$OH lines in the 3~mm band were carried out with the 45~m telescope at the Nobeyama
Radio Observatory (NRO) during 2014 January and 2015 March toward the target sources except for NGC~1333-16 (IRAS~4A) and NGC~1333-17 (SVS~13A).
These two sources were not observed due to the limited observation time.
The side-band-separating (2SB) mixer receiver T100HV was used as the front end with the typical
system noise temperature ranging from 150 to 200~K.
The beam size (HPBW) is 21$\arcsec$ at 90~GHz, which corresponds to 4900~au at the distance of 235~pc. 
The back end was a bank of 16 SAM-45 auto-correlators, whose band width and frequency resolution each are 250~MHz and 122~kHz 
(velocity resolution of  $\sim$0.4~km~s$^{-1}$), respectively.

The telescope pointing was checked every hour by observing the SiO maser source, NML~Tau. 
The pointing accuracy was confirmed to be better than 7$\arcsec$. 
The position switching mode was employed for all the above sources, where the position with the C$^{18}$O integrated intensity lower than 1~K~km~s$^{-1}$
(Hatchell et al. 2005) near each target molecular cloud is taken as the off-position.
The offset of the off position relative to the target source position is ($\delta$~R.A., $\delta$~Decl.) = (-1200\arcsec, 0\arcsec) for the sources in the NGC1333 region, (-600\arcsec, 0\arcsec) for the sources in the L1448 region, (-600\arcsec, 0\arcsec) for the sources in the IC~348 and the Barnard 5 regions, (-850\arcsec, 0\arcsec) for the sources in the B1 region, and (0\arcsec, 780\arcsec) for the sources in the L1455 region.
The intensity scale was calibrated to the antenna temperature ($T$$_{\rm{A}}^{*}$) scale by using the chopper-wheel method. 
The antenna temperature was converted to the main-beam brightness temperature by using the main beam efficiency of 0.45 
provided by the observatory. The uncertainty of the intensity calibration is estimated to be better than 20~$\%$.
The observed data were reduced with the software package NEWSTAR developed at NRO.

\subsection{Observation with IRAM 30~m telescope}

Observations of the CH$_{3}$OH, C$_{2}$H, and c-C$_{3}$H$_{2}$ lines in the 1.3~mm band were carried out with the Institut de Radio Astronomie Millim{\'e}trique (IRAM) 30~m telescope at Pico Veleta.
The sources except for NGC~1333-16 (IRAS~4A) and NGC~1333-17 (SVS~13A) were observed in the period from between 2015 January to 2016 May.
For these two sources, we use the data taken by the ASAI (Astrochemical Survey At IRAM) project (Lefloch et al. 2018).
The Eight Mixer Receiver (EMIR), E230, was employed in the dual-polarization mode. 
The system temperatures ranged from 250 to 400~K.
HPBW is 10$\arcsec$ at 260~GHz, which corresponds to 2400~au at the distance of 235~pc. 
The back end consists of eight Fourier transform spectrometers (FTS), whose bandwidth and channel width each are 400~MHz and 200~kHz 
(velocity resolution of $\sim$0.3~km~s$^{-1}$), respectively. 
The telescope pointing was checked every hour by observing nearby continuum sources and was confirmed to be better than $\sim$5$\arcsec$. 
The position switching mode was employed for all the above sources.
As for the off-position, the same position as in the case of the Nobeyama observations was used, while the wobbler switching mode was employed for NGC~1333-16 and NGC~1333-17.
\footnote{For IC~348-3, the C$^{18}$O integrated intensity did not meet the above criteria.
Since the systemic velocity of the off-position is shifted by $\sim$1.5 km~s$^{-1}$ from the systemic velocity of IC~348-3,
the influence on the analysis in this study would be negligible.}
The intensity scale was calibrated to the antenna temperature scale by using the two temperature loads. 
$T$$_{\rm{A}}^{*}$ was then converted to the main beam temperature $T_{\rm{MB}}$ by multiplying $F_{\rm{eff}}/B_{\rm{eff}}$ 
(mean value between 240 and 260~GHz), where $F_{\rm{eff}}$ is the forward efficiency (0.92) and $B_{\rm{eff}}$ is the main beam efficiency (0.59).
The uncertainty of the intensity calibration is estimated to be better than 20~$\%$.
The data were reduced with the CLASS software of the GILDAS package.

\section{Results}\label{res}
\subsection{Data analyses}

\begin{figure}
\epsscale{0.9}
\plotone{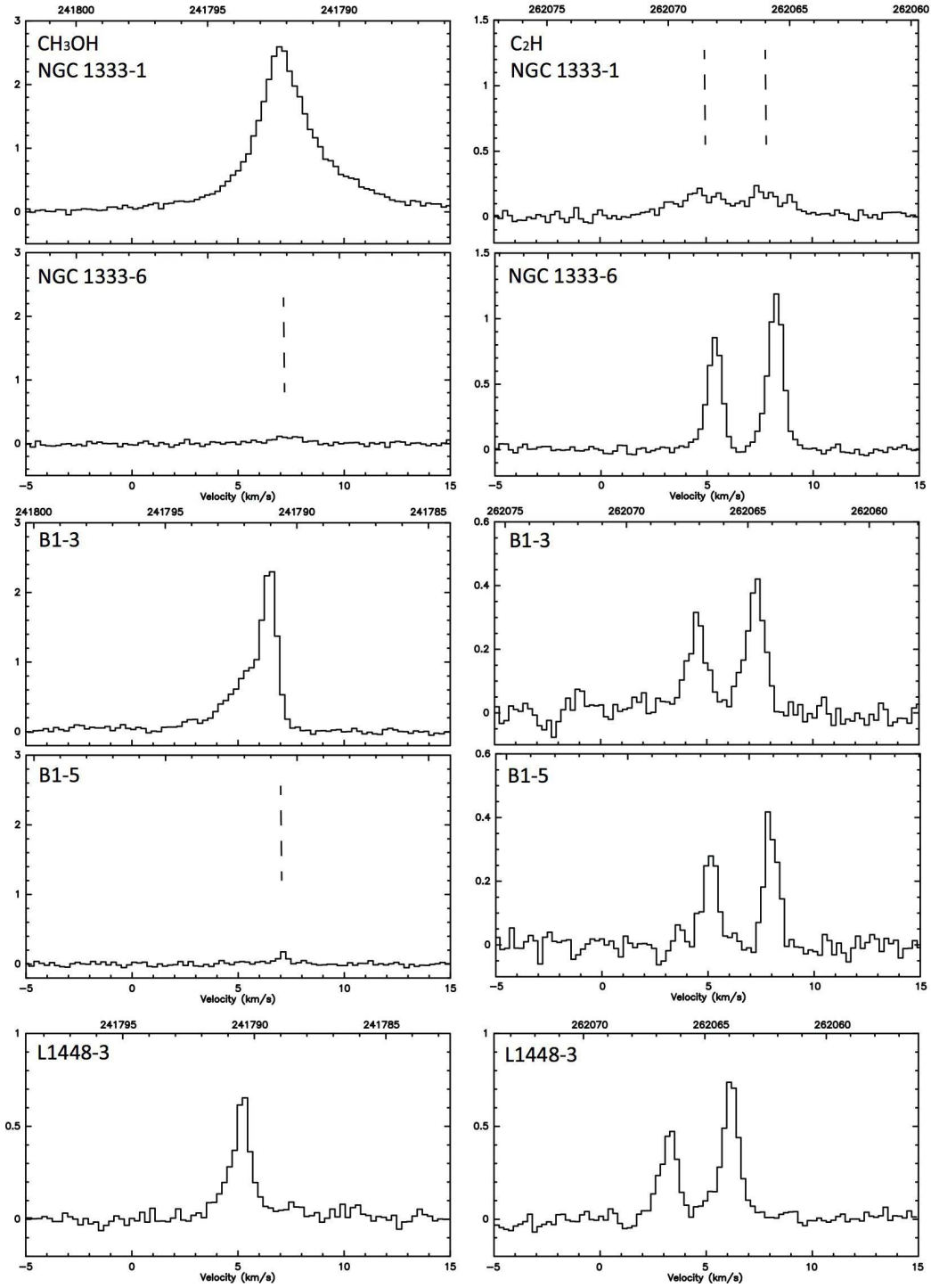}
\caption{Line profiles of CH$_{3}$OH ($J$=5--4, $K$=1, E$^{-}$) 
and C$_{2}$H ($N$=3--2, $J$=5/2--3/2) observed with IRAM~30~m toward NGC~1333-1, NGC~1333-6, B1-3, B1-5, and L1448-3.
Two hyperfine components, $F$=2--1 (left) and $F$=3--2 (right), are observed.}
\label{fg1}
\end{figure}

Figure~\ref{fg1} shows the observed spectral lines of CH$_{3}$OH ($J$=5--4, $K$=1, E$^{-}$, $E_{\rm{u}}$=40~K) and 
C$_{2}$H ($N$=3--2, $J$=5/2--3/2, $F$=3--2, $E_{\rm{u}}$=25~K) for a few selected sources.
The relative intensities between CH$_{3}$OH and C$_{2}$H are significantly different among the sources.
For instance, the CH$_{3}$OH line is strongly detected toward NGC~1333-1, whereas the intensity of the C$_{2}$H line is weak.
In contrast, NGC~1333-6 shows an opposite trend; CH$_{3}$OH is not detected.
A similar trend can be seen in B1-5: the C$_{2}$H lines are strong, while the CH$_{3}$OH lines are weak.
For B1-3 and L1448-3, both the CH$_{3}$OH and C$_{2}$H lines are moderately intense.
To quantify the trend, we evaluated 
the line parameters for each line by assuming that the line profile is approximated by a Gaussian function.

The CH$_{3}$OH lines at 242~GHz would likely trace a relatively dense and warm region rather than a cold ambient cloud because of their upper state energies 
(e.g., CH$_{3}$OH; $J$=5--4, $K$=1, E$^{-}$, $E_{\rm{u}}$=40~K) and their critical densities (10$^{5-6}$~cm$^{-3}$).
The CH$_{3}$OH ($J$=5--4) lines were detected toward 35 sources out of the 36 sources.
Their spectra are shown in Appendix (Figure~\ref{fg9})\footnote{Figures \ref{fg9}, \ref{fg10}, \ref{fg11}, 
\ref{fg12} and Tables \ref{t4}, \ref{t5}, \ref{t6}, and \ref{t7} are in the Appendix.}. 
Individual line parameters of CH$_{3}$OH ($J$=5--4) are listed in Table~\ref{t4}.
For NGC~1333-1 (IRAS~4B), and NGC~1333-2 (IRAS~2), nine and twelve K-structure lines of CH$_{3}$OH were detected, 
respectively, as also reported in Maret et al. (2005).
The CH$_{3}$OH ($J$=5--4) lines detected in L1448-5, B1-1 and B1-3 accompany strong wing components. 
On the other hand, the CH$_{3}$OH ($J$=2--1) lines (two or three K-structure) at 97~GHz 
were detected toward all the sources, whose line parameters are listed in Table~\ref{t5}.

The C$_{2}$H ($N$=3--2) lines at 262~GHz were detected toward all the sources, as shown in Figure~\ref{fg10}.
These lines also trace a relatively dense and warm region as in the case of the CH$_{3}$OH line.
Four hyperfine components were seen in all the sources.
Their individual line parameters obtained with the Gaussian fit are listed in Table~\ref{t6}.
The line parameters of the weakest hyperfine component are missing for some sources, because the Gaussian fit was unsuccessful due to a poor S/N ratio.
For NGC~1333-1 and NGC~1333-16, the line shapes of the C$_{2}$H lines are quite different from those of the other sources;
i.e., the intensities are weaker and the velocity widths are broader ($dv$$\sim$2.6~km~s$^{-1}$) than in the other sources.
The C$_{2}$H emission toward NGC~1333-1 and NGC~1333-16 may be affected by the protostellar activities within the cores (e.g., molecular outflows). 
For the other sources, the velocity widths of the C$_{2}$H emission mostly ranges from 0.6 to 1.5~km~s$^{-1}$,
indicating that the C$_{2}$H emission would mainly originate from protostellar envelopes or 
cavity walls of low-velocity outflows rather than main bodies of molecular outflows (e.g., Sakai et al. 2014; Oya et al. 2014).

The c-C$_{3}$H$_{2}$ (3$_{2,1}$--2$_{1,2}$) line at 244~GHz was detected toward 30 sources, as shown in Figure~\ref{fg11}.
The line parameters obtained with the Gaussian fit are summarized in Table~\ref{t7}.
For this line, the upper-state energy is 18~K, and the critical density is 10$^{6}$~cm$^{-3}$.
Hence, this line trace a moderately dense region.
c-C$_{3}$H$_{2}$ is a carbon-chain related molecule, and traces the protostellar envelope as C$_{2}$H 
(Sakai et al. 2010; Sakai et al. 2014; Yoshida et al. 2015).
The results show that the intensity of the 3$_{2,1}$--2$_{1,2}$ line differs from source to source.
The velocity widths of the c-C$_{3}$H$_{2}$ line are similar to those of C$_{2}$H.
Therefore, the c-C$_{3}$H$_{2}$ and C$_{2}$H emission likely comes from almost the same region in each source.

\subsection{Correlation of integrated intensities between C$_{2}$H and CH$_{3}$OH}

\begin{figure}
\epsscale{1}
\plotone{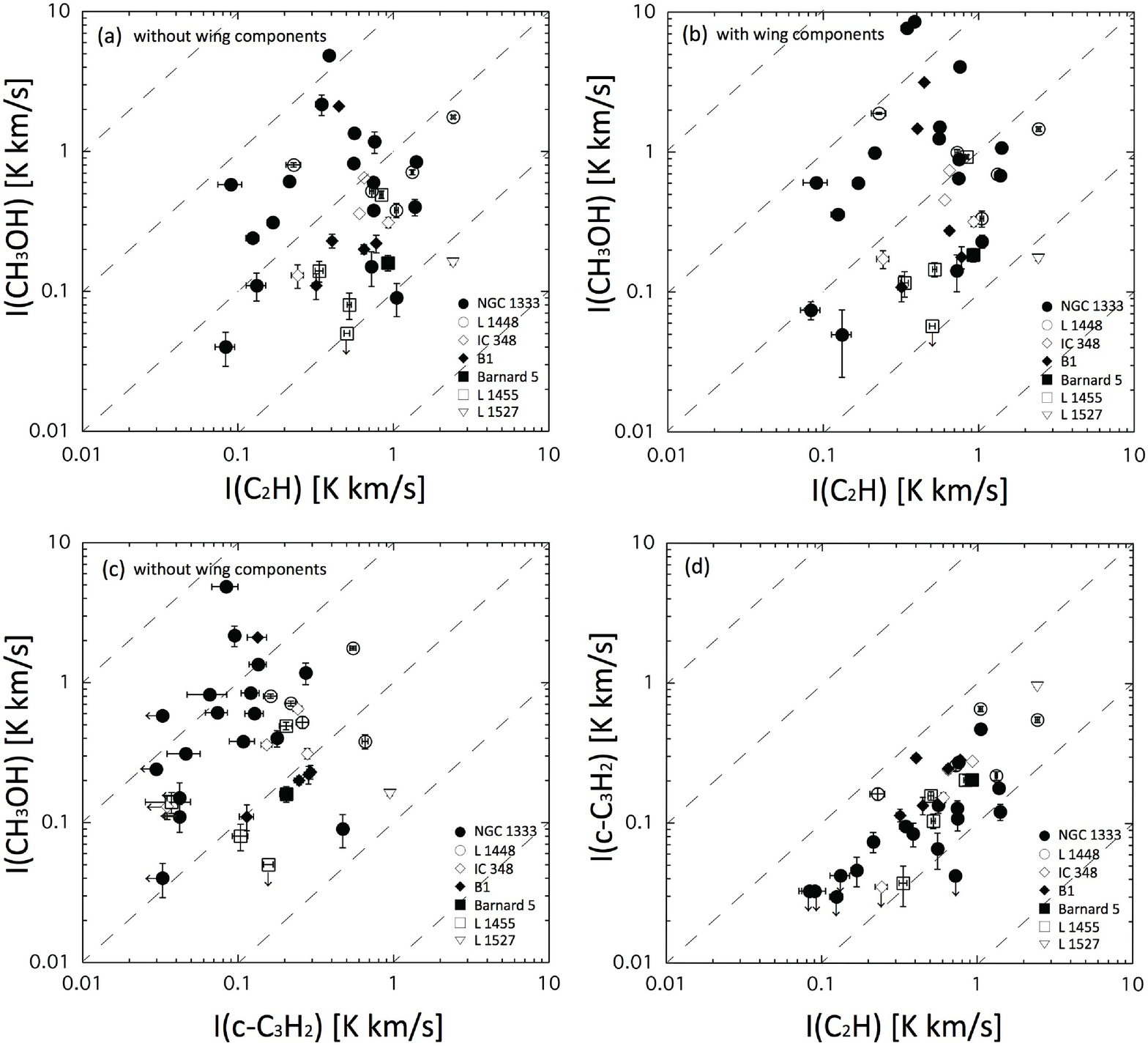}
\caption{
(a) The correlation plot between the integrated intensities of the C$_{2}$H ($N$=3--2, $J$=5/2--3/2, $F$=3--2) 
and CH$_{3}$OH ($J$=5--4, $K$=1, E$^{-}$) lines excluding wing components for the latter (correlation coefficient = 0.04). 
The dashed lines indicate the intensity ratios of 100, 10, 1, 0.1, and 0.01.
(b) The correlation plot between the integrated intensities of the C$_{2}$H ($N$=3--2, $J$=5/2--3/2, $F$=3--2) 
and CH$_{3}$OH ($J$=5--4, $K$=1, E$^{-}$) lines including the wing components for the latter (correlation coefficient = 0.04).
The dashed lines indicate the intensity ratios of 100, 10, 1, 0.1, and 0.01.
(c) The correlation plot between the integrated intensities of the c-C$_{3}$H$_{2}$ (3$_{2,1}$--2$_{1,2}$) 
and CH$_{3}$OH ($J$=5--4, $K$=1, E$^{-}$) lines (correlation coefficient = 0.17).
The dashed lines indicate the intensity ratios of 100, 10, 1, 0.1, and 0.01.
(d) The correlation plot between the integrated intensities of the C$_{2}$H ($N$=3--2, $J$=5/2--3/2, $F$=3--2) 
and c-C$_{3}$H$_{2}$ (3$_{2,1}$--2$_{1,2}$) lines (correlation coefficient = 0.75).
The dashed lines indicate the intensity ratios of 100, 10, 1, 0.1, and 0.01.
Data points with an arrow in CH$_{3}$OH and c-C$_{3}$H$_{2}$ intensities show the upper limits.
The correlation coefficient is defined as ${\Sigma}(x_{i}-\bar{x})(y_{i}-\bar{y})/\sqrt{{\Sigma}(x_{i}-\bar{x})^{2}{\Sigma}(y_{i}-\bar{y})^{2}}$,
where $\bar{x}$ and $\bar{y}$ represent the average values of the data $x_{i}$ and $y_{i}$, respectively.}
\label{fg2}
\end{figure}

A correlation plot between the integrated intensities of the C$_{2}$H ($N$=3--2, $J$=5/2--3/2, $F$=3--2, $E_{\rm{u}}$=25~K) 
and CH$_{3}$OH ($J$=5--4, $K$=1, E$^{-}$, $E_{\rm{u}}$=40~K) 
lines is then prepared to understand how the intensity ratios differ among the observed sources.
We employ the third weakest hyperfine component of C$_{2}$H in Table \ref{t2} in order to avoid the possible saturation 
effect as much as possible.  
Since a broad wing components of the CH$_{3}$OH lines would likely originate from outflow shocks,  
we need to exclude it to discuss the chemical composition of protostellar envelopes.
For this purpose, the C$_{2}$H velocity width (a full width at a half maximum: FWHM) 
is employed as the velocity range for the integrated intensities of the CH$_{3}$OH lines.
We use this simple procedure, because fitting by a double (or multiple) Gaussian function does not always work due to asymmetric line profiles.
The result is shown in Figure \ref{fg2}(a).
The intensities vary over one or two orders of magnitude among the sources, and no correlation can be seen between the C$_{2}$H and CH$_{3}$OH intensities.
Indeed, the correlation coefficient is 0.04 for Figure~\ref{fg2}(a), where the upper limits are not involved in the calculation of the correlation coefficient. 
The C$_{2}$H/CH$_{3}$OH integrated intensity ratio differs at most by a factor of 100.
Even if we focus on only the sources in the NGC~1333 cloud, the correlation plot still shows a large scatter (Figures~\ref{fg2}~(a)).

For references, we prepare the same plot by using the integrated intensities of CH$_{3}$OH including the wing components,
as shown in Figure \ref{fg2}(b).
The plots with and without the wing component of CH$_{3}$OH do not differ from each other as a whole.
In general, CH$_{3}$OH is not only abundant in hot inner envelopes but also in outflow-shocked regions (Bachiller et al. 1998).
CH$_{3}$OH is formed through hydrogenation of CO depleted on grain mantle in a cold starless phase 
(e.g., Tielens $\&$ Hagen 1982; Watanabe $\&$ Kouchi 2002; Soma et al. 2015), and is liberated into the gas phase in hot regions ($T$ $>$ 100~K) or in 
outflow-shocked regions (e.g., Bachiller \& P{\'e}rez Guti{\'e}rrez 1997; Saruwatari et al. 2011).
Furthermore, it can also be liberated even in cold regions to some extent through non-thermal desorption processes 
(e.g., Bizzocchi et al. 2014; Soma et al. 2015; Spezzano et al. 2016). 
For this reason, abundant CH$_{3}$OH in the gas phase means abundant CH$_{3}$OH in grain mantle just before the onset of star formation, 
whatever its liberation mechanism is.
Conversely, CH$_{3}$OH cannot be abundant in the gas phase, if it is deficient on grain mantle. 
Indeed, the CH$_{3}$OH emission is faint in the WCCC source, L1527, even for the outflow components (e.g., Sakai et al. 2014; Takakuwa et al. 2000).
Hence, the inclusion of the wing components originating from the outflow-shocked regions in the
integrated intensity of CH$_{3}$OH will not seriously affect the trend that CH$_{3}$OH is abundant in the source.
However, we use the integrated intensity without the wing components in the following discussion for fair comparison, as stated above.

In addition, the correlation plot of integrated intensities of the CH$_{3}$OH and c-C$_{3}$H$_{2}$ (3$_{2,1}$--2$_{1,2}$, $E_{\rm{u}}$=18~K) 
lines is shown (see Figure~\ref{fg2} (c)). 
No correlation can be found in this  plot as in the correlation plot between the integrated intensities of the C$_{2}$H and CH$_{3}$OH.
The correlation coefficient is 0.04.

In contrast, the integrated intensities of the C$_{2}$H ($N$=3--2, $J$=5/2--3/2, 
$F$=3--2, $E_{\rm{u}}$=25~K) and c-C$_{3}$H$_{2}$ (3$_{2,1}$--2$_{1,2}$, $E_{\rm{u}}$=18~K) lines are correlated with each other 
(see Figure~\ref{fg2} (d)).
The correlation coefficient is 0.75, where the upper limit values are not included.
Although C$_{2}$H is thought to be the photodissociation region (PDR) tracer (e.g., Cuadrado et al. 2015), 
the clear correlation between C$_{2}$H and c-C$_{3}$H$_{2}$ implies that the C$_{2}$H lines trace the dense core rather than the PDRs in this study 
(See Section 4.1).

The correlation of C$_{2}$H and c-C$_{3}$H$_{2}$ has been reported for diffuse clouds and photodissociation regions (e.g., Gerin et al. 2011; Guzman et al. 2015). 
In addition, C$_{2}$H and c-C$_{3}$H$_{2}$ exist in dense clouds including starless cores and protostellar cores 
(e.g., Sakai et al. 2008; Koumpia et al. 2016; Koumpia et al. 2017; Sakai et al. 2014b).
Since we observed the high excitation lines of C$_{2}$H and c-C$_{3}$H$_{2}$ toward the protostellar cores at almost the same distance, 
our result suggests that the correlation between the two lines holds in protostellar cores, as expected from carbon chemistry in dense clouds 
(Sakai $\&$ Yamamoto 2013; Yamamoto 2017).
In contrast, Fontani et al. (2012) present anti-correlation between the two molecules, 
C$_{2}$H and c-C$_{3}$H$_{2}$ in a UC~HII region, suggesting a complex physical structure of the source.

\subsection{Derivation of rotation temperatures and column densities}

\begin{figure}
\epsscale{1}
\plotone{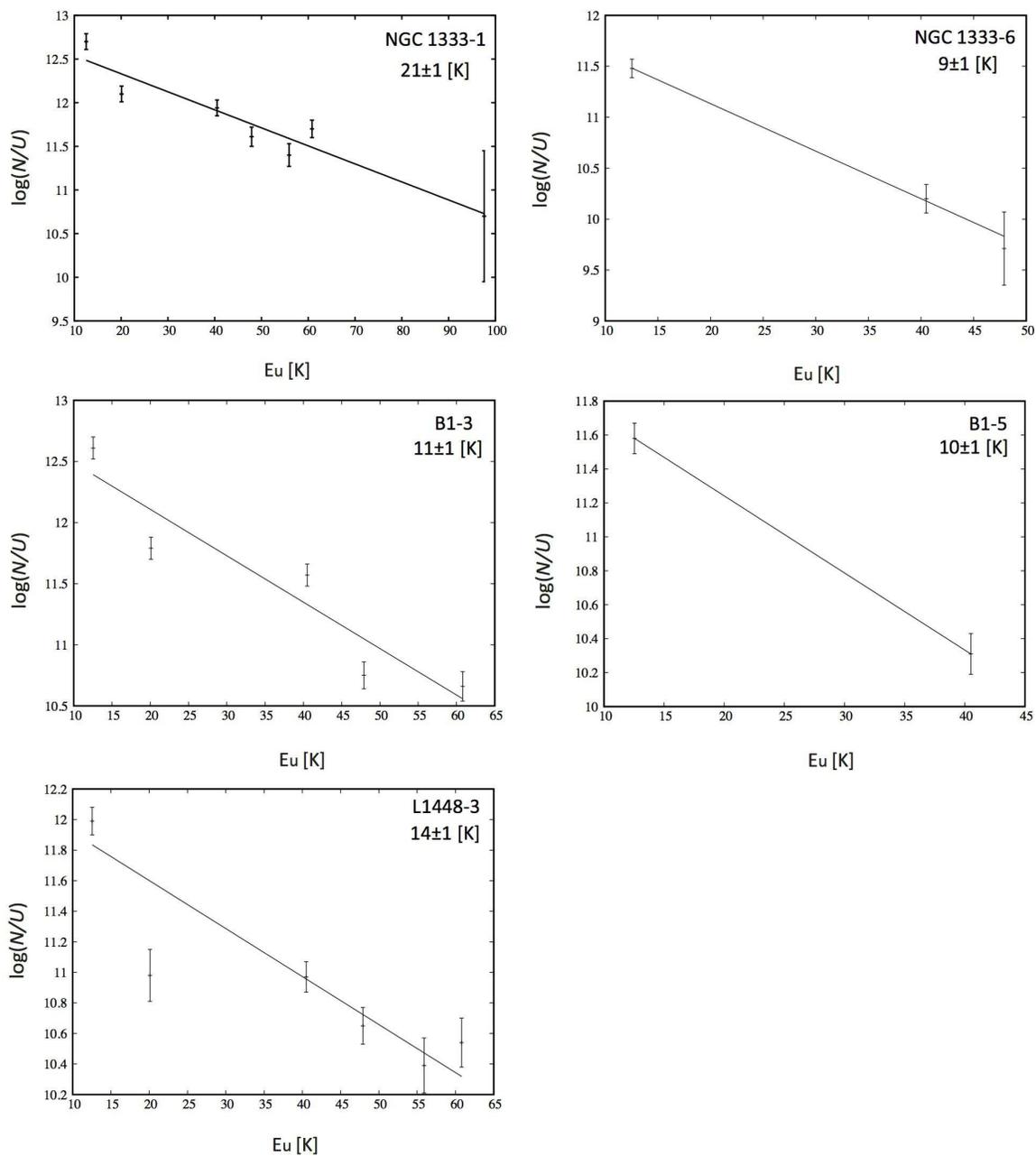}
\caption{Rotation diagrams of CH$_{3}$OH for NGC~1333-1, NGC~1333-6, B1-3, B1-5 and L1448-3.}
\label{fg3}
\end{figure}

To investigate the chemical diversity, we derive the beam-averaged column densities of CH$_{3}$OH, C$_{2}$H and c-C$_{3}$H$_{2}$
under the assumption of local thermodynamic equilibrium (LTE) conditions.
The rotation temperature of CH$_{3}$OH ($E$ state) was evaluated from the multiple transition lines with different upper-state energies,
where a rotational diagram method assuming optically thin emission is used (e.g., Goldsmith $\&$ Langer; Watanabe et al. 2015).
We used the $E$ state lines, because more lines are available than for the $A$ state.
Examples of the rotation diagrams prepared in our analyses are shown in Figure~\ref{fg3} (also see Figure \ref{fg10}).

The derived rotation temperature ranges from 8 to 21~K.
The rotation temperatures of CH$_{3}$OH derived for NGC~1333-1 (IRAS~4B), NGC~1333-2 (IRAS~2), 
L1448-2, and L1448-3 by using the K structure lines of the $J$=5--4 and $J$=7--6 transition are lower 
by 6 -- 82~K than those reported by Maret et al. (2005).
Maret et al. (2005) only employed the high excitation lines ($J$=5--4 and $J$=7--6) of CH$_{3}$OH in thier analysis.  
In contrast, we employ the $J$=2--1 lines instead of higher excitation lines, which would likely trace CH$_{3}$OH 
not only in a warm and dense part, but also in a colder envelope part.  
This seems to be a reason for the lower temperature obtained in our study.

\begin{figure}
\epsscale{0.5}
\plotone{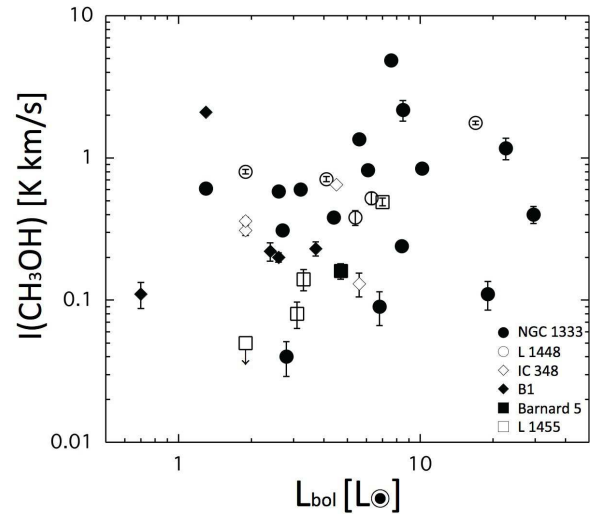}
\caption{The correlation plot between the integrated intensities of  CH$_{3}$OH ($J$=5--4, $K$=1, E$^{-}$) lines excluding wing components 
and the protostellar luminosity (correlation coefficient = 0.19). 
The data point with arrowed line in the CH$_{3}$OH intensity shows the upper limit.
The correlation coefficient is defined in the caption in Fugure \ref{fg2}.}
\label{fg33}
\end{figure}

It should be noted that we are observing the CH$_{3}$OH emission in the protostellar envelope. 
One may think that the CH$_{3}$OH emission mainly comes from the small hot region ($\sim$100~K) near the protostar.  
However, its contribution may not be dominant in our observation, because the rotation temperature is as low as 8 -- 21~K.
The rotation temperature of 8 -- 21~K are low to trace hot corinos even considering that CH$_{3}$OH emission is sub-thematically excited 
(e.g., Bachiller et al. 1998 and references therein).
Moreover, we do not find any correlation between the CH$_{3}$OH intensity and the protostellar luminosity (Figure \ref{fg33});
higher luminosity sources do not always give stronger CH$_{3}$OH emission in our observation.

The rotation temperature derived by the rotation diagram analysis depends on the assumed source size.
If the source size is smaller than the observation beams both for $J$=2--1 (97~GHz) and $J$=5--4 (242~GHz) lines, 
the rotation temperature derived above would be lower than our estimate, because the beam dilution effect is larger for the 97~GHz observation.
If the emitting region is smaller for the 242~GHz line than for the 97~GHz line due to the higher critical density, 
the beam dilution effect can be larger for the 242~GHz line.
In this case, the rotation temperature would be higher than our estimate.
Although these two situations may be the case, we do not know the internal structure within the observation beam for individual sources.  
When the two above limitations are considered, assuming that the beam filling factor is unity provides a moderate estimate of the rotation temperature.
Moreover, we discuss the results within our sample sources in the following sections, which are almost equally distant from the Sun.
Thus the systematic errors due to the beam dilution effect would be mitigated to some extent for the column density ratios, which are mainly used in our discussions.

By assuming that the abundance of the $A$ state is the same as that of the $E$ state, 
the total beam-averaged column density of CH$_{3}$OH for each source is determined from the integrated intensities 
within the velocity range of the C$_{2}$H line averaged for the four hyperfine components in order to eliminate the outflow component as much as possible.
Since the actual source size is unknown for most of the sources, the beam filling factor of unity is used for simplicity, as discussed above.
Thus the beam-averaged column density is derived in this study.
Here, uncertainties of the derived column densities are evaluated from the rms noise. 
For C$_{2}$H and c-C$_{3}$H$_{2}$, we used the $N$=3--2, $J$=5/2--3/2, $F$=3--2 line and the 3$_{2,1}$--2$_{1,2}$ line to 
derive the beam-averaged column density, respectively,
where we assume the rotation temperature of CH$_{3}$OH derived for each sources.
Note that for NGC1333-16 and NGC1333-17, the rotation temperatures of CH$_{3}$OH for the nearby sources, 
NGC1333-1 and NGC1333-2, are employed, respectively, because of the lack of the Nobeyama data.
Table \ref{t3} summarizes the derived beam-averaged column densities.

NGC~1333-1 (IRAS~4B), NGC~1333-16 (IRAS~4A), NGC~1333-7, and B1-3 show high CH$_{3}$OH column densities.
On the other hand, L1448-2 shows the highest column densities of C$_{2}$H and c-C$_{3}$H$_{2}$, 
which are about an order of magnitude higher than those in NGC~1333-1, and about a half of those in L1527 (Sakai et al. 2008).

\section{Discussion} \label{dis}
\subsection{Chemical variation between C$_{2}$H and CH$_{3}$OH}

In this section, we use the beam-averaged column densities summarized in Table \ref{t3} to characterize the chemical composition 
of protostellar sources at a few 1000~au scale. Smaller scale chemical variation is averaged out, and hence, a spatial attention 
is needed when they are compared with the column densities in other studies at a higher angular resolution or with source-size corrections.
Nevertheless, the beam-averaged column densities can be used for mutual comparison among our samples in the Perseus molecular cloud
complex, because the sources are at the similar distances and the column densities are derived in a uniform way.

\begin{figure}
\epsscale{1}
\plotone{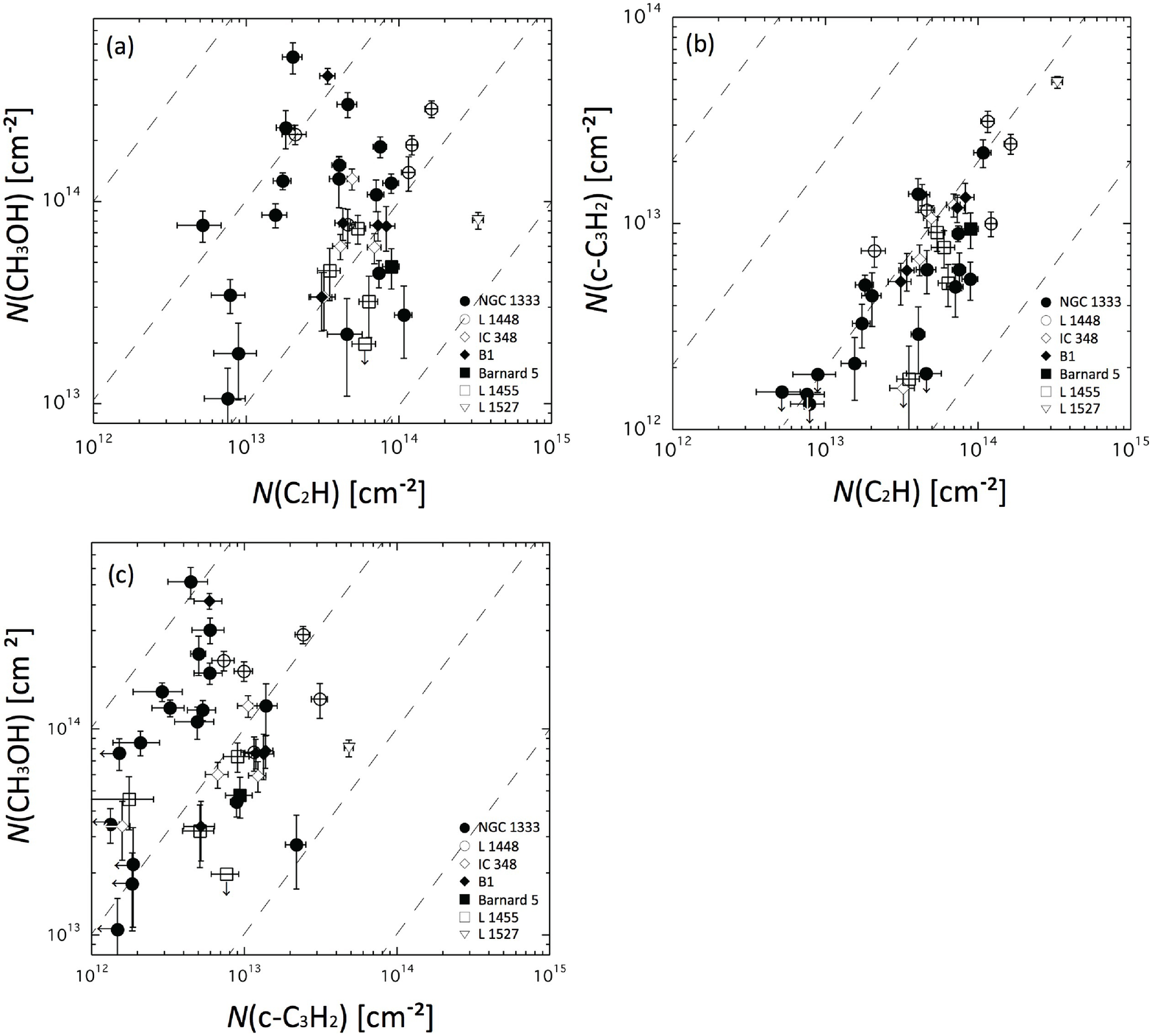}
\caption{(a) The correlation plot between the column densities of C$_{2}$H and CH$_{3}$OH (correlation coefficient = 0.01). 
The dashed lines indicate intensity ratios of 100, 10, 1, and 0.1.
(b) The correlation plot between the column densities of C$_{2}$H and c-C$_{3}$H$_{2}$ (correlation coefficient = 0.88).
The dashed lines indicate intensity ratios of 100, 10, 1, and 0.1.
(c) The correlation plot between the column densities of c-C$_{3}$H$_{2}$ and CH$_{3}$OH (correlation coefficient = 0.1).
The dashed lines indicate intensity ratios of 100, 10, 1, and 0.1.
Data points with an arrow in CH$_{3}$OH and c-C$_{3}$H$_{2}$ column densities show the upper limits.
The correlation coefficient is defined in the caption in Fugure \ref{fg2}.}
\label{fg4}
\end{figure}

Figure~\ref{fg4}(a) shows the correlation plot of the beam-averaged column densities between C$_{2}$H and CH$_{3}$OH, 
while Figure~\ref{fg4}(b) depicts those between C$_{2}$H and c-C$_{3}$H$_{2}$.
The column densities of C$_{2}$H and c-C$_{3}$H$_{2}$ correlate with each other, because these two molecules 
would be produced in related pathways (Sakai $\&$ Yamamoto 2013).
In contrast, no correlation can be seen between C$_{2}$H and CH$_{3}$OH in the column densities, 
as in the integrated intensities (Figures~\ref{fg2}~(a) and (b)).
Similarly, no correlation is found between c-C$_{3}$H$_{2}$ and CH$_{3}$OH, either, as shown in Figure~\ref{fg2} (c).

These results clearly indicate the chemical diversity at a few 1000~au scale around the protostar.
The C$_{2}$H/CH$_{3}$OH ratios range over almost two orders of magnitude even among the Class 0/I sources within the same molecular cloud complex.
A similar diversity of the C$_{4}$H/CH$_{3}$OH ratio among various protostellar sources is also reported by 
Graninger et al. (2016) and Lindberg et al. (2016).
However, our result is the first one based on the unbiased samples in the single molecular cloud complex.

It is likely that the above chemical diversity at a few 1000~au scale is related to the chemical diversity 
identified at a smaller scale; namely hot corino chemistry and WCCC.
As discussed by Sakai et al. (2009) and Sakai $\&$ Yamamoto (2013), definitive identification of hot corino chemistry and WCCC
requires the confirmation of the central concentration of COMs and carbon-chain molecules, respectively.
Such a concentration is not confirmed for C$_2$H, c-C$_{3}$H$_{2}$, and CH$_{3}$OH in our single-point observations.
Nevertheless, it is likely that the beam-averaged chemical composition does reflect the chemical composition of 
the protostellar core to some extent, because the high excitation lines of CH$_{3}$OH, C$_{2}$H, and c-C$_{3}$H$_{2}$ are 
employed in this study to trace dense regions rather than the component extended over parent molecular clouds.

For NGC~1333-1 (IRAS~4B) and NGC~1333-16 (IRAS~4A), which have previously been identified as hot corino sources, 
we indeed see abundant CH$_{3}$OH but deficient C$_{2}$H, their C$_{2}$H/CH$_{3}$OH column density ratios being almost the lowest among the observed sources (Table \ref{t3}).
On the other hand, the WCCC source L1527 in Taurus, which is employed as a reference, 
shows abundant C$_{2}$H and deficient CH$_{3}$OH with the single dish observation in the ASAI project (Lefloch et al. 2018), as shown in Figure \ref{fg5}(a).
Importantly, most of the sources show the ratios between those of the hot corino sources and the WCCC source.  
Thus, the hot corino sources and the WCCC source are certainly the two extreme cases of chemical variation.

Sakai et al. (2009) and Sakai $\&$ Yamamoto (2013) suggested that the difference of the chemical composition found between 
WCCC sources and hot corino sources could originate from the different duration time of the starless phase after shielding 
of the interstellar UV radiation in the parent molecular cloud.
For efficient formation of various COMs, a significant amount of CH$_{3}$OH is necessary as a parent molecule (e.g., Garrod $\&$ Herbst 2006).
CH$_{3}$OH is mainly formed by hydrogenation of CO on grain mantle (e.g., Tielens $\&$ Hagen 1982; Watanabe $\&$ Kouchi 2002; Soma et al. 2015).  
On the other hand, it takes about 10$^{5-6}$ years for the formation of CH$_{3}$OH from CO, 
if the H$_{2}$ density of the parent cloud is as high as 10$^{5}$~cm$^{-3}$ (Taquet et al. 2012).
When the core collapse starts well after the shielding of the interstellar UV radiation in the parent cloud, 
most of the carbon atoms are fixed into CO by gas-phase reactions, and CO is depleted onto dust grains in dense and cold regions to form CH$_{3}$OH.

On the other hand, there is not enough time for the C to CO conversion in the gas phase, if the core collapse starts just after the UV shielding.
Carbon atom can still be abundant in such cores.
The timescale for the depletion of atoms and molecules onto dust grains is roughly 10$^{5}$/($n$/10$^{4}$~cm$^{-3}$)~yr 
(e.g., Burke $\&$ Hollenbach 1983), which is comparable to the dynamical timescale.
Hence, carbon atom is depleted onto dust grains before it is converted to CO by gas phase reactions. 
Hydrogenation of C on grain surface forms CH$_{4}$ efficiently.
After the onset of star formation, CH$_{4}$ is evaporated into the gas phase in a warm region (T $>$ 25 K) to 
form various carbon-chain molecules through gas-phase reactions (i.e., the WCCC mechanism).
In this case, the core collapse occurs in chemically young clouds, and hence, carbon-chain molecules produced in the early evolutionary stage
of cold starless cores would also survive in the gas phase to some extent.

In L1527, a cold envelope with abundant carbon-chain molecules surrounds the dense 
($n$ $\sim$ 10$^{6}$~cm$^{-3}$) and warm ($T$ $>$ 25 K) core where the enhancement of carbon-chain molecules due to WCCC is occurring (Sakai et al. 2010). 
In contrast, carbon-chain molecules are relatively deficient in hot corino sources even in the surrounding component (e.g., Sakai $\&$ Yamamoto 2013).
Therefore, the ratios between carbon-chain molecules, especially C$_{\rm{n}}$H and C$_{\rm{n}}$H$_{2}$, and CH$_{3}$OH are
expected to represent chemical characteristics of protostellar cores,
even if a part of the observed emission comes from outflows or cold dense envelopes in addition to that from an inner part of protostellar cores.

It should be noted that C$_{2}$H is known to be abundant in PDRs (e.g., Pety et al. 2007; Cuadrado et al. 2015).  
In such regions, C$_{2}$H is efficiently produced from C$^{+}$ in the gas-phase reaction and/or is 
formed by destruction of very small grains and polycyclic aromatic hydrocarbons (PAHs) (e.g., Cuadrado et al. 2015).
However, we observed the regions with high extinction (A$_{\rm{v}}$ $>$ 5--7 mag; Kirk et al. 2006) in relatively high critical density lines.  
Hence, the detected C$_{2}$H emissions would mainly originate from protostellar cores rather than surrounding diffuse parts to which the UV radiation can well penetrate.
Nevertheless, C$_{2}$H would also exist in the cavity walls of low-velocity outflows, 
where C$_{2}$H could be formed by the UV radiation from the central protostar (e.g., Oya et al. 2014).
To assess this effect, we also observed the c-C$_{3}$H$_{2}$ lines. 
As demonstrated in Figure~\ref{fg4} (b), the column densities of C$_{2}$H and c-C$_{3}$H$_{2}$ show a good correlation.
The C$_{2}$H/c-C$_{3}$H$_{2}$ ratio is about 10, which is lower than those found in PDRs and diffuse clouds and is rather close to those of dense cores 
(Gerin et al. 2011; Cuadrado et al. 2015).  
In fact, Lindberg et al. (2015) also reported that c-C$_{3}$H$_{2}$ is not affected by the UV radiation in the R CrA region.
Although c-C$_{3}$H$_{2}$ is detected in the outflow-shocked region L1157 B1 (Yamaguchi et al. 2012), 
its abundance is not as high as that in the protostellar core of L1157 mm (Bachiller $\&$ Perez Gutierezz 1997).
Above all, the diversity seen in the C$_{2}$H/CH$_{3}$OH column density ratio most likely reflects 
the chemical diversity of protostellar cores.

\subsection{Effect from evolutionary stage of the source}

Figures~\ref{fg5}(a) (b) and (c) show the correlation plots of the C$_{2}$H/CH$_{3}$OH ratio against the envelope mass $M_{\rm{env}}$,  
the ratio of the bolometric luminosity to the sub-mm wavelength luminosity $L_{\rm{bol}}$/$L_{\rm{smm}}$,
and the bolometric temperature $T_{\rm{bol}}$, respectively.
The envelope mass ($M_{\rm{env}}$) represents the amount of the gas associated with the protostar, whereas
$L_{\rm{bol}}$/$L_{\rm{smm}}$ and $T_{\rm{bol}}$ are known as an evolutionary indicator of the 
protostellar source (Hatchell et al. 2005; 2007).

In Figure~\ref{fg5}(a), the $M_{\rm{env}}$ values are taken from Hatchell et al. (2007).
It is derived from the 850~$\micron$ dust continuum flux observed with JCMT 
which has a beam size similar to that of our IRAM~30~m observations in the 1.3~mm band ($\sim$ 15\arcsec).
The correlation coefficient is 0.02, indicating no correlation. 
Hence, the result indicates that there is no clear relationship between the chemical composition and the mass to be accreted.

Likewise, $L_{\rm{bol}}$/$L_{\rm{smm}}$ and $T_{\rm{bol}}$ do not show a correlation with 
the C$_{2}$H/CH$_{3}$OH ratio either (Figure~\ref{fg5}~(b) and (c)), where $L_{\rm{bol}}$/$L_{\rm{smm}}$ and $T_{\rm{bol}}$ are also taken from Hatchell et al. (2007). 
The correlation coefficients for the $L_{\rm{bol}}$/$L_{\rm{smm}}$ and $T_{\rm{bol}}$ plots are as small as 0.26 and 0.31, respectively.
Thus, the evolutionary stage does not show correlation with the C$_{2}$H/CH$_{3}$OH ratio.
This is because the chemical composition of the grain mantles, which characterizes the gas-phase chemical composition 
after the onset of star formation has already been determined 
by the processes during the starless core phase.

\subsection{Positional effect in the parent cloud}

\begin{figure}
\epsscale{0.8}
\plotone{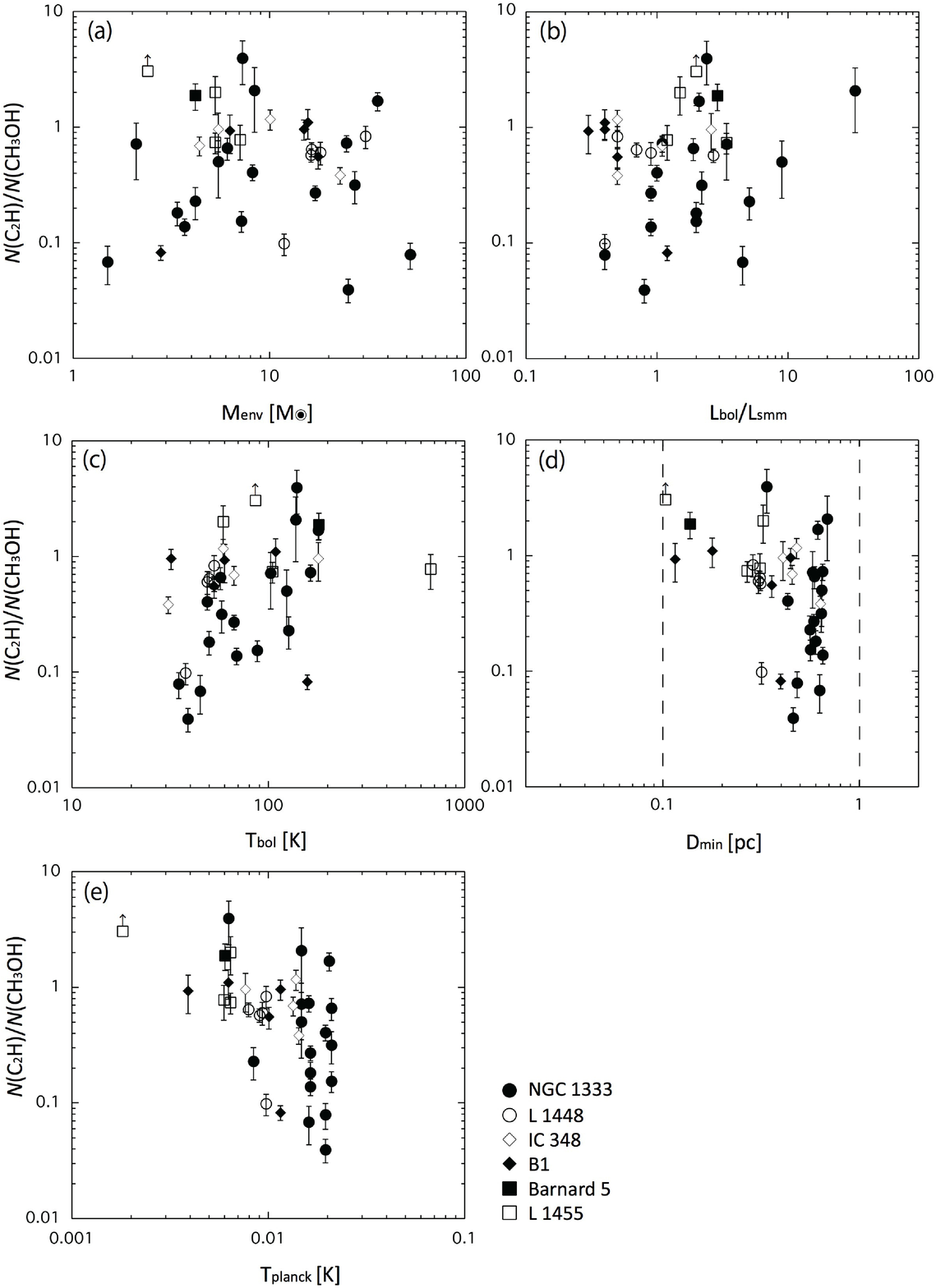}
\caption{The correlation plot between the C$_{2}$H/CH$_{3}$OH ratio vs. 
(a) envelope mass $M_{\rm{env}}$(correlation coefficient = 0.02), 
(b) $L_{\rm{bol}}$/$L_{\rm{smm}}$ (correlation coefficient = 0.26),
(c) $T_{\rm{bol}}$ (correlation coefficient = 0.31),
(d) $D_{\rm{min}}$, the minimum distance from the source position and cloud edge (correlation coefficient= 0.27), and 
(e) the peak intensity of the source position from the Planck map (correlation coefficient= 0.42). 
A dashed line of 0.1~pc shows the spatial resolution of observations, and the line of 1~pc shows the largest cloud-size for (d).
Data points with an arrow show the lower limits.
The correlation coefficient is defined in the caption in Fugure \ref{fg2}.
The correlation coefficient is derived excluding the lower limit data. 
If we include the data point, the correlation coefficient is changed to be 0.12 for (a), 0.13 for (b), 0.28 for (c), 0.45 for (d) and 0.58 for (e), respectively.}
\label{fg5}
\end{figure}

\begin{figure}
\rotate
\epsscale{0.73}
\plotone{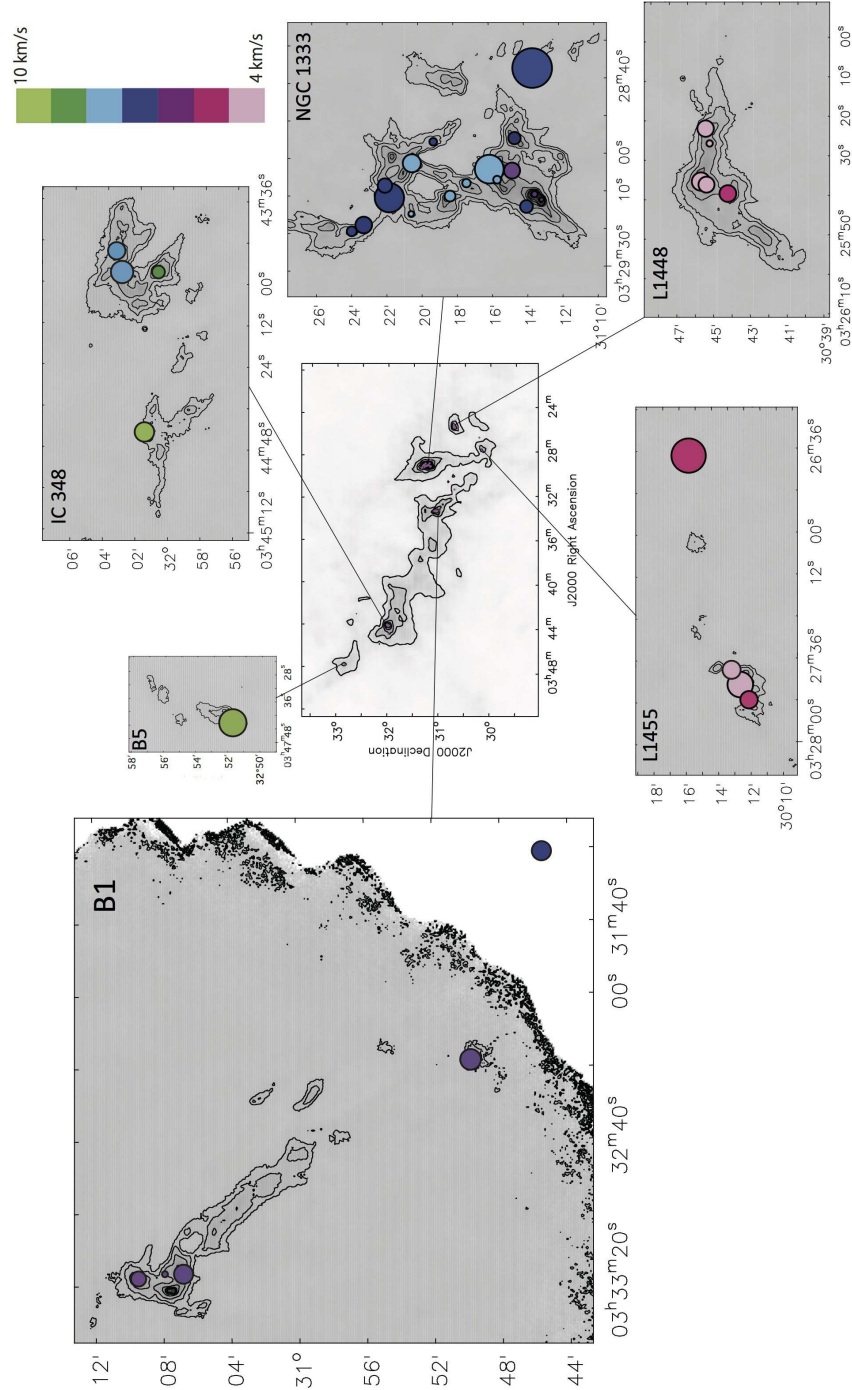}
\caption{
Planck 217~GHz map (central part; contours are 5, 10, 15, 20, 25, 30, 35, and 40 $\sigma$ levels, 1~$\sigma$=0.5~mK) 
and SCUBA 850$\micron$ maps of the NGC~1333, L1448, IC~348, B5, B1, and L1455 regions. 
Greyscale is 850$\micron$ continuum.
The flux is shown on a log scale from $\-$0.5~Jy~beam$^{-1}$ to 1.5~Jy~beam$^{-1}$.
Contours are 5, 10, 20, 40, 60, 80, 100, 200, and 400 $\sigma$ levels (1~$\sigma$=0.01~Jy~beam$^{-1}$).
The radius of each circle overlaid on the JCMT~850~$\mu$m images of the Perseus clouds (Chen et al. 2016) 
is proportional to the C$_{2}$H/CH$_{3}$OH ratio.
Color shows the systemic velocity of the individual position.}
\label{fg6}
\end{figure}

To test the relation between the positions of protostars and the C$_{2}$H/CH$_{3}$OH ratios, 
we introduce the minimum projected distance from the molecular cloud edge, $D_{\rm{min}}$, i.e., small $D_{\rm{min}}$ indicates a source closer to the edge of the cloud.
$D_{\rm{min}}$ is calculated by using the Planck 217~GHz continuum map (see Figure~\ref{fg6}), where the cloud edge is arbitrarily defined as the 10~$\sigma$ contour.
Note that the 10~$\sigma$ contour of the Planck continuum map, which we employ as the edge of the cloud, 
corresponds to A$_{\rm{v}}$  of 1~mag (Kirk et al. 2006).
Figure~\ref{fg5} (d) shows the correlation plot of the C$_{2}$H/CH$_{3}$OH ratio against $D_{\rm{min}}$. 
Note that a typical size of the cloud is 1~pc (e.g., Hacar et al. 2017), and hence, there is a sharp limit of $D_{\rm{min}}$ around 0.6--0.7~pc.

$D_{\rm{min}}$ is a projected distance, and the line-of-sight depth from the cloud surface is unknown.
Hence, sources that appear near the center of the cloud are not always embedded deeply in the clouds, 
but some of them may be close to the cloud periphery along the line of sight.
Although $D_{\rm{min}}$ is affected by this projection effect,  we can obtain clues of the origin of the chemical diversity from Figure \ref{fg5}(d).
These results suggest that the sources with lower ratios tend to appear only at larger $D_{\rm{min}}$.
Namely, they are likely more embedded in the central part of the large molecular clouds.

Figure \ref{fg5}(d) shows a kind of ``right angle" distribution of the points: sources with small $D_{\rm{min}}$ have only high C$_{2}$H/CH$_{3}$OH ratios.
On the other hand, sources with large $D_{\rm{min}}$ have both high and low C$_{2}$H/CH$_{3}$OH ratios, indicating a large scatter of the ratios.
Although this scatter mainly comes from the NGC~1333 region, the similar trend can be seen in the plots without the NGC~1333 sources.
It should be stressed that there is a blank area in the bottom-left corner, where $D_{\rm{min}}$ is small and the ratio is low.

One may think that this is caused by the source selection effect.
In our source sample, the very low luminosity ($L_{\rm{bol}}$ $<$ 1~$L_{\odot}$) sources are missing.
If these sources had a low C$_{2}$H/CH$_{3}$OH ratio, the blank area might be filled up.
However, this would not be the case, because there is no correlation between the luminosity and the C$_{2}$H/CH$_{3}$OH ratio.

Most of the chemical diversity in our sample is thus concentrated in the large $D_{\rm{min}}$ region.
Since all the sources near the cloud edge (small $D_{\rm{min}}$ sources) have the high ratios,
the sources with high ratios at large $D_{\rm{min}}$ may have a small line-of-sight depth from the cloud surface.
Alternatively, substructure (clumpy and/or filamentary structure) in the molecular clouds (e.g., velocity-coherent structures found in NGC~1333 by Hacar et al. 2017)
might also contribute to the high ratios, because such substructures allow the interstellar UV radiation to penetrate into the cloud 
(e.g., Stutzki et al. 1988; Meixner $\&$ Tielens 1993).

In order to investigate the relation between the C$_{2}$H/CH$_{3}$OH ratio and the line-of sight depth of the cloud,
we prepare the correlation of the ratio against the peak intensity of the Planck continuum map at the source positions, as shown in Figure \ref{fg5}(e).
The distribution of the points reveals a trend similar to that found in Figure \ref{fg5}(d).  
Again, there is a blank area in the bottom-left corner, where the peak intensity is low and the ratio is low.
The scatter in the high peak intensity region can be interpreted in the same way for that in the large $D_{\rm{min}}$ case.


In Figure~\ref{fg6},  the C$_{2}$H/CH$_{3}$OH ratio is represented by a radius of the circle, which is overlaid on
the JCMT~850~$\mu$m images of the Perseus clouds (Chen et al. 2016). Indeed, the high C$_{2}$H/CH$_{3}$OH
ratio sources, i.e., the WCCC-type sources (larger circles), seem to be isolated or located at the edge　of the cloud, although NGC~1333-17 (SVS 13A) is an exception. 
In contrast, the low C$_{2}$H/CH$_{3}$OH ratio sources, i.e., sources whose chemical character is close to the hot corino sources, 
tend to be concentrated in the inner regions of the cluster forming clouds (NGC~1333 and IC~348). 
This result implies that the isolated sources and the sources in cloud peripheries tend to have the WCCC character in the Perseus molecular cloud complex.

In the central part of the cloud, the time after the UV shielding would be longer than that in the cloud peripheries, 
because the surrounding gas gradually contracting to the main body of the cloud can shield the UV radiation.
In the cloud center, the carbon atom has well been converted to CO, and CO depletion has already occurred in dense cores.
In this case, CH$_{3}$OH can be abundant on dust grains, which is reflected in the gas-phase abundance through the thermal and/or non-thermal desorption processes.
On the other hand, the carbon atom can be abundant in the cloud peripheries.
It can be depleted directly onto dust grains to form CH$_{4}$, which leads to WCCC.
This picture is consistent with the result that the higher C$_{2}$H/CH$_{3}$OH and c-C$_{3}$H$_{2}$/CH$_{3}$OH ratios are seen 
in protostellar cores at the cloud peripheries, whereas low values only appear in the cores at the cloud center.

However, it is not clear at present whether the regeneration mechanism of carbon-chain molecules expected for WCCC is 
actually occurring in the sources having the high C$_{2}$H/CH$_{3}$OH and c-C$_{3}$H$_{2}$/CH$_{3}$OH ratios.
To assess this point, the central concentration of carbon-chain molecules should be confirmed for each source by high-resolution observations.
Moreover, it has recently been revealed that the situation may be more complicated at a smaller scale ($<$ 100~au).
For instance, the isolated Bok globule B335 shows WCCC in the outer envelope ($\sim$ 1000~au), 
while it harbors a hot corino in the central 10~au region (Imai et al. 2016). 
A similar structure can also be seen in L483 (Oya et al. 2017).  
Therefore, it is not obvious whether high ratios observed at a 1000 au scale is brought into a smaller scale ($<$ 100~au), as they are.  
On the other hand, the WCCC sources L1527 and TMC-1A indeed show faint emission of CH$_{3}$OH even at a 100~au scale.
Although the chemical composition at a 1000 au scale likely affects that at a smaller scale to some extent, the high resolution 
observations are essential to confirm the situation for each source.

\section{CONCLUSIONS}

We present the results of the unbiased survey of the chemical composition toward the 36 Class 0 and Class I protostars 
in the Perseus molecular cloud complex.
The results are summarized as follows:

\begin{enumerate}

\item 
Multiple transition lines of C$_{2}$H, c-C$_{3}$H$_{2}$ and CH$_{3}$OH were detected toward most of the target sources.
The CH$_{3}$OH ($J$=5--4) lines were detected toward 35 sources, 
the CH$_{3}$OH ($J$=2--1) and the C$_{2}$H ($N$=3--2) lines toward all the sources, 
and the c-C$_{3}$H$_{2}$ (3$_{2,1}$--2$_{1,2}$) lines toward 30 sources.

\item The correlation plot between the integrated intensities of the C$_{2}$H and CH$_{3}$OH lines is 
prepared for the two cases with and without the wing components.
In both cases, no correlation is seen between them. Similarly, no correlation is found between the integrated intensities of the c-C$_{3}$H$_{2}$ and CH$_{3}$OH lines. 
In contrast, the integrated intensities of the C$_{2}$H and c-C$_{3}$H$_{2}$ lines show a positive correlation, 
because these two species are thought to be produced through related chemical pathways.

\item The column density ratios of C$_{2}$H/CH$_{3}$OH show a significant diversity by 2 orders of magnitude.
The hot corino sources show the highest ratio, while the WCCC source L1527, 
employed as a reference, shows the lowest ratio.
The ratios of most sources are in between these two distinct cases.

\item 
The C$_{2}$H/CH$_{3}$OH ratio does not correlate with the evolutionary indicators ($L_{\rm{bol}}$/$L_{\rm{smm}}$ and $T_{\rm{bol}}$), nor the envelope mass.  
On the other hand, we find that the isolated sources and the sources located in cloud peripheries tend to have chemical 
characteristics of WCCC (i.e., high C$_{2}$H/CH$_{3}$OH ratios).
In the Perseus molecular cloud complex, the hot corino like sources (i.e., low C$_{2}$H/CH$_{3}$OH ratios)
do not exist in such regions, but are concentrated in the central parts of the cluster forming regions.  
This result is qualitatively consistent with the idea that the chemical diversity would originate from the different 
duration time of the starless core phase after the shielding of the interstellar UV radiation. 

\item 
It is important to study with high angular resolution observations whether the chemical composition seen in this study is brought into inner-envelope/disk system
for each sources.

\end{enumerate}

\bigskip
\acknowledgments
We thank the referee for the thoughtful and constructive comments. 
We are grateful to the staff of NRO and IRAM for excellent supports in observations.  
This study is supported by KAKENHI (25108005 and 16H03964).
\software{NEWSTAR, GILDAS}

\clearpage

\startlongtable
\begin{deluxetable}{c c c c c c c c }
\tabletypesize{\footnotesize}
\tablecaption{Source list \label{t1}}
\tablehead{
\colhead{IDs} & \colhead{Source Names}  & \colhead{R.A.} & \colhead{Decl.} \\
\colhead{} & \colhead{}  & \colhead{(J2000)}& \colhead{(J2000)}}
\startdata
NGC1333-1 & IRAS~4B  & 03:29:12.01 & 31:13:08.2 &  \\
NGC1333-2 & IRAS~2A  & 03:28:55.57 & 31:14:37.1 &   \\
NGC1333-3 & IRAS~6; SK-24  & 03:29:01.66 & 31:20:28.5 \\
NGC1333-4 & IRAS~7; SK-20  & 03:29:10.72 & 31:18:20.5  \\
NGC1333-5 & IRAS~4C; SK-5  & 03:29:13.62 & 31:13:57.9 \\
NGC1333-6 & IRAS~1; SK-6  & 03:28:37.11 & 31:13:28.3   \\
NGC1333-7 & HH7-11 MMS 1; SK-15  & 03:29:06.50 & 31:15:38.6  \\
NGC1333-8 & HH7-11 MMS 6; SK-14  & 03:29:04.09 & 31:14:46.6  \\
NGC1333-9 & SVS~3; SK-28  & 03:29:10.70 & 31:21:45.3  \\
NGC1333-10 & SK-29  & 03:29:07.70 & 31:21:56.8   \\
NGC1333-11 & SK-18  & 03:29:07.10 & 31:17:23.7  \\
NGC1333-12 & SK-32  & 03:29:18.25 & 31:23:16.9    \\
NGC1333-13 &  & 03:29:19.70 & 31:23:56.0  \\
NGC1333-14 & No SMM/MM source & 03:28:56.20 & 31:19:12.5    \\
NGC1333-15 & SK-22  & 03:29:15.30 & 31:20:31.2 \\
NGC1333-16 & IRAS~4A & 03:29:10.53 & 31:13:31.0  \\ 
NGC1333-17 & SVS~13A & 03:29:03.75 & 31:16:03.76  \\
L1448-1 & L1448 NW; IRS3C & 03:25:35.66 & 30:45:34.2     \\
L1448-2 & L1448 NB; IRS3  & 03:25:36.33 & 30:45:14.8     \\
L1448-3 & L1448 MM  & 03:25:38.87 & 30:44:05.3  \\
L1448-4 &  IRS2  & 03:25:22.38 & 30:45:13.3  \\
L1448-5 &  IRS2E & 03:25:25.90 & 30:45:02.7    \\
IC~348-1 & HH211  & 03:43:56.80 & 32:00:50.3  \\
IC~348-2 & IC 348 MMS & 03:43:57.05 & 32:03:05.0 \\
IC~348-3 &  & 03:44:43.32 & 32:01:31.6   \\
IC~348-4 &  & 03:43:50.99 & 32:03:24.7   \\
Barnard 5 & IRS 1 & 03:47:41.61 & 32:51:43.9    \\
B1-1  & B1-c  & 03:33:17.87 & 31:09:32.3  \\
B1-2  & B1-d  & 03:33:16.49 & 31:06:52.3    \\
B1-3  & B1-a  & 03:33:16.67 & 31:07:55.1   \\
B1-4  &  & 03:32:18.03 & 30:49:46.9   \\
B1-5  &  & 03:31:20.94 & 30:45:30.3   \\
L1455-1 & IRAS 03235+3004 & 03:26:37.46 & 30:15:28.2    \\
L1455-2 & IRS1; RNO 15 FIR; IRAS 03245+3002 & 03:27:39.11 & 30:13:02.8   \\
L1455-3 & IRS4  & 03:27:43.25 & 30:12:28.9   \\
L1455-4 & IRS2  & 03:27:47.69 & 30:12:04.4    \\
\hline
\enddata
\tablecomments{
All sources are listed in Hatchell et al. (2005). 
The distance of 235~pc is adopted for all the sources.}
\end{deluxetable}

\clearpage

\begin{deluxetable}{c c c c c c c c c }
\tabletypesize{\footnotesize}
\tablecaption{List of Observed Molecules\label{t2}}
\tablewidth{0pt}
\tablehead{
\colhead{Molecule} & \colhead{Transition} & \colhead{Frequency}  &  \colhead{E$_{u}$} 
& \colhead{S${\mu}^2$}  \\
\colhead{} & \colhead{} & \colhead{[GHz]}& \colhead{[K]} & \colhead{[D$^{2}$]}}
\startdata
C$_{2}$H & $N$=3--2, $J$=5/2--3/2, $F$=2--1 & 262.06746 & 25.2  & 1.1  \\
 & $N$=3--2, $J$=5/2--3/2, $F$=3--2 & 262.06498 & 25.2 & 1.6  \\
 & $N$=3--2, $J$=7/2--5/2, $F$=3--2 & 262.00648 & 25.1  & 1.7  \\
 & $N$=3--2, $J$=7/2--5/2, $F$=4--3 & 262.00426 & 25.2  & 2.3   \\
CH$_{3}$OH & $J$=5--4, $K$=2 E$^{-}$  & 241.90415 & 60.8  & 3.4   \\
		   & $J$=5--4, $K$=2 A$^{+}$  & 241.88770 & 72.6  &  3.4  \\
           & $J$=5--4, $K$=1 E$^{+}$ & 241.87907 & 55.9  & 4.0  \\
           & $J$=5--4, $K$=3 E$^{-}$  & 241.85235 & 97.6  & 2.6   \\
           & $J$=5--4, $K$=2 A$^{-}$  & 241.84232 & 72.5  &  3.4  \\
           & $J$=5--4, $K$=3 A$^{\pm}$ & 241.83291 & 84.7  & 2.6   \\             
           & $J$=5--4, $K$=4 E$^{+}$  & 241.82964 & 130.8  &  1.5  \\
           & $J$=5--4, $K$=4 E$^{-}$  & 241.81325 & 122.7  &  1.4  \\         
           & $J$=5--4, $K$=4 A$^{\pm}$  & 241.80650 & 115.2  & 1.5    \\
           & $J$=5--4, $K$=0 A$^{+}$  & 241.79143 & 34.8  &  4.0   \\
           & $J$=5--4, $K$=1 E$^{-}$  & 241.76722 & 40.4  & 3.9   \\
           & $J$=5--4, $K$=0 E$^{+}$  & 241.70021 & 47.9  & 4.0     \\                   
CH$_{3}$OH 
		   & $J$=2--1, $K$=0 E$^{+}$ & 96.74455 & 20.1  & 1.6  \\
		   & $J$=2--1, $K$=0 A$^{+}$ & 96.74137 & 7.0 & 1.6 \\
           & $J$=2--1, $K$=1 E$^{-}$ & 96.73936 & 12.5 & 1.2   \\
c-C$_{3}$H$_{2}$ & 3$_{2,1}$--2$_{1,2}$ (ortho) & 244.22216 & 18.2 & 7.3 \\
\hline
\enddata
\end{deluxetable}

\startlongtable
\begin{deluxetable}{ccccccccccccc}
\tabletypesize{\footnotesize}
\tablecaption{Physical Parameters\label{t3}}
\tablewidth{0pt}
\tablehead{
\colhead{IDs} & \colhead{$T$$_{\rm{rot}}$} & \colhead{$N$(CH$_{3}$OH)} & 
\colhead{$N$({C$_{2}$H)}} & \colhead{$N$(c-C$_{3}$H$_{2}$)}  & \colhead{$N$({C$_{2}$H})/$N$(CH$_{3}$OH)} & \colhead{$N$(c-C$_{3}$H$_{2}$)/$N$(CH$_{3}$OH)} \\
\colhead{} & \colhead{[K]} & \colhead{[10$^{14}$~cm$^{-2}$]}& \colhead{[10$^{13}$~cm$^{-2}$]}& 
\colhead{[10$^{12}$~cm$^{-2}$]} & \colhead{} & \colhead{}}
\startdata
NGC~1333-1  & 21$\pm$2    & 5.2$\pm$0.9      & 2.0$\pm$0.3	 & 4.5$\pm$1.3    & 0.04$\pm$0.01 & 0.009$\pm$0.003  \\
NGC~1333-2  & 19$\pm$2    & 1.3$\pm$0.4      & 4.1$\pm$0.5	 & 14$\pm$3       & 0.32$\pm$0.10 & 0.11$\pm$0.04   \\
NGC~1333-3  & 14$\pm$1    & 1.2$\pm$0.2      & 9.0$\pm$1.1   & 5.4$\pm$1.1    & 0.72$\pm$0.11 & 0.04$\pm$0.01   \\
NGC~1333-4  & 12$\pm$1    & 1.5$\pm$0.2      & 4.1$\pm$0.4	 & 2.9$\pm$1.0    & 0.27$\pm$0.04 & 0.019$\pm$0.007	\\
NGC~1333-5  & 9$\pm$1     & 1.9$\pm$0.4      & 7.6$\pm$0.8	 & 6.0$\pm$1.2    & 0.41$\pm$0.06 & 0.032$\pm$0.008	\\
NGC~1333-6  & 9$\pm$1     & 0.3$\pm$0.1      & 11$\pm$1	     & 22$\pm$3       & 3.9$\pm$1.6   & 0.80$\pm$0.34	\\
NGC~1333-7  & 11$\pm$1    & 3.0$\pm$0.8      & 4.6$\pm$0.7	 & 6.0$\pm$1.4    & 0.15$\pm$0.03 & 0.020$\pm$0.005	\\	
NGC~1333-8  & 10$\pm$1    & 1.1$\pm$0.4      & 7.1$\pm$0.9	 & 5.0$\pm$1.4    & 0.66$\pm$0.14 & 0.05$\pm$0.02	\\
NGC~1333-9  & 14$\pm$3    & 0.2$\pm$0.1      & 4.6$\pm$1.2	 & $<$ 1.9	      & 2.1$\pm$1.2   & 0.09$\pm$0.05   \\
NGC~1333-10	& 13$\pm$2    & 0.2$\pm$0.07     & 0.9$\pm$0.3	 & $<$ 1.9        & 0.50$\pm$0.26 & 0.11$\pm$0.05	\\
NGC~1333-11 & 10$\pm$1    & 0.9$\pm$0.1      & 1.6$\pm$0.3	 & 2.1$\pm$0.7    & 0.18$\pm$0.04 & 0.024$\pm$0.009	\\
NGC~1333-12 & 10$\pm$1    & 0.1$\pm$0.04     & 0.8$\pm$0.2	 & $<$ 1.5        & 0.72$\pm$0.37 & 0.14$\pm$0.06	\\
NGC~1333-13 & 14$\pm$2    & 0.3$\pm$0.07     & 0.8$\pm$0.2	 & $<$ 1.3        & 0.23$\pm$0.07 & 0.039$\pm$0.009   \\
NGC~1333-14 & 11$\pm$1    & 1.3$\pm$0.1      & 1.7$\pm$0.2	 & 3.3$\pm$0.8    & 0.14$\pm$0.02 & 0.026$\pm$0.007  \\
NGC~1333-15	& 16$\pm$2    & 0.8$\pm$0.1      & 0.5$\pm$0.2	 & $<$ 1.5   	  & 0.07$\pm$0.03 & 0.020$\pm$0.005   \\
NGC~1333-16 & 21\tablenotemark{a} & 2.3$\pm$0.5 & 1.8$\pm$0.2 & 5.0$\pm$0.6   & 0.08$\pm$0.02 & 0.022$\pm$0.005  \\
NGC~1333-17 & 19\tablenotemark{b} & 0.5$\pm$0.1 & 7.4$\pm$0.6 & 8.9$\pm$0.8   & 1.7$\pm$0.3   & 0.20$\pm$0.04   \\
L1448-1     & 9$\pm$1     & 1.4$\pm$0.3  & 12$\pm$1    & 31$\pm$4       & 0.84$\pm$0.18 & 0.23$\pm$0.05   \\
L1448-2     & 13$\pm$1    & 2.9$\pm$0.3  & 16$\pm$2	   & 24$\pm$3	    & 0.57$\pm$0.08 & 0.09$\pm$0.01   \\
L1448-3     & 14$\pm$1    & 0.8$\pm$0.2  & 4.6$\pm$0.6 & 12$\pm$2　     & 0.60$\pm$0.14 & 0.15$\pm$0.04  \\
L1448-4     & 10$\pm$1    & 1.9$\pm$0.2  & 12$\pm$1    & 10$\pm$1	    & 0.64$\pm$0.09 & 0.05$\pm$0.01   \\
L1448-5     & 10$\pm$1    & 2.1$\pm$0.2  & 2.1$\pm$0.4 & 7.4$\pm$1.2	& 0.10$\pm$0.02 & 0.034$\pm$0.007  \\
IC~348-1    & 11$\pm$1    & 1.3$\pm$0.2 & 5.0$\pm$0.5	 & 11$\pm$2　  & 0.38$\pm$0.06 & 0.08$\pm$0.02 \\
IC~348-2    & 12$\pm$1    & 0.6$\pm$0.1 & 7.0$\pm$0.7	 & 12$\pm$2　  & 1.2$\pm$0.23  & 0.21$\pm$0.04 \\
IC~348-3    & 10$\pm$1    & 0.3$\pm$0.1 & 3.2$\pm$0.6	 & $<$ 1.6 	   & 0.96$\pm$0.35 & 0.05$\pm$0.02 \\
IC~348-4    & 13$\pm$1    & 0.6$\pm$0.1 & 4.2$\pm$0.5	 & 6.7$\pm$1.1 & 0.69$\pm$0.13 & 0.11$\pm$0.02 \\
Barnard 5   & 10$\pm$1    & 0.5$\pm$0.1 & 9.0$\pm$1.1	 & 9.4$\pm$1.9 & 1.9$\pm$0.5   & 0.20$\pm$0.06  \\
B1-1        & 9$\pm$1     & 0.8$\pm$0.1 & 4.3$\pm$0.5	 & 14$\pm$2　   & 0.55$\pm$0.12 & 0.18$\pm$0.04  \\
B1-2        & 9$\pm$1     & 0.8$\pm$0.1 & 7.3$\pm$0.8    & 12$\pm$2	    & 0.96$\pm$0.19 & 0.16$\pm$0.03  \\
B1-3        & 11$\pm$1    & 4.2$\pm$0.4 & 3.4$\pm$0.4	 & 5.9$\pm$1.2	& 0.08$\pm$0.01 & 0.014$\pm$0.003  \\
B1-4        & 9$\pm$1     & 0.8$\pm$0.2 & 8.3$\pm$1.2	 & 13$\pm$2	    & 1.1$\pm$0.3   & 0.18$\pm$0.05  \\
B1-5        & 10$\pm$1    & 0.4$\pm$0.1 & 3.1$\pm$0.5	 & 5.2$\pm$1.2	& 0.93$\pm$0.34 & 0.16$\pm$0.06   \\
L1455-1     & 9$\pm$1     & $<$ 0.2     & 6.0$\pm$1.1	 & 7.7$\pm$1.6	& 3.1$\pm$1.5   & 0.39$\pm$0.20 \\
L1455-2     & 14$\pm$1    & 0.7$\pm$0.1 & 5.4$\pm$0.7	 & 9.0$\pm$1.7　& 0.74$\pm$0.15 & 0.12$\pm$0.03 \\
L1455-3     & 8$\pm$1     & 0.3$\pm$0.1 & 6.4$\pm$0.9	 & 5.1$\pm$1.2	& 2.0$\pm$0.7   & 0.16$\pm$0.07 \\
L1455-4     & 9$\pm$1     & 0.5$\pm$0.1 & 3.5$\pm$0.6	 & 1.8$\pm$0.8	& 0.78$\pm$0.26 & 0.04$\pm$0.02 \\
\hline
Reference &&&&&& \\
L1527  & 8$\pm$1 & 0.8$\pm$0.1 & 33$\pm$3 & 49$\pm$3 & 4.1$\pm$0.5 &  0.60$\pm$0.07 \\
\enddata
\tablenotetext{a}{Rotation temperature of NGC~1333-1 is applied due to the lack of NRO data. 
The column density changes within 10$\%$ for the change in the assumed rotation temperature of $\pm$5~K.}
\tablenotetext{b}{Rotation temperature of NGC~1333-2 is applied due to the lack of NRO data.
The column density changes within 10$\%$ for the change in the assumed rotation temperature of $\pm$5~K.}
\tablecomments{Rotation temperatures are derived by the rotation diagram method from the CH$_{3}$OH ($J$=2--1) and CH$_{3}$OH ($J$=5--4) lines.
Column densities are derived from the CH$_{3}$OH ($J$=5--4, $K$=1, E$^{-}$), 
C$_{2}$H ($N$=3--2, $J$=5/2--3/2, $F$=3--2), and c-C$_{3}$H$_{2}$ (3$_{2,1}$--2$_{1,2}$) lines. 
Error bar are calculated for the rms noise and do not included calibration uncertainty (20~$\%$).
As a reference, the column densities of L1527 are derived with the available IRAM~30m/NRO~45m dataset.}
\end{deluxetable}

\clearpage

\clearpage

\begin{longrotatetable}
\begin{deluxetable}{ccccccccc}
\tabletypesize{\scriptsize}
\tablecaption{Line Parameters of the CH$_{3}$OH ($J$=5--4) lines Observed with IRAM 30~m \label{t4}}
\tablewidth{0pt}
\tablehead{
\colhead{IDs} & \colhead{Transition} & \colhead{$T$$_{\rm{MB}}$\tablenotemark{a}} & \colhead{$dv$\tablenotemark{b}} 
& \colhead{$\int$$T$$_{\rm{MB}}$d$v_{\rm{eff}}$\tablenotemark{c}}
& \colhead{$\int$$T$$_{\rm{MB}}$$dv$\tablenotemark{d}} & 
\colhead{rms\tablenotemark{e}} & \colhead{$V$$_{\rm{LSR}}$} \\
\colhead{} & \colhead{$J,K$} & \colhead{[K]} & \colhead{[km~s$^{-1}$]} & \colhead{[K km~s$^{-1}$]} & \colhead{[K km~s$^{-1}$]}
& \colhead{[K]} & \colhead{[km~s$^{-1}$]}}
\startdata
NGC~1333-1  & $J$=5--4, $K$=2 E$^{-}$ & 0.98 &	3.23 (0.49)	&  2.45	(0.33) & 5.04 (0.49) &0.13 & 6.6 \\
			& $J$=5--4, $K$=2 A$^{+}$ & 0.22 &	2.53 (1.94)	&  0.54	(0.33)	& 0.85	(0.43) &0.13 & 6.2 \\
			& $J$=5--4, $K$=1 E$^{+}$ & 0.58 &	2.91 (0.81)	&  1.44	(0.33)	& 2.61 (0.47) &0.13 & 7.0 \\
			& $J$=5--4, $K$=3 E$^{-}$ & 0.08 & 3.66 (6.23)\tablenotemark{f}	&  0.19	(0.33)\tablenotemark{f}	& 0.38 (0.45)\tablenotemark{f} &0.13 & 7.1 \\
			& $J$=5--4, $K$=2 A$^{-}$ & 0.27 & 4.40 (2.32)\tablenotemark{f}&  0.67 (0.33)\tablenotemark{f}	& 1.65 (0.59)\tablenotemark{f} &0.13 & 6.2 \\
			& $J$=5--4, $K$=3 A$^{\pm}$ & 0.34	& 3.98 (0.84) & 0.87 (0.33)	& 2.18 (0.08) &0.13 & 5.8 \\
            & $J$=5--4, $K$=0 A$^{+}$ & 2.20 &	2.87 (0.10)	& 5.41 (0.33) & 10.4 (0.08) &0.13 & 6.7 \\
            & $J$=5--4, $K$=1 E$^{-}$ & 1.94	& 2.82 (0.11) & 4.84 (0.33) & 8.56 (0.08)	&0.13 & 6.8 \\
            & $J$=5--4, $K$=0 E$^{+}$ & 0.94	& 3.68 (0.31) & 2.38 (0.33) & 4.89 (0.08)	&0.13 & 6.9 \\     
NGC~1333-2  & $J$=5--4, $K$=2 E$^{-}$ & 0.49	& 3.05 (0.44)	& 0.67 (0.20) & 2.24 (0.22)	& 0.15 & 6.5 \\
			& $J$=5--4, $K$=2 A$^{+}$ & 0.27	& 2.79 (0.70)	& 0.37 (0.20) & 	1.08 (0.21) & 0.15 & 5.9 \\
			& $J$=5--4, $K$=1 E$^{+}$ & 0.34	& 2.95 (0.62)	& 0.47 (0.20)	& 1.47	(0.22) & 0.15 & 6.8 \\
			& $J$=5--4, $K$=3 E$^{-}$ & 0.23	& 2.54 (0.65)	& 0.32 (0.20) & 0.79 (0.19)  & 0.15 & 6.2 \\
			& $J$=5--4, $K$=2 A$^{-}$ & 0.30	& 3.51 (0.62)	& 0.42 (0.20) & 	1.47 (0.22) & 0.15 & 5.6  \\
			& $J$=5--4, $K$=3 A$^{\pm}$& 0.31	& 2.71 (0.62)	& 0.43 (0.20) & 1.14 (0.23) & 0.15 & 5.5\\
			& $J$=5--4, $K$=4 E$^{+}$ & 0.18	& 2.02 (1.0)	& 0.25 (0.20) & 0.60 (0.22)	& 0.15 & 6.1 \\
			& $J$=5--4, $K$=4 E$^{-}$ & 0.19	& 2.58 (0.79)	& 0.26 (0.20) & 0.67 (0.17)	& 0.15 & 6.4 \\
			& $J$=5--4, $K$=4 A$^{\pm}$ & 0.22	& 2.48 (0.65) 	& 0.30 (0.20) & 0.75 (0.16)	& 0.15 & 6.0  \\
            & $J$=5--4, $K$=0 A$^{+}$ & 1.02	& 2.68 (0.22)	& 1.41 (0.20) & 4.53 (0.20)	& 0.15 & 7.3  \\
            & $J$=5--4, $K$=1 E$^{-}$ & 0.84	& 3.05 (0.28)	& 1.17 (0.20) & 4.08 (0.23)	& 0.15 & 7.4 \\
            & $J$=5--4, $K$=0 E$^{+}$ & 0.41	& 3.04 (0.53)	& 0.57 (0.20) & 1.99 (0.21)	& 0.15 & 6.9 \\
NGC~1333-3  & $J$=5--4, $K$=2 E$^{-}$ & 0.16 &	0.92 (0.11)	& 0.16 (0.03) & 0.18 (0.02) &0.03 & 7.3 \\
			& $J$=5--4, $K$=0 A$^{+}$ & 1.02 &	1.06 (0.02) & 1.03 (0.03) & 1.33 (0.02)	&0.03 & 7.5 \\
			& $J$=5--4, $K$=1 E$^{-}$ & 0.84 & 0.98 (0.03)  & 0.84 (0.03) & 1.07 (0.02) & 0.03 & 7.7 \\
			& $J$=5--4, $K$=0 E$^{+}$ & 0.18 &	0.92 (0.13) & 0.18 (0.03) & 0.18 (0.02) &0.03 & 7.8 \\
NGC~1333-4  & $J$=5--4, $K$=2 E$^{-}$ & 0.14 & 0.63 (0.12)  &  0.15 (0.03) & 0.14 (0.02) &0.03 & 7.9\\
			& $J$=5--4, $K$=0 A$^{+}$ & 0.95 & 1.24 (0.02)  &  0.95	(0.03) & 1.56 (0.02)	&0.03 & 8.2\\
			& $J$=5--4, $K$=1 E$^{-}$ & 0.81 & 1.17 (0.03)  &  0.82 (0.03) & 1.25 (0.02)	&0.03 & 8.4\\
			& $J$=5--4, $K$=0 E$^{+}$ & 0.15 & 1.04 (0.18)  & 0.15 (0.03)  & 0.19 (0.02) & 0.03 & 8.4\\
NGC~1333-5  & $J$=5--4, $K$=0 A$^{+}$ & 0.68 &0.76 (0.04)  & 0.67 (0.03) & 0.69 (0.02)	& 0.03 & 7.5  \\
			& $J$=5--4, $K$=1 E$^{-}$ & 0.61 &0.72 (0.04)  & 0.60 (0.03)	& 0.65 (0.02) & 0.03 &  7.6 \\
			& $J$=5--4, $K$=0 E$^{+}$ & 0.14 &0.62 (0.12)  & 0.14 (0.03)	& 0.21 (0.02) & 0.03 & 7.6 \\
NGC~1333-6 & $J$=5--4, $K$=0 A$^{+}$ & 0.11	& 1.82 (0.31)  & 0.09 (0.02) & 0.24 (0.03)  & 0.03 & 6.8 \\
           & $J$=5--4, $K$=1 E$^{-}$ & 0.10	 & 2.27 (0.32) & 0.09 (0.02) & 0.23 (0.03)	& 0.03 & 7.2 \\
		   & $J$=5--4, $K$=0 E$^{+}$ & 0.04	& 4.00 (1.08)  & 0.03 (0.02) & 0.17 (0.04)	& 0.03 & 7.0 \\
NGC~1333-7 & $J$=5--4, $K$=0 A$^{+}$ & 1.13	& 1.24 (0.02) & 1.75 (0.04) & 1.86 (0.02) & 0.03 & 7.3 \\
           & $J$=5--4, $K$=1 E$^{-}$ & 0.87	& 1.28 (0.03) &	1.35 (0.04) & 1.51 (0.02) & 0.03 & 7.4 \\
           & $J$=5--4, $K$=0 E$^{+}$ & 0.11	& 1.52 (0.27) & 0.17 (0.04) & 0.22 (0.03) & 0.03 & 7.3 \\
NGC~1333-8 & $J$=5--4, $K$=0 A$^{+}$ & 0.43 &	2.03 (0.07) & 0.49 (0.03) & 1.11 (0.03)	& 0.03 & 6.8  \\
           & $J$=5--4, $K$=1 E$^{-}$ & 0.33 &	2.04 (0.10) & 0.38 (0.03) & 0.89 (0.03)	& 0.03 & 6.9 \\
           & $J$=5--4, $K$=0 E$^{+}$ & 	0.06 &  2.11 (0.42) & 0.07 (0.03) & 0.17 (0.03) & 0.03 & 7.0 \\
NGC~1333-9  & $J$=5--4, $K$=0 A$^{+}$ & 0.08 & 1.00 (0.68) & 0.08 (0.04) & 0.16 (0.04)	& 0.03 & 7.2 \\
           & $J$=5--4, $K$=1 E$^{-}$ & 0.09	& 0.89 (0.16)  & 0.15 (0.04) & 0.14 (0.02)	& 0.03 & 7.1 \\
NGC~1333-10	& $J$=5--4, $K$=0 A$^{+}$ & 0.07 & 1.87 (0.26) & 0.09 (0.03) & 0.12 (0.02)	& 0.02 & 7.3  \\
           & $J$=5--4, $K$=1 E$^{-}$ & 0.08 & 1.00 (0.18) &  0.11 (0.03) & 0.05 (0.01) & 0.02 & 7.4 \\
NGC~1333-11 & $J$=5--4, $K$=0 A$^{+}$ & 0.67 & 1.08 (0.02) & 0.40 (0.01) & 0.92 (0.01)	&0.02 & 8.2 \\
            & $J$=5--4, $K$=1 E$^{-}$ & 0.52 & 1.12 (0.03) & 0.31 (0.01) & 0.60 (0.02)	&0.02 & 8.4 \\
           & $J$=5--4, $K$=0 E$^{+}$ & 0.07	&  1.43 (0.22) & 0.04 (0.01) & 0.09 (0.02)      & 0.03 & 8.3 \\
NGC~1333-12 & $J$=5--4, $K$=0 A$^{+}$ & 0.10	& 0.48 (0.10) & 0.06 (0.01)	& 0.04 (0.01) & 0.02 & 7.2 \\
           & $J$=5--4, $K$=1 E$^{-}$  & 0.07	& 0.77 (0.22) & 0.04 (0.01) &  0.07 (0.01)	& 0.02 & 7.4 \\
NGC~1333-13 & $J$=5--4, $K$=0 A$^{+}$ & 0.50	& 0.69 (0.03) & 0.28 (0.01) & 0.43 (0.01)	& 0.02 & 7.2\\
           & $J$=5--4, $K$=1 E$^{-}$  & 0.43	& 0.69 (0.03) & 0.24 (0.01) & 0.43 (0.01)	& 0.02 & 7.3 \\
NGC~1333-14 &  $J$=5--4, $K$=2 E$^{-}$ & 0.09	& 0.94 (0.18)  & 0.07 (0.02)	& 0.06 (0.02) & 0.02 & 7.1 \\
           & $J$=5--4, $K$=0 A$^{+}$  & 0.91	& 1.17 (0.02) & 0.80 (0.02) & 1.27 (0.02) & 0.02 & 7.6 \\ 
           & $J$=5--4, $K$=1 E$^{-}$ & 0.70	&1.20 (0.02)  & 0.61 (0.02) & 0.99 (0.02) &0.02 & 7.7 \\
		   & $J$=5--4, $K$=0 E$^{+}$ & 0.12	&1.17 (0.18)  & 0.10 (0.02) & 0.18 (0.02) & 0.02 & 7.8\\
NGC~1333-15	& $J$=5--4, $K$=0 A$^{+}$ & 0.81 & 0.72 (0.02)  & 0.70 (0.02) & 0.76 (0.01) &	0.02 & 7.8 \\
           & $J$=5--4, $K$=1 E$^{-}$ & 0.67  & 0.68 (0.02)  & 0.58 (0.02) & 0.61 (0.01) & 0.02 & 8.0 \\
		   & $J$=5--4, $K$=0 E$^{+}$ & 0.07	 & 1.15 (0.31) 	& 0.06 (0.02) & 0.11 (0.02) &	0.02 & 8.0\\
NGC~1333-16  & $J$=5--4, $K$=2 E$^{-}$& 0.36 &	4.91 (1.96)	& 0.77 (0.36) & 2.23 (0.52)  & 0.11 & 6.2 \\	
			& $J$=5--4, $K$=1 E$^{+}$ & 0.21 &	4.72 (3.29)	& 0.45 (0.36) & 1.17 (0.53)  & 0.11 &  6.7 \\
			& $J$=5--4, $K$=2 A$^{-}$ & 0.08 &	4.43 (7.02)\tablenotemark{f} & 0.16 (0.36)\tablenotemark{f} & 0.35 (0.46)\tablenotemark{f}  & 0.11 &  6.3 \\
			& $J$=5--4, $K$=3 A$^{\pm}$ & 0.09 & 5.50 (6.57)\tablenotemark{f} & 0.20 (0.36)\tablenotemark{f} & 0.56 (0.50)\tablenotemark{f} & 0.11 &  6.1 \\
            & $J$=5--4, $K$=0 A$^{+}$ & 1.21 &	4.88 (0.72)	& 2.57 (0.36) & 8.44 (0.59)  & 0.11 &  6.4 \\
            & $J$=5--4, $K$=1 E$^{-}$ & 1.02 & 5.11 (0.78) &  2.17 (0.36) & 7.76 (0.71)  & 0.11 &  6.5 \\
            & $J$=5--4, $K$=0 E$^{+}$ & 0.37 & 5.01 (0.78) &  0.79 (0.36) & 2.31 (0.61)  & 0.11 &  6.5 \\     
NGC~1333-17 & $J$=5--4, $K$=2 E$^{-}$ & 0.17	& 3.11 (0.39)	&  0.20 (0.05) & 0.64 (0.22)  & 0.03 & 7.5 \\
			& $J$=5--4, $K$=2 A$^{+}$ & 0.11	& 3.62 (0.55)	&  0.13 (0.05) & 0.46 (0.21)  &  0.03 & 7.0 \\
			& $J$=5--4, $K$=1 E$^{+}$ & 0.15	& 3.01 (0.39)   &  0.18 (0.05) & 0.51 (0.22)  & 0.03 & 7.9 \\
			& $J$=5--4, $K$=3 E$^{-}$ & 0.10	& 3.43 (0.61)	&  0.12 (0.05) & 0.41 (0.19)  &  0.03 & 7.5 \\
			& $J$=5--4, $K$=2 A$^{-}$ & 0.12	& 4.61 (0.57)	&  0.14 (0.05) & 0.63 (0.22)  &  0.03 &  6.7 \\
			& $J$=5--4, $K$=3 A$^{\pm}$& 0.12	& 6.00 (0.55)	&  0.14 (0.05) & 0.78 (0.23)  &  0.03 & 8.1 \\
			& $J$=5--4, $K$=4 E$^{+}$ & 0.09	& 3.15 (0.50) 	&  0.11 (0.05) & 0.32 (0.17) &  0.03 &  7.7 \\
            & $J$=5--4, $K$=0 A$^{+}$ & 0.41	& 1.29 (0.08)	&  0.48 (0.05) & 0.71 (0.20) & 0.03 & 7.8 \\
            & $J$=5--4, $K$=1 E$^{-}$ & 0.34	& 1.32 (0.10)	&  0.40 (0.05) & 0.68 (0.23) & 0.03 & 7.9\\
            & $J$=5--4, $K$=0 E$^{+}$ & 0.16	& 2.84 (0.50)	&  0.19 (0.05) & 0.61 (0.21) & 0.03 & 7.8 \\
L1448-1    & $J$=5--4, $K$=0 A$^{+}$  & 0.28	& 1.08 (0.12) &	0.43 (0.05) & 0.35 (0.01) & 0.03 & 4.1 \\
		   & $J$=5--4, $K$=1 E$^{-}$  & 0.25	& 1.17 (0.11)  & 0.38 (0.05) & 0.34 (0.01) & 0.03 & 4.3 \\
		   & $J$=5--4, $K$=0 E$^{+}$  & 0.12	& 0.30 (0.10)  & 0.19 (0.05) & 0.08 (0.01) & 0.03 & 4.5 \\
L1448-2    & $J$=5--4, $K$=2 E$^{-}$ & 0.24	& 1.46 (0.15)  & 0.45 (0.05) & 0.44 (0.02)	&0.03 & 4.2 \\
		   & $J$=5--4, $K$=1 E$^{+}$ & 0.16	& 1.14 (0.16)  & 0.30 (0.05) & 0.29 (0.02)	&0.03 & 4.7 \\
		   & $J$=5--4, $K$=0 A$^{+}$ & 1.15	& 1.28 (0.03) &	2.16 (0.05) & 2.22 (0.03)	&0.03 & 4.2 \\
		   & $J$=5--4, $K$=1 E$^{-}$ & 0.94	& 1.21 (0.03) &	1.76 (0.05) & 1.46 (0.02)	&0.03 & 4.3 \\
		   & $J$=5--4, $K$=0 E$^{+}$ & 0.34	& 1.33 (0.09) &  0.63 (0.05) & 0.54 (0.03)	&0.03 & 4.5  \\
L1448-3   & $J$=5--4, $K$=2 E$^{-}$ & 0.16	& 2.65 (0.48)  &  0.17 (0.05) & 0.67 (0.06)	&0.05 & 4.6 \\
		  & $J$=5--4, $K$=1 E$^{+}$ & 0.13	& 1.30 (0.36)  &  0.14 (0.05) & 0.27 (0.04)	&0.05 & 5.0 \\
          & $J$=5--4, $K$=0 A$^{+}$ & 0.58	& 1.18 (0.09)  &  0.63 (0.05) & 1.04 (0.04) & 0.05 & 4.6 \\
          & $J$=5--4, $K$=1 E$^{-}$ & 0.48	& 1.14 (0.11)  &  0.52 (0.05) & 0.99 (0.04) & 0.05 & 4.8 \\
          & $J$=5--4, $K$=0 E$^{+}$ & 0.24	& 1.16 (0.26)  &  0.26 (0.05) & 0.56 (0.05) & 0.05 & 4.9 \\
L1448-4     & $J$=5--4, $K$=2 E$^{-}$ & 0.10	& 1.21 (0.33)  & 0.10 (0.03) &  0.21 (0.03) &	0.03 &  3.3\\
	        & $J$=5--4, $K$=0 A$^{+}$ & 0.82    & 0.72 (0.03)  & 0.82 (0.03) & 0.78 (0.02) & 0.03 & 3.7 \\
            & $J$=5--4, $K$=1 E$^{-}$ & 0.71	& 0.68 (0.03)  & 0.71 (0.03) & 0.69 (0.02)  & 0.03 & 3.9 \\
            & $J$=5--4, $K$=0 E$^{+}$ & 0.09	& 1.38 (0.69)  & 0.09 (0.03) & 0.13 (0.04) & 0.03 & 3.8 \\
L1448-5     & $J$=5--4, $K$=2 E$^{-}$ & 0.06    &3.27 (0.91)  & 0.05 (0.03) & 0.19 (0.03)	& 0.04 & 3.6 \\
			& $J$=5--4, $K$=1 E$^{+}$ & 0.12	& 0.56 (0.25) &	0.08 (0.03) & 0.04 (0.03)	& 0.04 & 4.1\\
			& $J$=5--4, $K$=0 A$^{+}$ & 1.32	& 0.97 (0.03) & 0.87 (0.03) & 2.37 (0.03)	&0.04 & 3.8 \\
			& $J$=5--4, $K$=1 E$^{-}$ & 1.20	& 0.86 (0.03) & 0.80 (0.03) & 1.89 (0.03)	&0.04 & 3.9\\
			& $J$=5--4, $K$=0 E$^{+}$ & 0.27	& 0.60 (0.13) & 0.18 (0.03) & 0.27 (0.03)	&0.04 & 3.9 \\
IC~348-1     & $J$=5--4, $K$=2 E$^{-}$ & 0.07	& 0.96 (0.40) & 0.07	(0.03) & 0.09 (0.02)    &0.03 & 8.3 \\
			& $J$=5--4, $K$=0 A$^{+}$ & 0.93	& 0.63 (0.02) & 0.81	(0.03) & 0.81 (0.02)	&0.03 & 8.6 \\
			& $J$=5--4, $K$=1 E$^{-}$ & 0.76	& 0.62 (0.03) & 0.65	(0.03) & 0.74 (0.02)	&0.03 & 8.7 \\
			& $J$=5--4, $K$=0 E$^{+}$ & 0.13	& 1.04 (0.17) & 0.10	(0.03) & 0.13 (0.02)	&0.03 & 8.8 \\
IC~348-2     & $J$=5--4, $K$=2 E$^{-}$ & 0.07    & 1.59 (0.29) & 0.07 (0.03) & 0.12 (0.02)	& 0.03 & 8.2 \\
			& $J$=5--4, $K$=0 A$^{+}$ & 0.36	& 0.80 (0.06) & 0.40 (0.03) & 0.37 (0.02)	& 0.02 & 8.4\\
            & $J$=5--4, $K$=1 E$^{-}$ & 0.28    & 0.74 (0.07) & 0.31 (0.03) & 0.32 (0.02)	& 0.02 & 8.5 \\
            & $J$=5--4, $K$=0 E$^{+}$ & 0.06	& 1.29 (0.47) & 0.07 (0.03) & 0.10 (0.02)	&0.03 & 8.5 \\
IC~348-3     & $J$=5--4, $K$=0 A$^{+}$ & 0.16	& 1.16 (0.10)  & 0.18 (0.03) & 0.15 (0.02)	& 0.02 & 9.9\\
            & $J$=5--4, $K$=1 E$^{-}$ & 0.11	& 1.38 (0.16)  & 0.13 (0.03) & 0.17 (0.02)	& 0.02 & 10 \\
IC~348-4     & $J$=5--4, $K$=2 E$^{-}$ & 0.09	& 0.36 (0.11)  	& 0.08 (0.02) & 0.06 (0.01)	& 0.02 & 7.6\\
            & $J$=5--4, $K$=0 A$^{+}$ & 0.46	& 0.76 (0.04)  	& 0.40 (0.02) & 0.52 (0.01)	& 0.02 & 8.0 \\
            & $J$=5--4, $K$=1 E$^{-}$ & 0.41	& 0.69 (0.04)  	& 0.36 (0.02) & 0.46 (0.02) & 0.02 & 8.2 \\
Barnard~5    & $J$=5--4, $K$=0 A$^{+}$  & 0.26	& 0.80 (0.07)  	& 0.22 (0.02) & 0.26 (0.02)	&0.02 & 9.6 \\
            & $J$=5--4, $K$=1 E$^{-}$  & 0.19	& 0.80 (0.08)	& 0.16 (0.02) & 0.18 (0.02)	&0.02 & 9.8 \\
B1-1   & $J$=5--4, $K$=2 E$^{-}$ & 0.07	& 4.64 (0.24) & 0.07 (0.03) & 0.46 (0.09)	& 0.02 & 7.0 \\
       & $J$=5--4, $K$=1 E$^{+}$  & 0.05 & 3.33(0.24) & 0.06 (0.03) & 0.20 (0.03)	& 0.02 & 6.5 \\
	   & $J$=5--4, $K$=0 A$^{+}$ & 0.25 & 4.52 (0.24) & 0.39 (0.03) & 1.75 (0.07)	& 0.02 & 6.2 \\
       & $J$=5--4, $K$=1 E$^{-}$ & 0.21	& 4.54 (0.24) & 0.23 (0.03) & 1.47 (0.11)	& 0.02 & 6.9 \\
       & $J$=5--4, $K$=0 E$^{+}$ & 0.08	& 2.66 (0.24) & 0.08 (0.03) & 0.21  (0.04)	& 0.02 & 6.5 \\
B1-2   & $J$=5--4, $K$=0 A$^{+}$ & 0.42 & 0.66 (0.04) & 0.29 (0.02) & 0.33 (0.01)	& 0.02  & 6.2 \\
       & $J$=5--4, $K$=1 E$^{-}$ & 0.30	& 0.68 (0.05) & 0.20 (0.02) & 0.27 (0.01)	& 0.02 & 6.1 \\
%
B1-3   & $J$=5--4, $K$=2 E$^{-}$  & 0.18	& 1.99 (0.22) & 0.22 (0.05) & 0.40 (0.04)	& 0.04 & 5.2 \\
       & $J$=5--4, $K$=1 E$^{+}$  & 0.10	& 1.89 (0.36) & 0.12 (0.05) & 0.20 (0.04)	& 0.04 & 5.7 \\
       & $J$=5--4, $K$=0 A$^{+}$  & 1.96	& 1.45 (0.07) & 2.50 (0.05) & 3.99 (0.04)	& 0.04 & 5.8  \\
       & $J$=5--4, $K$=1 E$^{-}$  & 1.65	& 1.37 (0.03) & 2.10 (0.05) & 3.16 (0.03)	& 0.04 & 5.9  \\
       & $J$=5--4, $K$=0 E$^{+}$  & 0.26	& 2.31 (0.18) & 0.33 (0.05) & 0.66 (0.04)	& 0.04 & 5.6 \\
B1-4   & $J$=5--4, $K$=0 A$^{+}$ & 0.29	 & 0.78 (0.10) & 0.31 (0.03) & 0.33 (0.02) & 0.03 & 6.5 \\
       & $J$=5--4, $K$=1 E$^{-}$  & 0.21 & 0.71 (0.09) & 0.22 (0.03) & 0.18 (0.02)	& 0.03 & 6.5 \\
B1-5   
       & $J$=5--4, $K$=0 A$^{+}$ &  0.17 & 0.49 (0.11) 	& 0.16 (0.02) & 0.16 (0.02) & 0.03 & 5.6 \\
       & $J$=5--4, $K$=1 E$^{-}$ &  0.12 & 0.82 (0.21)  & 0.11 (0.02) & 0.11 (0.01) & 0.03 & 6.6  \\
%
L1455-1  & ...  &  ... & ...  & 0.05 (0.02) & 0.06 & 0.02 & ... \\
L1455-2  & $J$=5--4, $K$=2 E$^{-}$ &  0.09	& 4.65 (0.66) & 0.11 (0.03) & 0.59 (0.04) & 0.03 & 4.1\\
         & $J$=5--4, $K$=0 A$^{+}$ &  0.47	& 1.38 (0.07) & 0.57 (0.03) & 0.89 (0.02) & 0.03 &  4.4 \\
         & $J$=5--4, $K$=1 E$^{-}$  &  0.40	& 1.26 (0.07) & 0.49 (0.03) & 0.92 (0.02) & 0.03 & 4.6 \\
         & $J$=5--4, $K$=0 E$^{+}$ &  0.07	& 5.97 (0.85) & 0.08 (0.03) & 0.44 (0.04) & 0.03 & 4.1 \\
L1455-3  & $J$=5--4, $K$=0 A$^{+}$ & 0.20	& 0.87 (0.09) & 0.16 (0.02) & 0.18 (0.02)	&0.02 & 4.4 \\
         & $J$=5--4, $K$=1 E$^{-}$ & 0.10	& 1.69 (0.23) & 0.08 (0.02)	& 0.15 (0.02) &0.02 & 4.7  \\
L1455-4  & $J$=5--4, $K$=0 A$^{+}$  & 0.14	& 1.06 (0.13) & 0.16 (0.02) & 0.16 (0.02) & 0.02 & 4.7 \\
         & $J$=5--4, $K$=1 E$^{-}$  & 0.12	& 0.78 (0.15) & 0.14 (0.02) & 0.12 (0.02) & 0.02 & 4.8 \\        
\hline
\enddata
\tablenotetext{a}{Obtained by the Gaussian fit.}
\tablenotetext{b}{The wing components are excluded in the Gaussian fit.}
\tablenotetext{c}{Derived by using the C$_{2}$H velocity width.}
\tablenotetext{d}{The wing components are included in calculating the integrated intensity.}
\tablenotetext{e}{The rms noise averaged over the line width.}
\tablenotetext{f}{The error in the Gaussian fitting is large.}
\tablecomments{The errors are 1$\sigma$. The upper limit to the integrated intensity is calculated as 
$\int$$T$$_{\rm{MB}}$$dv$ $<$ 3~$\sigma$ $\times$ $\sqrt{({dv/dv_{\rm{res}}})}$$dv_{\rm{res}}$ where $dv$ is the assumed line width 
(0.8~km~s$^{-1}$) and $dv_{\rm{res}}$ is the velocity resolution per channel.}
\end{deluxetable}
\end{longrotatetable}

\begin{longrotatetable}
\begin{deluxetable}{ccc c c c c c}
\tabletypesize{\scriptsize}
\rotate
\tablecaption{Line Parameters of the CH$_{3}$OH ($J$=2--1) lines Observed with Nobeyama 45m telescope \label{t5}}
\tablewidth{0pt}
\tablehead{
\colhead{IDs} & \colhead{Transition} & \colhead{$T$$_{\rm{MB}}$\tablenotemark{a}} & \colhead{$dv$\tablenotemark{b}} & 
\colhead{$\int$$T$$_{\rm{MB}}$d$v_{\rm{eff}}$\tablenotemark{c}} 
& \colhead{$\int$$T$$_{\rm{MB}}$$dv$\tablenotemark{d}} & \colhead{rms\tablenotemark{e}} & \colhead{$V$$_{\rm{LSR}}$} \\
\colhead{} & \colhead{$J,K$} & \colhead{[K]} & \colhead{[km~s$^{-1}$]} & \colhead{[K km~s$^{-1}$]} & \colhead{[K km~s$^{-1}$]}
& \colhead{[K]} & \colhead{[km~s$^{-1}$]}}
\startdata
NGC~1333-1  & $J$=2--1, $K$=1 E$^{-}$   & 1.40 &  2.16 (0.04) & 3.50 (0.04) & 3.53 (0.04) & 0.02 & 7.7 \\
			& $J$=2--1, $K$=0 A$^{+}$ & 1.97 & 2.15 (0.04) & 4.92 (0.04) & 5.94 (0.04) & 0.02 & 7.5 \\
			& $J$=2--1, $K$=0 E$^{+}$  & 0.47 & 2.56 (0.04) & 1.18 (0.04) & 1.73 (0.04) & 0.02 & 6.8 \\
NGC~1333-2  & $J$=2--1, $K$=1 E$^{-}$   & 1.38 & 1.32 (0.02) & 1.91 (0.02) & 2.57 (0.02) & 0.02 & 7.7 \\
			& $J$=2--1, $K$=0 A$^{+}$ & 1.73 & 1.37 (0.02) & 2.39 (0.02) & 3.32 (0.02) & 0.02 & 7.9 \\
			& $J$=2--1, $K$=0 E$^{+}$  & 0.27 & 1.56 (0.02) & 0.38 (0.02) & 0.62 (0.02) & 0.02 & 7.6 \\
NGC~1333-3  & $J$=2--1, $K$=1 E$^{-}$   & 0.82 & 1.21 (0.02) & 0.82 (0.02) & 1.15 (0.02) & 0.02 & 7.8 \\
			& $J$=2--1, $K$=0 A$^{+}$ & 1.27 & 1.14 (0.02) & 1.27 (0.02) & 1.69 (0.02) & 0.02 & 7.9 \\
			& $J$=2--1, $K$=0 E$^{+}$  & 0.22 & 1.13 (0.02) & 0.22 (0.02) & 0.23 (0.02) &  0.02 & 7.9 \\
NGC~1333-4  & $J$=2--1, $K$=1 E$^{-}$   & 1.18 & 1.19 (0.02) & 1.19 (0.02) & 1.68 (0.02) & 0.02 & 8.5 \\
			& $J$=2--1, $K$=0 A$^{+}$ & 1.74 & 1.18 (0.02) & 1.76 (0.02) & 2.52 (0.02) & 0.02 & 8.6 \\
			& $J$=2--1, $K$=0 E$^{+}$  & 0.35 & 1.01 (0.02) & 0.35 (0.02) & 0.42 (0.02) & 0.02 & 8.7\\
NGC~1333-5  & $J$=2--1, $K$=1 E$^{-}$   & 1.54 & 0.98 (0.01) & 1.52 (0.01) & 1.83 (0.01) & 0.01 & 7.7 \\
			& $J$=2--1, $K$=0 A$^{+}$ & 2.03 & 1.04 (0.02) & 2.00 (0.01) & 2.45 (0.02) & 0.01 &  7.9 \\
			& $J$=2--1, $K$=0 E$^{+}$  & 0.32 & 0.92 (0.01) & 0.31 (0.01) & 0.38 (0.01) & 0.01 & 7.9 \\
NGC~1333-6 & $J$=2--1, $K$=1 E$^{-}$  & 0.23 & 1.17 (0.02) & 0.21 (0.01) & 0.32 (0.02)  & 0.02 & 7.4 \\
                        & $J$=2--1, $K$=0 A$^{+}$ & 0.29 & 1.02 (0.02) & 0.26 (0.01) & 0.39 (0.02) & 0.02 & 7.2 \\
NGC~1333-7 & $J$=2--1, $K$=1 E$^{-}$  & 2.01 & 1.34 (0.02) & 3.11 (0.02) & 3.03 (0.02) & 0.02 & 7.7 \\
                        & $J$=2--1, $K$=0 A$^{+}$ & 2.64 & 1.42 (0.02) & 4.08 (0.02) & 4.31 (0.02) & 0.02 & 7.5 \\
                        & $J$=2--1, $K$=0 E$^{+}$  & 0.38 & 1.32 (0.02) & 0.58 (0.02) & 0.58 (0.02) & 0.02 & 7.6 \\
NGC~1333-8 & $J$=2--1, $K$=1 E$^{-}$  & 1.02 & 1.61 (0.02) & 1.17 (0.02) & 1.89 (0.02) & 0.02 & 7.4  \\
                        & $J$=2--1, $K$=0 A$^{+}$ & 1.45 & 1.54 (0.02) & 1.68 (0.02) & 2.54 (0.02) & 0.02 & 7.5 \\
                        & $J$=2--1, $K$=0 E$^{+}$  & 0.15 & 1.78 (0.03) & 0.17 (0.02) & 0.31 (0.03) & 0.02 & 7.6 \\
NGC~1333-9 & $J$=2--1, $K$=1 E$^{-}$  & 0.09 & 1.98 (0.03) & 0.14 (0.03) & 0.18 (0.03)  & 0.02 & 8.1 \\
                        & $J$=2--1, $K$=0 A$^{+}$ & 0.11 & 1.92 (0.03) & 0.18 (0.03) & 0.19 (0.03)  & 0.02 & 7.2 \\
NGC~1333-10 & $J$=2--1, $K$=1 E$^{-}$  & 0.09 & 1.91 (0.03) & 0.12 (0.02) & 0.20 (0.03) & 0.02 & 8.1  \\
                          & $J$=2--1, $K$=0 A$^{+}$ & 0.12 & 2.36 (0.04) & 0.16 (0.02) & 0.25 (0.04) & 0.02 & 7.9 \\
NGC~1333-11 & $J$=2--1, $K$=1 E$^{-}$  & 1.41  & 1.21 (0.02) & 0.85 (0.01) & 1.90 (0.02) & 0.01 & 8.5 \\
                          & $J$=2--1, $K$=0 A$^{+}$ & 1.84  & 1.24 (0.02) & 1.12 (0.01) & 2.64 (0.02) & 0.01 & 8.6 \\
                          & $J$=2--1, $K$=0 E$^{+}$  & 0.33  & 1.14 (0.02) & 0.20 (0.01) & 0.43 (0.02) & 0.01 & 8.7 \\
NGC~1333-12 & $J$=2--1, $K$=1 E$^{-}$   & 0.15 & 0.86 (0.01) & 0.08 (0.01) &  0.17 (0.01) & 0.03 & 7.8 \\
                          & $J$=2--1, $K$=0 A$^{+}$  & 0.20 & 1.02 (0.02) & 0.12 (0.01) &  0.29 (0.02) & 0.03 & 7.5 \\
NGC~1333-13 & $J$=2--1, $K$=1 E$^{-}$   & 0.39 & 0.96 (0.02) & 0.22 (0.01) & 0.41 (0.01) & 0.02 & 7.7\\
                          & $J$=2--1, $K$=0 A$^{+}$  & 0.62 & 0.95 (0.02) & 0.35 (0.01) & 0.59 (0.01) & 0.02 & 7.5 \\
NGC~1333-14 & $J$=2--1, $K$=1 E$^{-}$   & 1.08 & 1.32 (0.02) & 0.95 (0.01) & 1.57 (0.02) & 0.02 & 7.7 \\
                          & $J$=2--1, $K$=0 A$^{+}$  & 1.57 & 1.32 (0.02) & 1.38 (0.01) & 2.31 (0.02) & 0.02 & 7.9 \\ 
                          & $J$=2--1, $K$=0 E$^{+}$   & 0.24 & 1.65 (0.03) & 0.21 (0.01) & 0.40 (0.02) & 0.02 & 7.9 \\
NGC~1333-15 &$J$=2--1, $K$=1 E$^{-}$    & 0.24 & 0.95 (0.02) & 0.21 (0.01) &  0.24 (0.01) & 0.02 & 7.4 \\
                          & $J$=2--1, $K$=0 A$^{+}$  & 0.30 & 1.03 (0.02) & 0.26 (0.01) &  0.35 (0.02) & 0.02 & 7.5 \\          
L1448-1      & $J$=2--1, $K$=1 E$^{-}$   & 0.93 & 1.17	(0.02) & 1.45 (0.03) & 1.28 (0.02)	& 0.02 & 4.5 \\
		   & $J$=2--1, $K$=0 A$^{+}$  & 1.22 & 1.18 (0.02) & 1.96 (0.03) & 1.80 (0.02) & 0.02 & 4.6 \\
		   & $J$=2--1, $K$=0 E$^{+}$   & 0.23 & 1.14 (0.02) & 0.35 (0.03) & 0.39 (0.02) & 0.02 & 4.7 \\
L1448-2    & $J$=2--1, $K$=1 E$^{-}$  & 1.34 & 1.03 (0.02) & 2.52 (0.03) & 1.74 (0.02)	& 0.02 & 4.5 \\
		 & $J$=2--1, $K$=0 A$^{+}$ & 1.73 & 1.09 (0.02) & 3.26 (0.03) & 2.99 (0.02)	& 0.02 & 4.3 \\
		 & $J$=2--1, $K$=0 E$^{+}$  & 0.39 & 1.56 (0.03) & 0.74 (0.03) & 1.32 (0.03)   & 0.02 & 4.7 \\
L1448-3   & $J$=2--1, $K$=1 E$^{-}$  & 0.64 & 1.77 (0.03) & 0.69 (0.02) & 2.58 (0.03) & 0.02 & 4.8 \\
		& $J$=2--1, $K$=0 A$^{+}$ & 0.74 & 1.71 (0.03) & 0.80 (0.02) & 1.70 (0.03) & 0.02 & 5.0 \\
                 & $J$=2--1, $K$=0 E$^{+}$  & 0.08 & 1.30 (0.02) & 0.09 (0.02) & 0.11 (0.02) & 0.02 & 5.0 \\
L1448-4  & $J$=2--1, $K$=1 E$^{-}$  & 1.92 & 0.81 (0.01) & 1.91 (0.02) &  1.74 (0.01) & 0.02 &  4.1 \\
	        & $J$=2--1, $K$=0 A$^{+}$ & 2.39 & 0.87 (0.01) & 2.38 (0.02) &  2.34 (0.01) & 0.02 & 4.3  \\
                 & $J$=2--1, $K$=0 E$^{+}$ & 0.35 & 0.84 (0.01) & 0.35 (0.02) &  0.34 (0.01)  & 0.02 & 4.3  \\
L1448-5   & $J$=2--1, $K$=1 E$^{-}$  & 2.31 & 1.05 (0.02) & 1.57 (0.01) & 5.19 (0.02) & 0.02 & 4.1 \\
	         & $J$=2--1, $K$=0 A$^{+}$ & 2.65 & 1.10 (0.02) & 1.80 (0.01) & 6.27 (0.02) & 0.02 & 4.3\\
		& $J$=2--1, $K$=0 E$^{+}$  & 0.49 & 0.92 (0.02) & 0.33 (0.01) & 1.12 (0.02) & 0.02 & 4.3 \\
%
IC~348-1    & $J$=2--1, $K$=1 E$^{-}$  & 0.93 & 1.25 (0.02) & 0.80 (0.01)  & 1.21 (0.02)  & 0.02 & 9.1\\
		   & $J$=2--1, $K$=0 A$^{+}$ & 1.31 & 1.24	(0.02) & 1.13 (0.01) &  1.81 (0.02) & 0.02 & 9.3 \\
		   & $J$=2--1, $K$=0 E$^{+}$  & 0.25 & 1.14 (0.02) & 0.22 (0.01) &  0.40 (0.02) & 0.02 & 9.3 \\
IC~348-2    & $J$=2--1, $K$=1 E$^{-}$  & 0.49 & 0.90	(0.01) & 0.53 (0.02) & 0.49 (0.01) & 0.02 & 8.8 \\
		  & $J$=2--1, $K$=0 A$^{+}$ & 0.69 & 0.89	(0.01) & 0.76 (0.02) & 0.77 (0.01) & 0.02 & 8.5\\
                   & $J$=2--1, $K$=0 E$^{+}$  & 0.13 & 0.98	(0.02) & 0.15 (0.02) & 0.16 (0.01) & 0.02 & 8.9 \\
IC~348-3   & $J$=2--1, $K$=1 E$^{-}$  & 0.24 &  1.48	(0.02) & 0.26 (0.02) & 0.40 (0.05) & 0.02 & 10 \\
                   & $J$=2--1, $K$=0 A$^{+}$ & 0.31 & 1.59 (0.02)  & 0.35 (0.02) & 0.57 (0.05) & 0.02 & 10 \\
IC~348-4    & $J$=2--1, $K$=1 E$^{-}$  & 0.72 & 0.88 (0.01) & 0.64 (0.01) & 0.76 (0.01) & 0.02 & 8.4\\
                    & $J$=2--1, $K$=0 A$^{+}$ & 1.11 & 0.84 (0.01) & 0.98 (0.01) & 1.07 (0.01) & 0.02 & 8.5 \\
                    & $J$=2--1, $K$=0 E$^{+}$  & 0.20 & 0.83	(0.01) & 0.18 (0.01) & 0.20 (0.01) & 0.02 & 8.6 \\
Barnard~5   & $J$=2--1, $K$=1 E$^{-}$  & 0.43 & 1.00 (0.02) & 0.36 (0.01) & 0.43 (0.02) & 0.02 & 9.9 \\
                      & $J$=2--1, $K$=0 A$^{+}$ & 0.60 & 1.02 (0.02) & 0.51 (0.01) & 0.63 (0.02) & 0.02 & 9.7 \\
%
B1-1    & $J$=2--1, $K$=1 E$^{-}$  & 1.16 & 1.35 (0.02) &  1.25 (0.02) & 1.94 (0.02) & 0.02 & 6.5 \\
            & $J$=2--1, $K$=0 A$^{+}$ & 1.60 & 1.27	(0.02) &  1.72 (0.02) & 2.44 (0.02) & 0.02 & 6.3 \\
	   & $J$=2--1, $K$=0 E$^{+}$  & 0.18 & 1.33 (0.02) &  0.19 (0.02) & 0.25 (0.02) & 0.02 & 6.3 \\
B1-2   & $J$=2--1, $K$=1 E$^{-}$  & 1.33  & 0.86 (0.01) & 0.90 (0.01) & 1.25 (0.01) & 0.02 & 6.5 \\
           & $J$=2--1, $K$=0 A$^{+}$ & 1.62  & 0.93 (0.02) & 1.10 (0.01) & 1.64 (0.01) & 0.02  & 6.7 \\
           & $J$=2--1, $K$=0 E$^{+}$  & 0.18  & 0.89 (0.01) & 0.12 (0.01) & 0.15 (0.01)	& 0.02 & 6.7 \\
B1-3   & $J$=2--1, $K$=1 E$^{-}$  & 2.24 & 1.16 (0.02) & 2.86 (0.02) & 3.23 (0.02) & 0.02 & 6.5 \\
           & $J$=2--1, $K$=0 A$^{+}$ & 2.97 & 1.19	(0.02) & 3.80 (0.02) & 4.47 (0.02) & 0.02 & 6.3 \\
           & $J$=2--1, $K$=0 E$^{+}$  & 0.45 & 1.22	(0.02) & 0.58 (0.02) & 0.66 (0.02) & 0.02 & 6.3  \\
B1-4    & $J$=2--1, $K$=1 E$^{-}$   & 0.66 & 0.90 (0.01) & 0.72 (0.02) & 0.63 (0.01) & 0.02 & 7.2 \\
            & $J$=2--1, $K$=0 A$^{+}$  & 0.98 & 0.88 (0.01) & 1.06 (0.02) & 0.86 (0.01) & 0.02 & 7.0 \\
            & $J$=2--1, $K$=0 E$^{+}$   & 0.10 & 0.80 (0.01) & 0.11 (0.02) & 0.09 (0.01) & 0.02 & 7.1 \\
B1-5   & $J$=2--1, $K$=1 E$^{-}$  & 0.29 & 0.92 (0.02) & 0.27 (0.01) & 0.34 (0.01) & 0.02 & 7.2\\
           & $J$=2--1, $K$=0 A$^{+}$ & 0.40 & 1.10 (0.02) & 0.37 (0.01) & 0.53 (0.02) & 0.02 & 7.0  \\
L1455-1 & $J$=2--1, $K$=1 E$^{-}$   & 0.19 & 0.61 (0.01) & 0.16 (0.01) & 0.11 (0.01) & 0.02 & 5.2\tablenotemark{f} \\
               & $J$=2--1, $K$=0 A$^{+}$  & 0.22 & 0.79 (0.01) & 0.18 (0.01) & 0.17 (0.01) & 0.02 & 5.4\tablenotemark{f} \\
%
L1455-2  & $J$=2--1, $K$=1 E$^{-}$  & 0.34 & 1.24 (0.02) & 0.41 (0.02)  & 0.55 (0.02) & 0.02 &  4.8\tablenotemark{f} \\
                & $J$=2--1, $K$=0 A$^{+}$ & 0.49 & 1.27 (0.02) & 0.59 (0.02) & 0.87 (0.02) & 0.02 &  5.0\tablenotemark{f}  \\
L1455-3  & $J$=2--1, $K$=1 E$^{-}$  & 0.33 & 1.67 (0.03) & 0.26 (0.01) & 0.54 (0.03) & 0.02 & 5.6\tablenotemark{f} \\
                & $J$=2--1, $K$=0 A$^{+}$ & 0.49 & 1.65 (0.03) & 0.40 (0.01) & 0.78(0.02) & 0.02 & 5.8\tablenotemark{f}  \\
L1455-4  & $J$=2--1, $K$=1 E$^{-}$  & 0.33 & 1.46 (0.02) & 0.36 (0.02) & 0.61 (0.02) & 0.02 & 5.2\tablenotemark{f} \\
                 & $J$=2--1, $K$=0 A$^{+}$ & 0.45 & 1.41 (0.02) & 0.50 (0.02) & 0.88 (0.02) & 0.02 & 5.0\tablenotemark{f} \\
\hline
\enddata
\tablenotetext{a}{Obtained by the Gaussian fit.}
\tablenotetext{b}{The wing components are excluded in the Gaussian fit.}
\tablenotetext{c}{Derived by using the C$_{2}$H velocity widths.}
\tablenotetext{d}{The wing components are included in calculating the integrated intensity.}
\tablenotetext{e}{The rms noise averaged over the line width.}
\tablenotetext{f}{Only for the L1455 region, $V$$_{\rm{LSR}}$ is corrected by 9~km~s$^{-1}$ due to the problem of the NRO 45~m.
It is recovered by using the $V$$_{\rm{LSR}}$ obtained by IRAM 30~m.}
\end{deluxetable}
\end{longrotatetable}

\begin{longrotatetable}
\begin{deluxetable}{l l l l l l l l l c c c c c c c c}
\tabletypesize{\scriptsize}
\rotate
\tablecaption{Line Parameters of the C$_{2}$H ($N$=3--2) lines Observed with IRAM 30~m \label{t6}}
\tablewidth{0pt}
\tablehead{
\colhead{IDs} & \colhead{Transition} & \colhead{$T$$_{\rm{MB}}$\tablenotemark{a}} & \colhead{$dv$} & 
\colhead{$\int$$T$$_{\rm{MB}}$$dv$} 
& \colhead{rms\tablenotemark{b}} & \colhead{$V$$_{\rm{LSR}}$} \\
\colhead{} & \colhead{$J,F$} & \colhead{[K]} & \colhead{[km~s$^{-1}$]} & \colhead{[K km~s$^{-1}$]} & \colhead{[K]} & \colhead{[km~s$^{-1}$]}}
\startdata
NGC~1333-1  & $J$=5/2--3/2, $F$=2--1 & 0.16 &  2.62 (0.20) & 0.45 (0.02)  & 0.03 & 6.4 \\
			& $J$=5/2--3/2, $F$=3--2 & 0.16 &  2.27 (0.20) & 0.39 (0.02)  & 0.03 & 6.9 \\
			& $J$=7/2--5/2, $F$=3--2 & 0.22 &  2.43 (0.20) & 0.57 (0.02)  & 0.03 & 6.6 \\
			& $J$=7/2--5/2, $F$=4--3 & 0.27 &  2.10 (0.20) & 0.60 (0.02)  & 0.03 & 6.9 \\
NGC~1333-2  & $J$=5/2--3/2, $F$=2--1 & 0.42 & 1.30 (0.07)  & 0.58	(0.02) & 0.03 & 7.6\\
			& $J$=5/2--3/2, $F$=3--2 & 0.60 & 1.19 (0.04)  & 0.76	(0.02) &  0.03 & 7.6 \\
			& $J$=7/2--5/2, $F$=3--2& 0.61 & 1.26 (0.05)  & 0.83	(0.02) &  0.03 & 7.5 \\
			& $J$=7/2--5/2, $F$=4--3& 0.72 & 1.45 (0.04)  & 1.10	(0.03) &  0.03 & 7.5\\
NGC~1333-3  & $J$=5/2--3/2, $F$=2--1 & 0.99 &  0.91 (0.22) & 0.96	(0.05) &  0.03 & 8.2 \\
			& $J$=5/2--3/2, $F$=3--2 & 1.37 &  0.97 (0.22) & 1.41	(0.05) &  0.03 & 8.2\\
			& $J$=7/2--5/2, $F$=3--2& 1.46 &  0.94 (0.22) & 1.46	(0.05) &  0.03 & 8.2 \\
			& $J$=7/2--5/2, $F$=4--3& 1.83 &  0.95 (0.22) & 1.85	(0.05)  & 0.03 & 8.1 \\
NGC~1333-4  & $J$=5/2--3/2, $F$=2--1 & 0.36 &  0.92 (0.06) & 0.36	(0.02) &  0.03 & 8.7 \\
			& $J$=5/2--3/2, $F$=3--2 & 0.62 &  0.85 (0.03) & 0.56	(0.02) &  0.03 & 8.7\\
			& $J$=7/2--5/2, $F$=3--2& 0.65 &  0.97 (0.03)  & 0.67	(0.02) &  0.03 & 8.6 \\
			& $J$=7/2--5/2, $F$=4--3& 0.76 &   1.05 (0.03)  & 0.85	(0.02)  & 0.03 & 8.6 \\
NGC~1333-5  & $J$=5/2--3/2, $F$=2--1 & 0.56 &  0.94 (0.04) & 0.56	(0.02)  & 0.03 & 7.7 \\
			& $J$=5/2--3/2, $F$=3--2 & 0.78 &  0.90 (0.03) & 0.75	(0.02) & 0.03 & 7.7 \\
			& $J$=7/2--5/2, $F$=3--2& 0.74 &  0.94 (0.03) & 0.74	(0.02)  & 0.03 & 7.7 \\
			& $J$=7/2--5/2, $F$=4--3& 0.97 &  0.93 (0.02) & 0.95	(0.02)  & 0.03 & 7.6 \\
NGC~1333-6  & $J$=5/2--3/2, $F$=2--1 & 0.85 &   0.80 (0.02)   & 0.72	(0.02)  & 0.02 & 7.2 \\
			& $J$=5/2--3/2, $F$=3--2 & 1.17 &   0.85 (0.02)   & 1.05	(0.02)  & 0.02 & 7.3\\
			& $J$=7/2--5/2, $F$=3--2& 1.21 &   0.85 (0.02)  & 1.09	(0.02)  & 0.02 & 7.2\\
			& $J$=7/2--5/2, $F$=4--3& 1.53 &   0.86 (0.01)  & 1.39	(0.02)  & 0.02 & 7.2 \\
NGC~1333-7  & $J$=5/2--3/2, $F$=2--1 & 0.26 &  1.76 (0.22)  & 0.49	(0.02)  & 0.02 & 8.0 \\
			& $J$=5/2--3/2, $F$=3--2 & 0.40 &  1.33 (0.22)  & 0.56	(0.02)  & 0.02 & 8.0 \\
			& $J$=7/2--5/2, $F$=3--2& 0.44 &  1.46 (0.22)  & 0.68	(0.02)  & 0.02 & 8.0 \\
			& $J$=7/2--5/2, $F$=4--3& 0.58 &  1.27 (0.22) & 0.78	(0.02)  & 0.02 & 7.9 \\
NGC~1333-8  & $J$=5/2--3/2, $F$=2--1 & 0.44 &  1.06 (0.04)  & 0.49	(0.02) & 0.02 & 6.6 \\
			& $J$=5/2--3/2, $F$=3--2 & 0.64 &  1.09 (0.03)   & 0.75	(0.02) & 0.02 & 6.6 \\
			& $J$=7/2--5/2, $F$=3--2& 0.66 &  1.06 (0.03)  & 0.75	(0.02) & 0.02 & 6.6 \\
			& $J$=7/2--5/2, $F$=4--3& 0.89 &  1.12 (0.02)  & 1.06	(0.02) & 0.02 & 6.6\\
NGC~1333-9  & $J$=5/2--3/2, $F$=2--1 & 0.33 &  1.54 (0.07)  & 0.54	(0.02) & 0.02 & 7.6 \\
			& $J$=5/2--3/2, $F$=3--2 & 0.47 &  1.45 (0.05)  & 0.72	(0.02) & 0.02 & 7.6 \\
			& $J$=7/2--5/2, $F$=3--2& 0.51 &  1.58 (0.06)  & 0.86	(0.02) & 0.02 & 7.6 \\
			& $J$=7/2--5/2, $F$=4--3& 0.65 &  1.47 (0.04)   & 1.01	(0.02) & 0.02 & 7.5 \\
%
NGC~1333-10	& $J$=5/2--3/2, $F$=3--2 & 0.10 &  1.25 (0.18)  & 0.13	(0.02) & 0.02 & 7.6 \\
			& $J$=7/2--5/2, $F$=3--2& 0.12 &  1.09 (0.18)  & 0.14	(0.02) & 0.02 & 7.5 \\
			& $J$=7/2--5/2, $F$=4--3& 0.13 &  1.49 (0.24)  & 0.21	(0.02) & 0.02 & 7.3 \\
NGC~1333-11 & $J$=5/2--3/2, $F$=2--1 & 0.18 &  0.63 (0.09)  & 0.12	(0.01) & 0.02 & 8.5 \\
			& $J$=5/2--3/2, $F$=3--2 & 0.27 &  0.58 (0.06)  & 0.17	(0.01) & 0.02 & 8.5\\
			& $J$=7/2--5/2, $F$=3--2& 0.25 &  0.56 (0.07)  & 0.15	(0.01) & 0.02 & 8.5\\
			& $J$=7/2--5/2, $F$=4--3& 0.31 &  0.50 (0.05)  & 0.17	(0.01) & 0.02 & 8.5\\
NGC~1333-12 & $J$=5/2--3/2, $F$=2--1 & 0.12 &  0.47 (0.14)   & 0.06	(0.01) & 0.02 & 7.5 \\
        	& $J$=5/2--3/2, $F$=3--2 & 0.14 &  0.55 (0.11)   & 0.08	(0.01) & 0.02 & 7.6\\
    		& $J$=7/2--5/2, $F$=3--2& 0.16 &  0.63 (0.09)   & 0.11	(0.01) & 0.02 & 7.5\\
			& $J$=7/2--5/2, $F$=4--3& 0.17 &  0.54 (0.07)   & 0.10	(0.01) & 0.02 & 7.5\\
NGC~1333-13 & $J$=5/2--3/2, $F$=2--1 & 0.14 &  0.53 (0.10)  & 0.08	(0.01) & 0.02 & 7.5\\
			& $J$=5/2--3/2, $F$=3--2 & 0.24 &  0.48 (0.06)  & 0.12	(0.01) & 0.02 & 7.5\\
			& $J$=7/2--5/2, $F$=3--2& 0.29 &  0.48 (0.05)  & 0.15	(0.01)&  0.02 & 7.5\\
			& $J$=7/2--5/2, $F$=4--3& 0.35 &  0.60 (0.05)  & 0.22	(0.01)&  0.02 & 7.5\\
NGC~1333-14 & $J$=5/2--3/2, $F$=2--1 & 0.13 &   1.15 (0.16)  & 0.16	(0.02) & 0.02 & 7.9 \\
			& $J$=5/2--3/2, $F$=3--2 & 0.29 &   0.70 (0.05)  & 0.21	(0.01) & 0.02 & 7.9 \\
			& $J$=7/2--5/2, $F$=3--2& 0.26 &   0.76 (0.06)  & 0.21 (0.01) & 0.02 & 7.8 \\
			& $J$=7/2--5/2, $F$=4--3& 0.35 &   0.68 (0.04)  & 0.25	(0.01) & 0.02 & 7.8 \\
%
NGC~1333-15	& $J$=5/2--3/2, $F$=3--2 & 0.10 &   0.87 (0.17) & 0.09 (0.02) & 0.02 & 8.3 \\
			& $J$=7/2--5/2, $F$=3--2& 0.10 &   0.96 (0.14)  & 0.10	(0.01) & 0.02 & 8.2\\
			& $J$=7/2--5/2, $F$=4--3& 0.12 &   0.61 (0.10)  & 0.08	(0.01)  & 0.02 & 8.2\\
NGC~1333-16 & $J$=5/2--3/2, $F$=2--1 & 0.10 &  1.36 (0.78) & 0.14 (0.03)  & 0.02 &  6.6 \\
			& $J$=5/2--3/2, $F$=3--2 & 0.14 &  2.34 (0.78) & 0.35 (0.03)  & 0.02 &  6.2 \\
			& $J$=7/2--5/2, $F$=3--2 & 0.14 &  1.85 (0.78) & 0.28 (0.03)  & 0.02 &  6.5 \\
			& $J$=7/2--5/2, $F$=4--3 & 0.20 &  2.41 (0.78) & 0.51 (0.03)  & 0.02 &  6.3 \\
NGC~1333-17 & $J$=5/2--3/2, $F$=2--1 & 0.86 & 1.01 (0.20)  & 0.97 (0.05) & 0.04 & 8.7 \\
			& $J$=5/2--3/2, $F$=3--2 & 1.15 & 1.13 (0.20)  & 1.38 (0.05) & 0.04 & 8.7 \\
			& $J$=7/2--5/2, $F$=3--2 & 1.24 & 1.07 (0.20)  & 1.40 (0.05) & 0.04 & 8.7 \\
			& $J$=7/2--5/2, $F$=4--3 & 1.54 & 1.12 (0.20)  & 1.84 (0.05) & 0.04 & 8.7 \\			
L1448-1     & $J$=5/2--3/2, $F$=2--1 & 0.44 &  1.44 (0.06)  & 0.68	(0.02) & 0.03 & 4.0\\
			& $J$=5/2--3/2, $F$=3--2 & 0.66 &  1.48 (0.05) & 1.05	(0.03)&  0.03 & 4.0\\
			& $J$=7/2--5/2, $F$=3--2& 0.74 &  1.38 (0.04)  & 1.09	(0.03)&  0.03 & 4.0\\
			& $J$=7/2--5/2, $F$=4--3& 0.86 &  1.56 (0.04)  & 1.42	(0.03)&  0.03 & 3.9 \\
L1448-2    & $J$=5/2--3/2, $F$=2--1 & 1.01 &   1.70 (0.22) & 1.83	(0.07) &  0.03 & 4.8 \\
			& $J$=5/2--3/2, $F$=3--2 & 1.33 &  1.72 (0.22) & 2.44	(0.07) & 0.03 & 4.8 \\
			& $J$=7/2--5/2, $F$=3--2& 1.36 &  1.80 (0.22)   & 2.61	(0.07) & 0.03 & 4.8 \\
			& $J$=7/2--5/2, $F$=4--3& 1.59 &  1.85 (0.22)  & 3.12	(0.07) & 0.03 & 4.7\\
L1448-3  & $J$=5/2--3/2, $F$=2--1 & 0.45 & 1.03 (0.06)  & 0.50	(0.02) & 0.03 & 5.1 \\
			& $J$=5/2--3/2, $F$=3--2 & 0.71 &  0.96 (0.04)  & 0.73	(0.02)& 0.03 & 5.2\\
			& $J$=7/2--5/2, $F$=3--2& 0.77 &  0.98 (0.03)   & 0.81	(0.02) & 0.03 & 5.2\\
			& $J$=7/2--5/2, $F$=4--3& 0.93 &  1.09 (0.03)   & 1.08	(0.02) & 0.03 & 5.1\\
L1448-4  & $J$=5/2--3/2, $F$=2--1 & 1.05 &  0.89 (0.02)  & 0.99	(0.02) &  0.03 & 4.0 \\
			& $J$=5/2--3/2, $F$=3--2 & 1.36 & 0.91 (0.02)   & 1.33 (0.02) & 0.03 & 4.1\\
			& $J$=7/2--5/2, $F$=3--2& 1.45 &  0.96 (0.02)  & 1.47 (0.02) &  0.03 & 4.0 \\
			& $J$=7/2--5/2, $F$=4--3& 1.58 &  0.98 (0.01)  & 1.65 (0.01) &  0.03 & 4.0 \\
L1448-5     & $J$=5/2--3/2, $F$=2--1 & 0.22 &  0.57 (0.12) & 0.13	(0.03)& 0.05 & 4.1 \\
			& $J$=5/2--3/2, $F$=3--2 & 0.35 &  0.61 (0.08)   & 0.23	(0.03) & 0.05 & 4.1\\
			& $J$=7/2--5/2, $F$=3--2& 0.38 &  0.50 (0.08)   & 0.20	(0.03) & 0.05 & 4.2\\
			& $J$=7/2--5/2, $F$=4--3& 0.40 &  0.87 (0.08)  & 0.37	(0.03) &  0.05 & 4.2\\
IC~348-1     & $J$=5/2--3/2, $F$=2--1 & 0.60 &   0.78 (0.04) & 0.49	(0.02) & 0.03 & 9.1\\
			& $J$=5/2--3/2, $F$=3--2 & 0.75 &   0.82 (0.03)  & 0.65	(0.02) & 0.03 & 9.1 \\
			& $J$=7/2--5/2, $F$=3--2& 0.81 &   0.80 (0.03)   & 0.69	(0.02) & 0.03 & 9.1 \\
			& $J$=7/2--5/2, $F$=4--3& 1.03 &  0.83 (0.02)  & 0.91	(0.02)& 0.03 & 9.0 \\
IC~348-2    & $J$=5/2--3/2, $F$=2--1 & 0.61 &  1.00 (0.03)  & 0.66	(0.02) & 0.02 & 8.6 \\
			& $J$=5/2--3/2, $F$=3--2 & 0.85 &  1.03 (0.02)  & 0.93	(0.02)& 0.02 & 8.7 \\
			& $J$=7/2--5/2, $F$=3--2& 0.89 &  1.03 (0.02)  & 0.97	(0.02) & 0.02 & 8.6 \\
			& $J$=7/2--5/2, $F$=4--3& 1.08 &  1.03 (0.02)  & 1.19	(0.02) & 0.02 & 8.6 \\
IC~348-3     & $J$=5/2--3/2, $F$=2--1 & 0.16 &   0.97 (0.11)  & 0.16 (0.02) & 0.02 & 10 \\
			& $J$=5/2--3/2, $F$=3--2 & 0.23 &  1.45 (0.12) & 0.24	(0.02) & 0.02 & 10 \\
			& $J$=7/2--5/2, $F$=3--2& 0.26 &  1.01 (0.06)   & 0.28	(0.02) & 0.02 & 10 \\
			& $J$=7/2--5/2, $F$=4--3& 0.32 &   1.19 (0.07)  & 0.41	(0.02) & 0.02 & 10 \\
IC~348-4  & $J$=5/2--3/2, $F$=2--1 & 0.45 &  0.82 (0.04)    & 0.39	(0.02) & 0.02 & 8.5 \\
			& $J$=5/2--3/2, $F$=3--2 & 0.70 & 0.82 (0.02)  & 0.61	(0.02)&  0.02 & 8.5 \\
			& $J$=7/2--5/2, $F$=3--2& 0.70 & 0.83 (0.03)  & 0.60	(0.02)&  0.02 & 8.5 \\
			& $J$=7/2--5/2, $F$=4--3& 0.95 & 0.84 (0.02)  & 0.85	(0.02) & 0.02 & 8.5 \\
Barnard~5     & $J$=5/2--3/2, $F$=2--1 & 0.81 &   0.77 (0.02)  & 0.66 (0.02)&  0.03 & 10 \\
			& $J$=5/2--3/2, $F$=3--2 & 1.13 &  0.77 (0.02)  & 0.93	(0.02)&  0.03 & 10 \\
			& $J$=7/2--5/2, $F$=3--2& 1.12 &  0.82 (0.02) & 0.98	(0.02) & 0.03 & 10 \\
			& $J$=7/2--5/2, $F$=4--3& 1.38 &  0.85 (0.02)  & 1.24	(0.02)&  0.03 & 10 \\
B1-1  & $J$=5/2--3/2, $F$=2--1 & 0.27 & 1.06 (0.09)  & 0.30 (0.02) &0.03 & 6.4 \\
			& $J$=5/2--3/2, $F$=3--2 & 0.41 &  0.92 (0.07)  & 0.40 (0.02)&  0.03 & 6.3 \\
			& $J$=7/2--5/2, $F$=3--2& 0.43 &  1.01 (0.07)  & 0.46	(0.02)&  0.03 & 6.4 \\
			& $J$=7/2--5/2, $F$=4--3& 0.54 &  1.04 (0.06)  & 0.60	(0.02)& 0.03 & 6.3\\
B1-2  & $J$=5/2--3/2, $F$=2--1 & 0.74 &  0.62 (0.02)  & 0.49	(0.01) & 0.03 & 6.6 \\
			& $J$=5/2--3/2, $F$=3--2 & 0.92 & 0.67 (0.02)  & 0.65	(0.02)&  0.03 & 6.6 \\
			& $J$=7/2--5/2, $F$=3--2& 0.97 & 0.61 (0.02)  & 0.63	(0.01)&  0.03 & 6.6 \\
			& $J$=7/2--5/2, $F$=4--3& 1.07 &  0.66 (0.02) & 0.76	(0.02) & 0.03 & 6.6\\
B1-3  & $J$=5/2--3/2, $F$=2--1 & 0.27 & 1.20 (0.10)   & 0.34	(0.02)&  0.03 & 6.4\\
			& $J$=5/2--3/2, $F$=3--2 & 0.39 & 1.08 (0.06) & 0.45	(0.02) & 0.03 & 6.3\\
			& $J$=7/2--5/2, $F$=3--2& 0.39 & 1.59 (0.11) & 0.65	(0.03) & 0.03 & 6.4 \\
			& $J$=7/2--5/2, $F$=4--3& 0.57 & 0.93 (0.04)  & 0.57	(0.02) & 0.03 & 6.3\\
B1-4  & $J$=5/2--3/2, $F$=2--1 & 0.50 &  0.98 (0.22) & 0.52	(0.03) & 0.03 & 6.9\\
			& $J$=5/2--3/2, $F$=3--2 & 0.71 & 1.04 (0.22)  & 0.78 (0.03) & 0.03 & 6.9 \\
			& $J$=7/2--5/2, $F$=3--2& 0.77 & 1.03 (0.22)  & 0.85 (0.03)&  0.03 & 6.9 \\
			& $J$=7/2--5/2, $F$=4--3& 0.92 & 1.05 (0.22) & 1.02 (0.03) & 0.03 & 6.8\\
B1-5  & $J$=5/2--3/2, $F$=2--1 & 0.29 & 0.82 (0.07)  & 0.25 (0.02) & 0.03 & 7.0 \\
			& $J$=5/2--3/2, $F$=3--2 & 0.40 & 0.76 (0.04)   & 0.32	(0.02) & 0.03 & 7.0 \\
			& $J$=7/2--5/2, $F$=3--2& 0.41 &  0.96 (0.06)   & 0.41	(0.02) & 0.03 &7.0 \\
			& $J$=7/2--5/2, $F$=4--3& 0.50 & 0.90 (0.05)  & 0.48	(0.02) & 0.03 & 7.0 \\
L1455-1  & $J$=5/2--3/2, $F$=2--1 & 0.43 & 0.72 (0.22)   & 0.33	(0.02)&  0.02 & 5.2\\
			& $J$=5/2--3/2, $F$=3--2 & 0.59 & 0.81 (0.22)  & 0.50 (0.02)& 0.02 & 5.2\\
			& $J$=7/2--5/2, $F$=3--2& 0.63 & 0.81 (0.22)  & 0.54 (0.02)&  0.02 & 5.2\\
			& $J$=7/2--5/2, $F$=4--3& 0.69 & 0.87 (0.22)  & 0.63 (0.02)&  0.02 & 5.2\\
L1455-2  & $J$=5/2--3/2, $F$=2--1 & 0.52 &  1.09 (0.04) & 0.61	(0.02)&  0.03 & 4.9 \\
			& $J$=5/2--3/2, $F$=3--2 & 0.69 & 1.15 (0.03)   & 0.84 (0.02)&  0.03 & 4.9\\
			& $J$=7/2--5/2, $F$=3--2& 0.80 & 1.12 (0.03)  & 0.96	(0.02)&  0.03 & 4.9\\
			& $J$=7/2--5/2, $F$=4--3& 0.96 & 1.20 (0.03)   & 1.23  (0.02)&  0.03 & 4.9\\
L1455-3  & $J$=5/2--3/2, $F$=2--1 & 0.43 &  0.79 (0.04)   & 0.36	(0.02)&  0.03 & 4.8 \\
			& $J$=5/2--3/2, $F$=3--2 & 0.66 &  0.75 (0.03) & 0.52	(0.02)&  0.03 & 4.8 \\
			& $J$=7/2--5/2, $F$=3--2& 0.67 & 0.73 (0.03)  & 0.52	(0.02)&  0.03 & 4.8 \\
			& $J$=7/2--5/2, $F$=4--3& 0.88 &  0.75 (0.02)  & 0.71	(0.02)&   0.03 & 4.7 \\
L1455-4  & $J$=5/2--3/2, $F$=2--1 & 0.23 &  1.01 (0.08)  & 0.25	(0.02) & 0.03 & 5.3  \\
			& $J$=5/2--3/2, $F$=3--2 & 0.32 &  1.00 (0.06)   & 0.33 (0.02)&  0.03 & 5.3 \\
			& $J$=7/2--5/2, $F$=3--2& 0.35 & 1.09 (0.06)  & 0.40	(0.02) & 0.03 & 5.3 \\
			& $J$=7/2--5/2, $F$=4--3& 0.44 & 1.08 (0.05) & 0.50	(0.02) & 0.03 & 5.2 \\
\hline
\enddata
\tablenotetext{a}{Obtained by the Gaussian fit.}
\tablenotetext{b}{The rms noise averaged over the line width.}
\end{deluxetable}
\end{longrotatetable}

\begin{longrotatetable}
\begin{deluxetable}{l l l l l l l l l c c c c c c c c}
\tabletypesize{\scriptsize}
\rotate
\tablecaption{Line Parameters of the c-C$_{3}$H$_{2}$ (3$_{2,1}$--2$_{1,2}$) lines Observed with IRAM 30~m \label{t7}}
\tablewidth{0pt}
\tablehead{
\colhead{IDs} & \colhead{$T$$_{\rm{MB}}$\tablenotemark{a}} & \colhead{$dv$} & 
\colhead{$\int$$T$$_{\rm{MB}}$$dv$} &  
\colhead{rms\tablenotemark{b}} &  \colhead{$V$$_{\rm{LSR}}$} \\
\colhead{} & \colhead{[K]} & \colhead{[km~s$^{-1}$]} & \colhead{[K km~s$^{-1}$]} & \colhead{[K]} & \colhead{[km~s$^{-1}$]}}
\startdata
NGC~1333-1   & 0.23 & 0.35 (0.08)  &	0.08 (0.02)   &	0.04 & 6.5 \\
NGC~1333-2   & 0.25 &	1.03 (0.10)  &	0.27 (0.02)  & 0.03 & 7.5\\
NGC~1333-3   & 0.18 &	0.64 (0.08)  &	0.12 (0.02)  &	0.03 & 8.1 \\
NGC~1333-4   & 0.09	& 0.72 (0.26)	& 0.07 (0.02)	& 0.03 & 8.6 \\
NGC~1333-5   & 0.17 &	0.69 (0.12) &	0.13 (0.02)	& 0.03 & 7.7 \\
NGC~1333-6   & 0.66 &	0.67 (0.03) &	0.47 (0.02)  &	0.03 & 7.3 \\
NGC~1333-7   & 0.21 &	0.60 (0.08)  &	0.14 (0.02)	& 0.03 & 8.1  \\	
NGC~1333-8   & 0.11 &	0.90 (0.21)  &	0.11 (0.02) &	0.03 & 6.6\\
NGC~1333-9   & ...	& ... &	 $<$ 0.04 & 0.03 & ... \\
NGC~1333-10	 & ...	& ... &  $<$ 0.04 & 0.02 & ... \\
NGC~1333-11  & 0.11 &  0.41 (0.11)  &	0.05 (0.01)&  0.02 & 8.4 \\
NGC~1333-12  & ...	& ... &	$<$ 0.03 	& 0.02 & ... \\
NGC~1333-13  & ...  & ... & $<$	0.03    & 0.02 & ... \\
NGC~1333-14  & 0.12 & 0.60 (0.11) & 	0.07 (0.01) &	 0.02 & 7.7 \\
NGC~1333-15	 & ...	& ...&	$<$ 0.03 & 0.02 & ... \\
NGC~1333-16   & 0.12 & 0.72 (0.08)  &	0.10 (0.01)  & 0.08 & 6.8 \\
NGC~1333-17   & 0.15 & 1.12 (0.04)  &	0.18 (0.01)  & 0.04 & 8.4 \\
L1448-1     & 0.50 &	1.22 (0.06) & 	0.66 (0.03) &	0.04 & 4.2\\
L1448-2     & 0.39 &	1.34 (0.07)  &	0.55 (0.03) &	  0.03 & 4.6\\
L1448-3     & 0.25 &	0.97 (0.10) &	0.26 (0.02) & 	0.03 & 4.8\\
L1448-4     & 0.39 &	0.52 (0.04) &	0.22 (0.02) &	 0.03 & 4.0 \\
L1448-5     & 0.33	& 0.46 (0.05) &	0.16 (0.02) &	0.03 & 4.1\\
IC~348-1     & 0.37 &	0.61 (0.05) &	0.24 (0.02) &  0.03 & 9.0 \\
IC~348-2     & 0.47 &	0.56 (0.03)	& 0.28 (0.01) &	0.02 & 8.8 \\
IC~348-3     & ...	& ...&	$<$ 0.04	& 0.02 & ...\\
IC~348-4     & 0.28	& 0.52 (0.04)  &	0.15 (0.01) &  0.02 & 8.5 \\
Barnard~5     & 0.22	& 0.89 (0.10)  &	0.21 (0.02)&	 0.02 & 10 \\
B1-1        & 0.31	& 0.89 (0.05)  & 0.29 (0.02)&	 0.02 & 6.3 \\
B1-2        & 0.36	& 0.65 (0.05)  & 0.25 (0.02)&	 0.02 & 6.5\\
B1-3        & 0.13	& 0.99 (0.15)  & 0.13 (0.02)&	 0.03 & 6.3\\
B1-4        & 0.36	& 0.75 (0.06)  & 0.28 (0.02)&  0.03 & 6.9\\
B1-5        & 0.20	& 0.53 (0.06)  & 0.11 (0.02)&	 0.03 & 6.9\\
L1455-1     & 0.28	& 0.53 (0.04)  & 0.16 (0.01)&	 0.02 & 5.2\\
L1455-2     & 0.19	& 1.01 (0.13)  & 0.20 (0.02)&	 0.02 & 5.0 \\
L1455-3     & 0.19	& 0.51 (0.07)  & 0.10 (0.01) & 0.03 & 4.8\\
L1455-4     & 0.10	& 0.35 (0.17)  & 0.03 (0.01)&	 0.02 & 5.1\\
\hline
\enddata
\tablenotetext{a}{Obtained by the Gaussian fit.}
\tablenotetext{b}{The rms noise averaged over the line width.}
\tablecomments{The errors are 1$\sigma$. The upper limit to the integrated intensity is calculated as 
$\int$$T$$_{\rm{MB}}$$dv$ $<$ 3~$\sigma$ $\times$ $\sqrt{({dv/dv_{\rm{res}}})}$$dv_{\rm{res}}$ where $dv$ is the assumed line width 
(0.8~km~s$^{-1}$) and $dv_{\rm{res}}$ is the velocity resolution per channel.}
\end{deluxetable}
\end{longrotatetable}


\begin{figure}
\epsscale{1}
\plotone{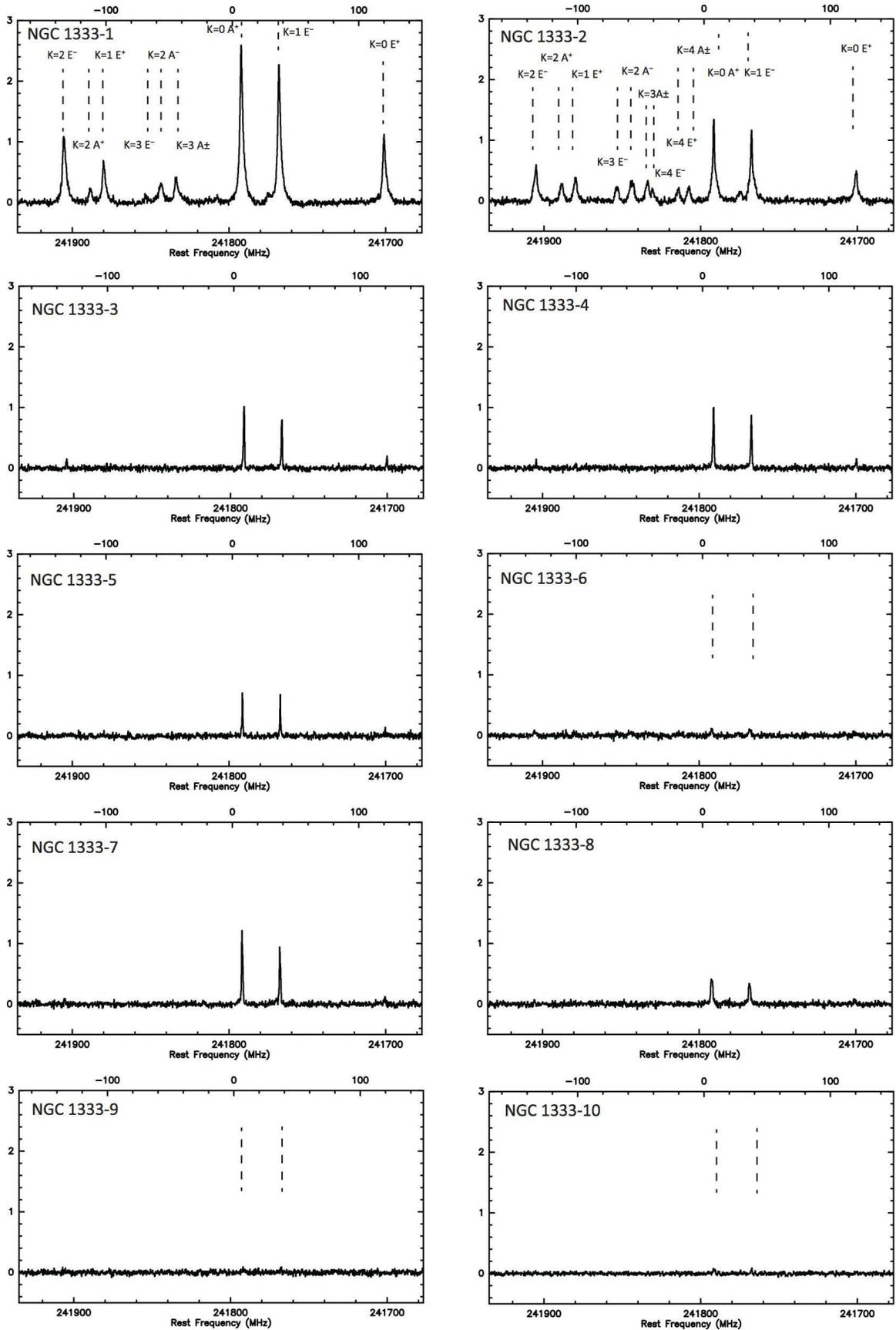}
\caption{Line profiles of CH$_{3}$OH ($J$=5--4) observed with IRAM 30~m.
The spectra taken with the wobbler switching are indicated with an asterisk (ASAI project).}
\label{fg9}
\end{figure}
\addtocounter{figure}{-1}

\begin{figure}
\epsscale{1}
\plotone{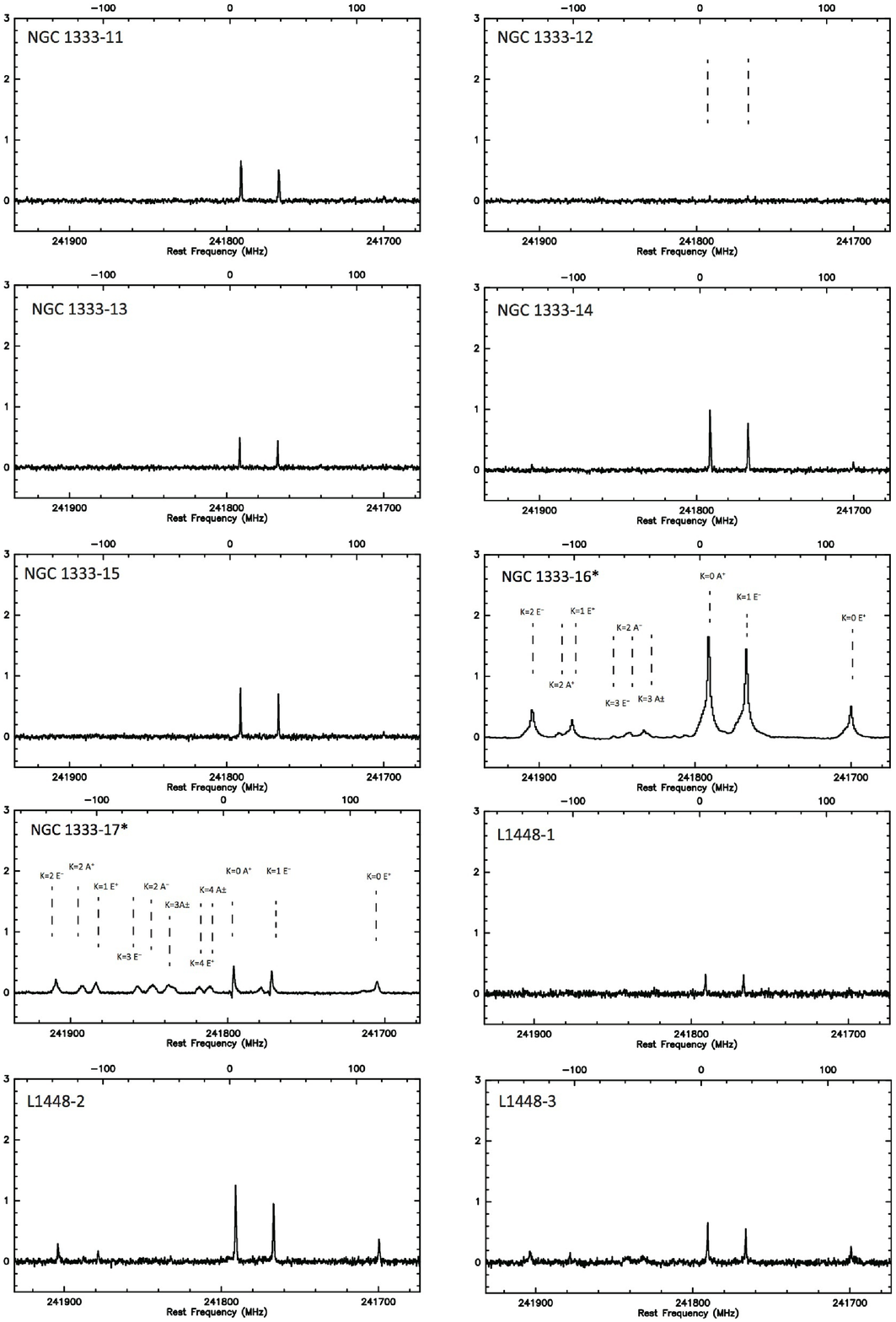}
\caption{Continued.}
\label{fg9}
\end{figure}
\addtocounter{figure}{-1}

\begin{figure}
\epsscale{1}
\plotone{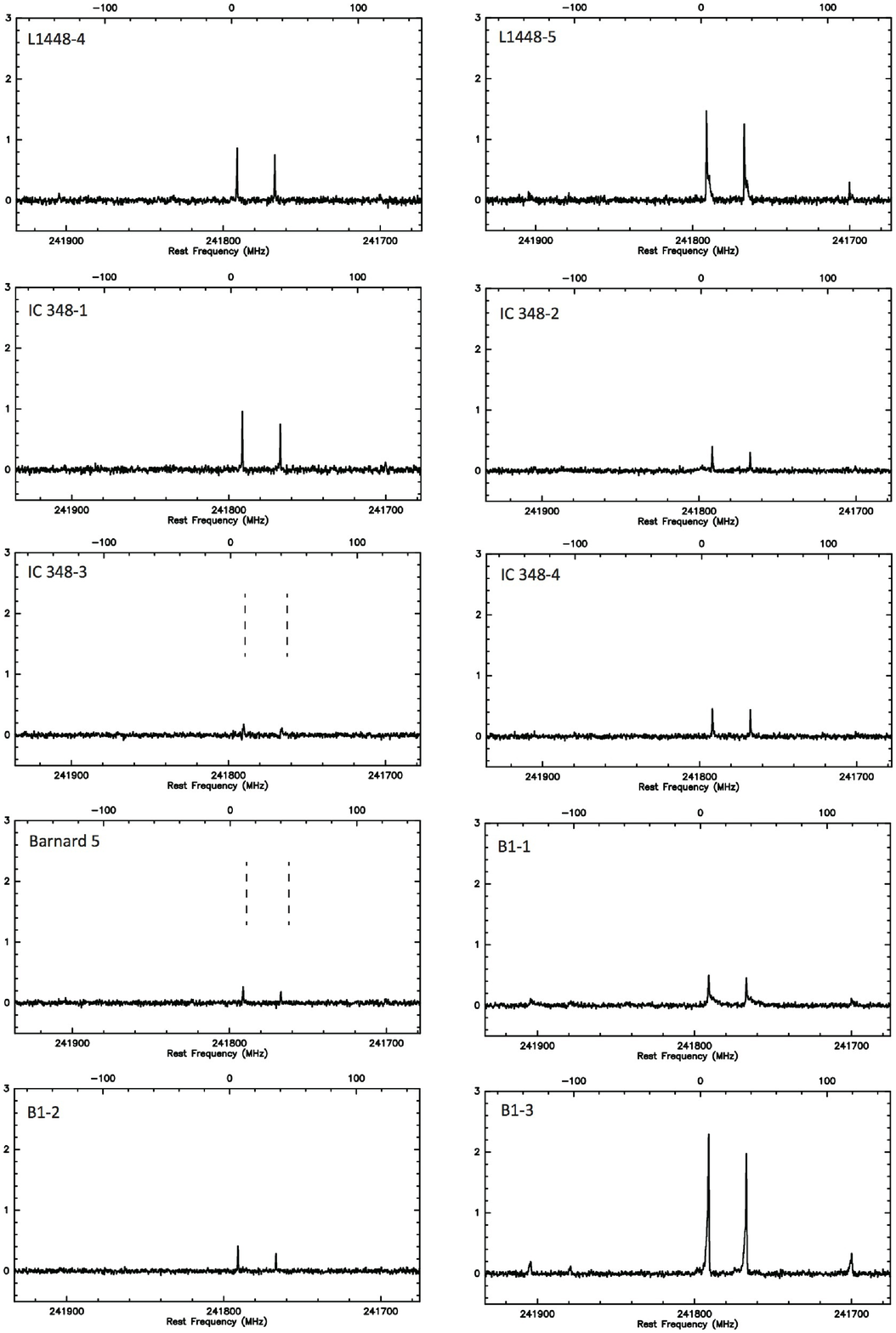}
\caption{Continued.}
\label{fg9}
\end{figure}
\addtocounter{figure}{-1}

\begin{figure}
\epsscale{1}
\plotone{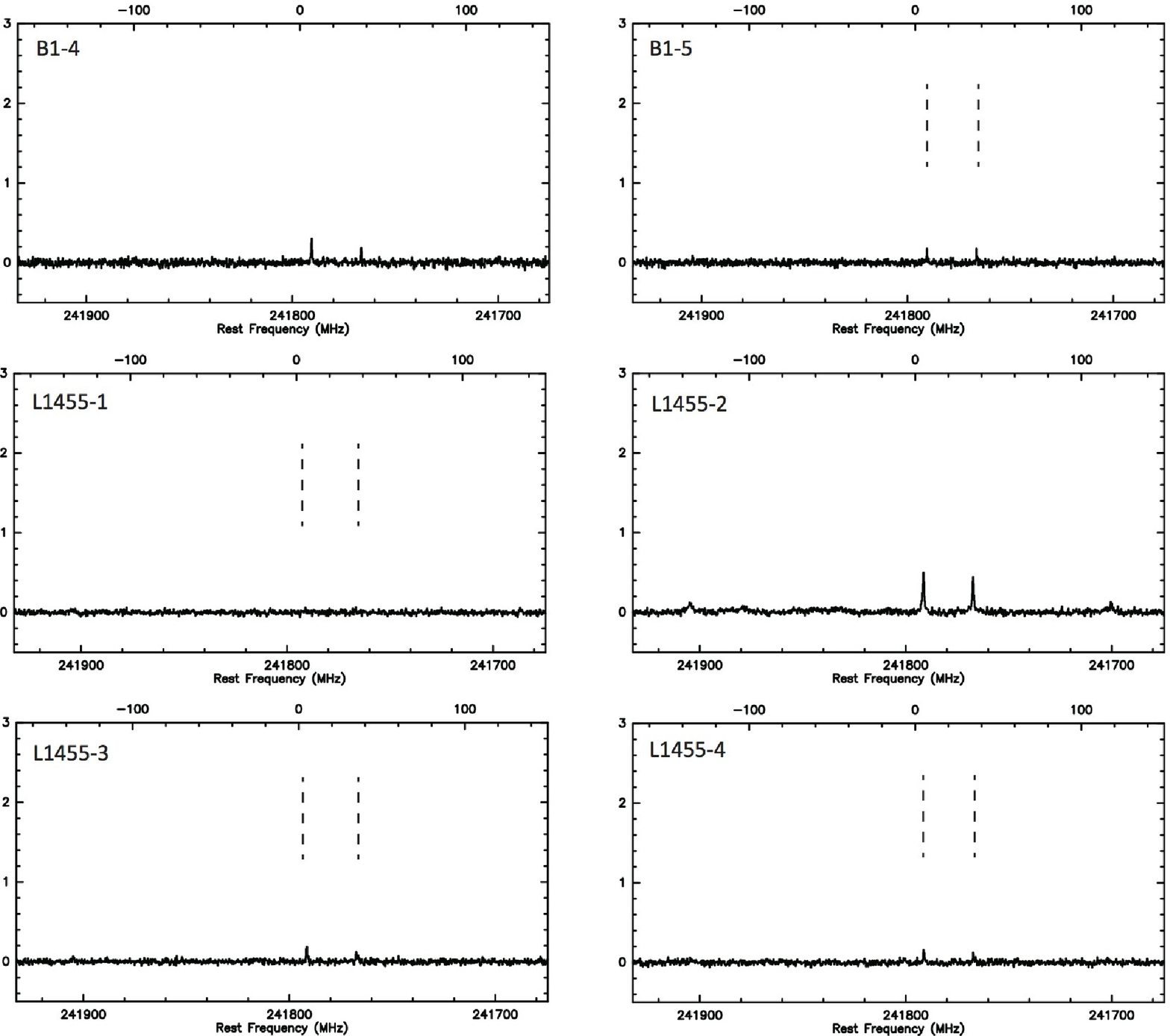}
\caption{Continued.}
\label{fg9}
\end{figure}

\begin{figure}
\epsscale{1}
\plotone{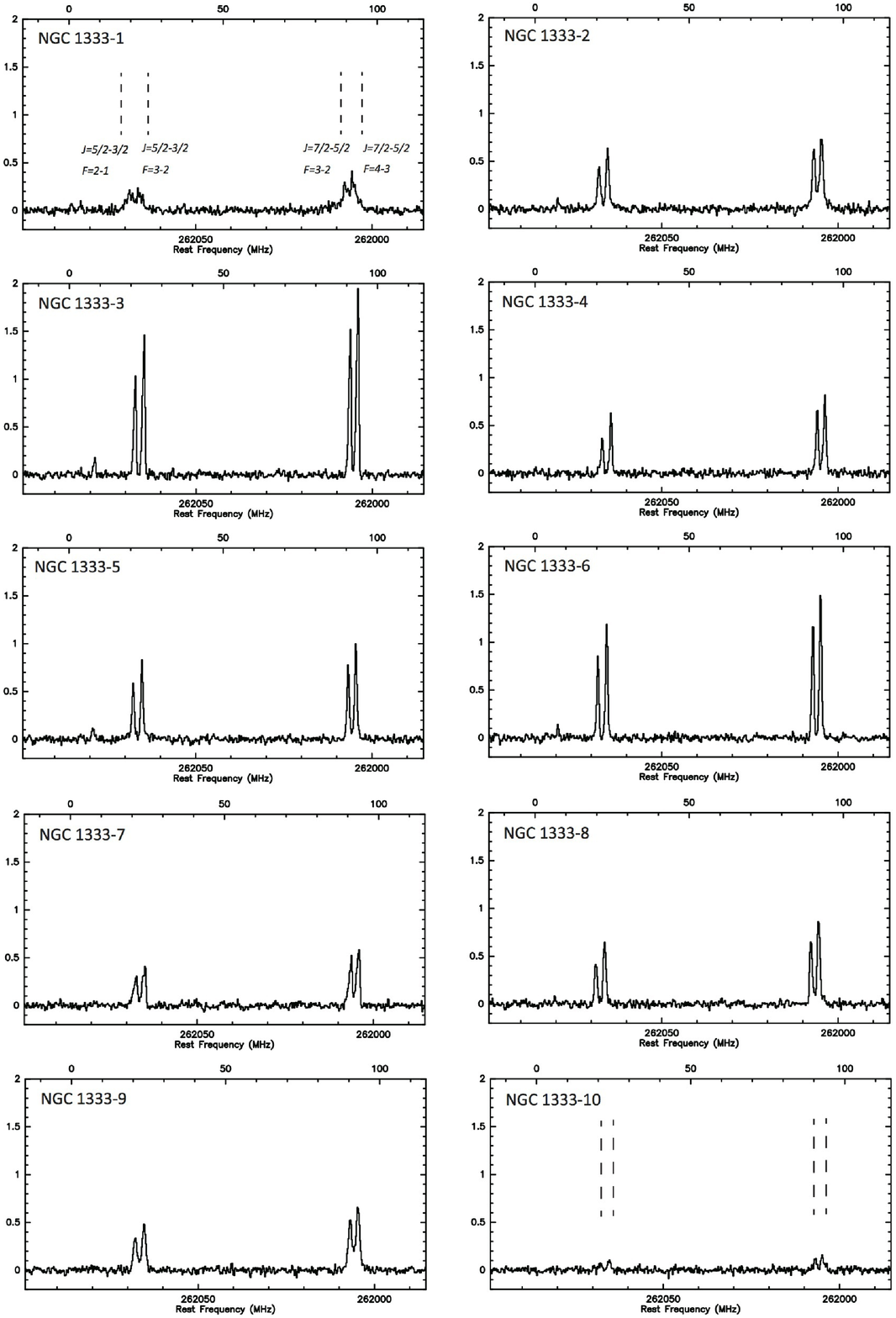}
\caption{Line profiles of C$_{2}$H ($N$=3--2) observed with IRAM 30~m.
The spectra taken with the wobbler switching are indicated with an asterisk (ASAI project).}
\label{fg10}
\end{figure}
\addtocounter{figure}{-1}

\begin{figure}
\epsscale{1}
\plotone{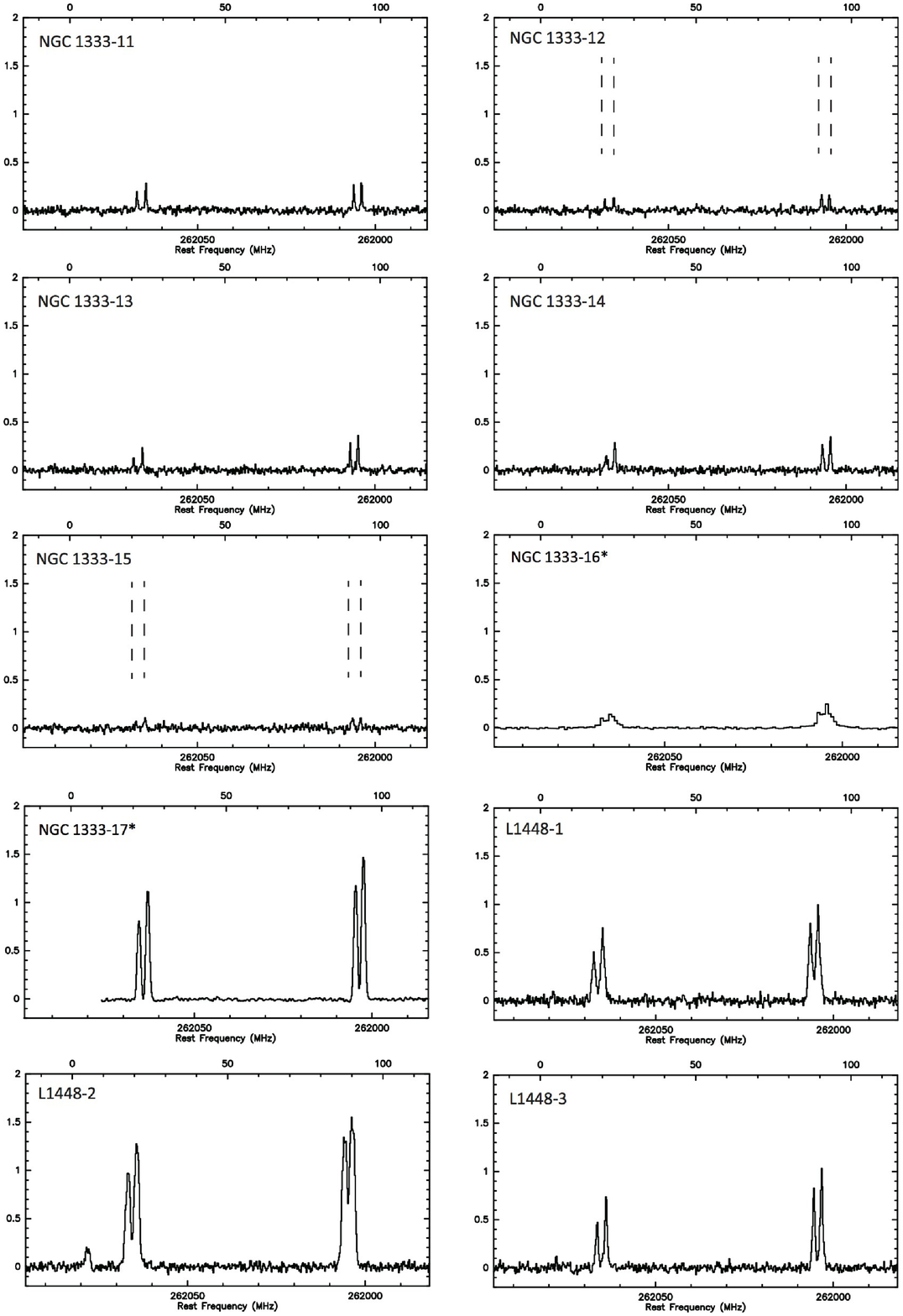}
\caption{Continued.}
\label{fg10}
\end{figure}
\addtocounter{figure}{-1}

\begin{figure}
\epsscale{1}
\plotone{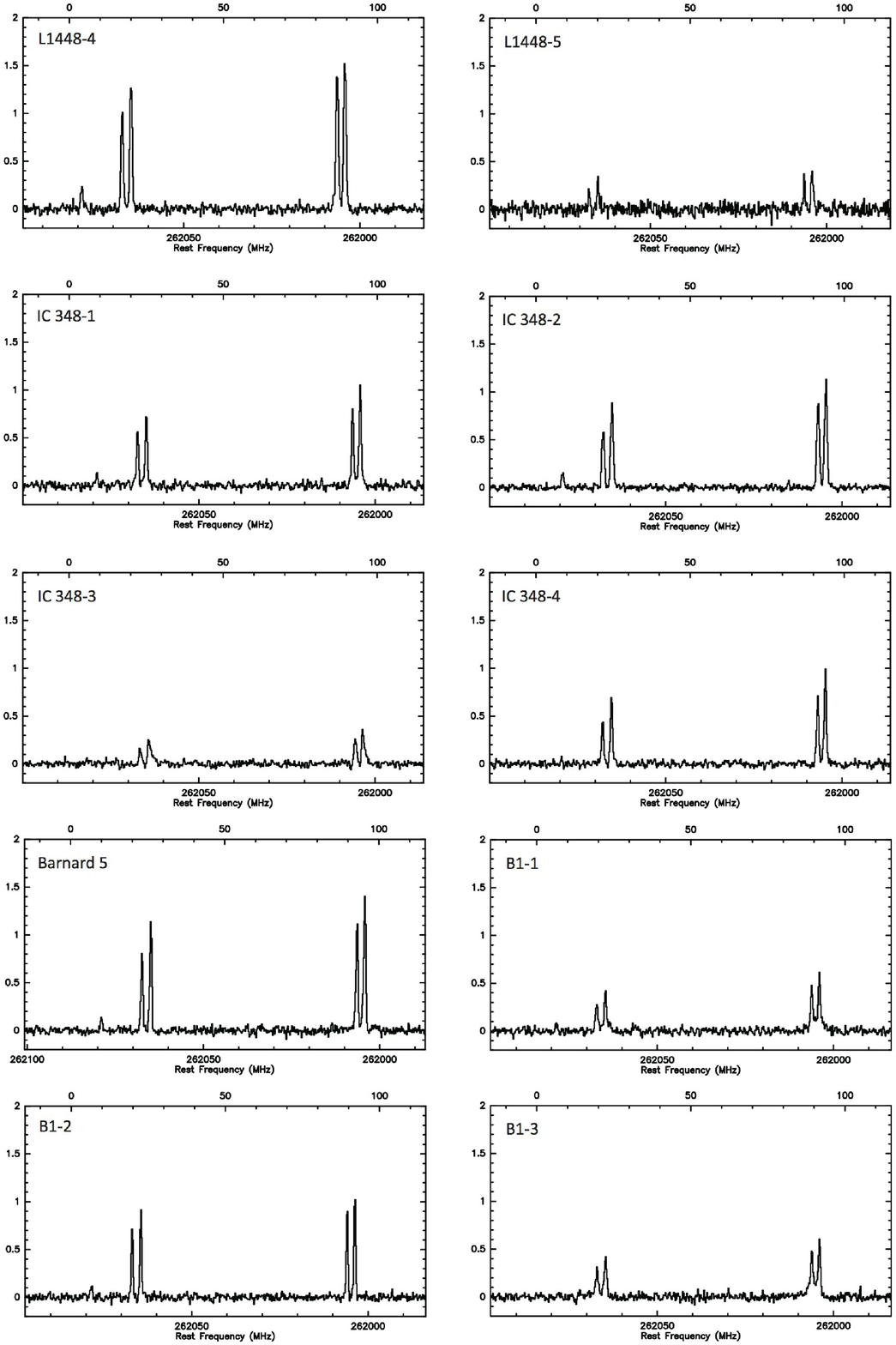}
\caption{Continued.}
\label{fg10}
\end{figure}
\addtocounter{figure}{-1}

\begin{figure}
\epsscale{1}
\plotone{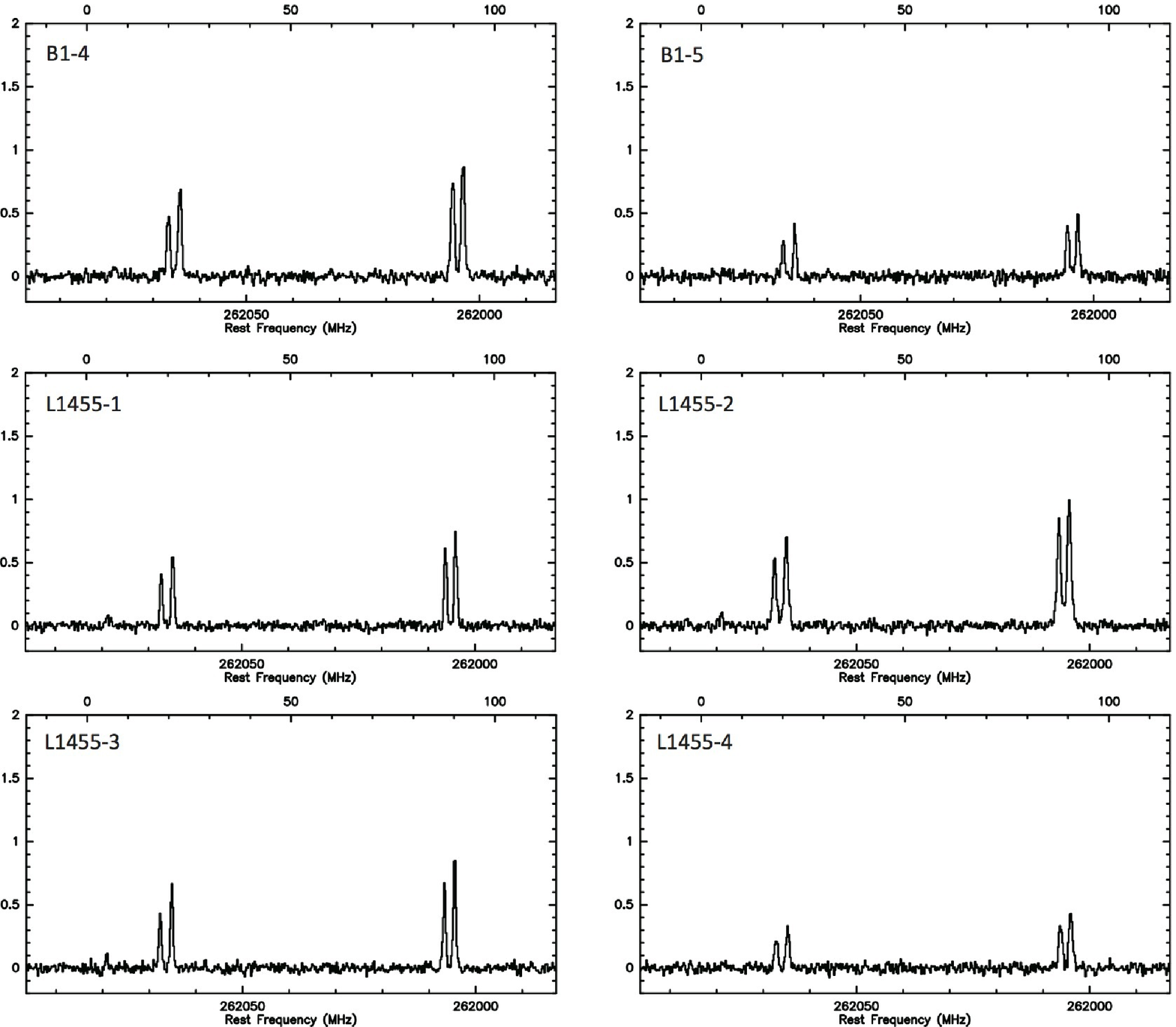}
\caption{Continued.}
\label{fg10}
\end{figure}

\begin{figure}
\epsscale{1}
\plotone{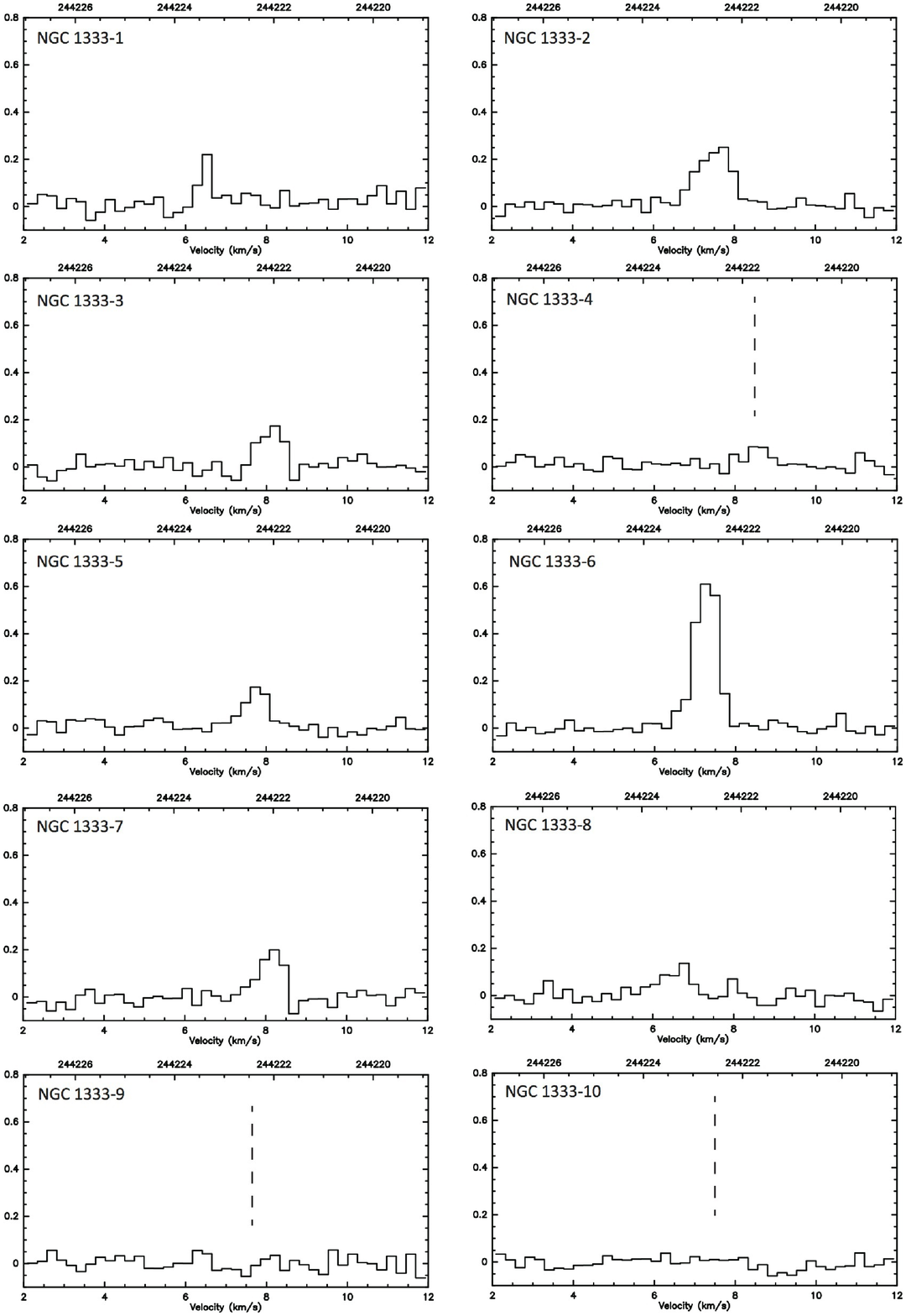}
\caption{Line profiles of c-C$_{3}$H$_{2}$ (3$_{2,1}$--2$_{1,2}$) observed with IRAM 30~m.
The spectra taken with the wobbler switching are indicated with an asterisk (ASAI project).}
\label{fg11}
\end{figure}
\addtocounter{figure}{-1}

\begin{figure}
\epsscale{1}
\plotone{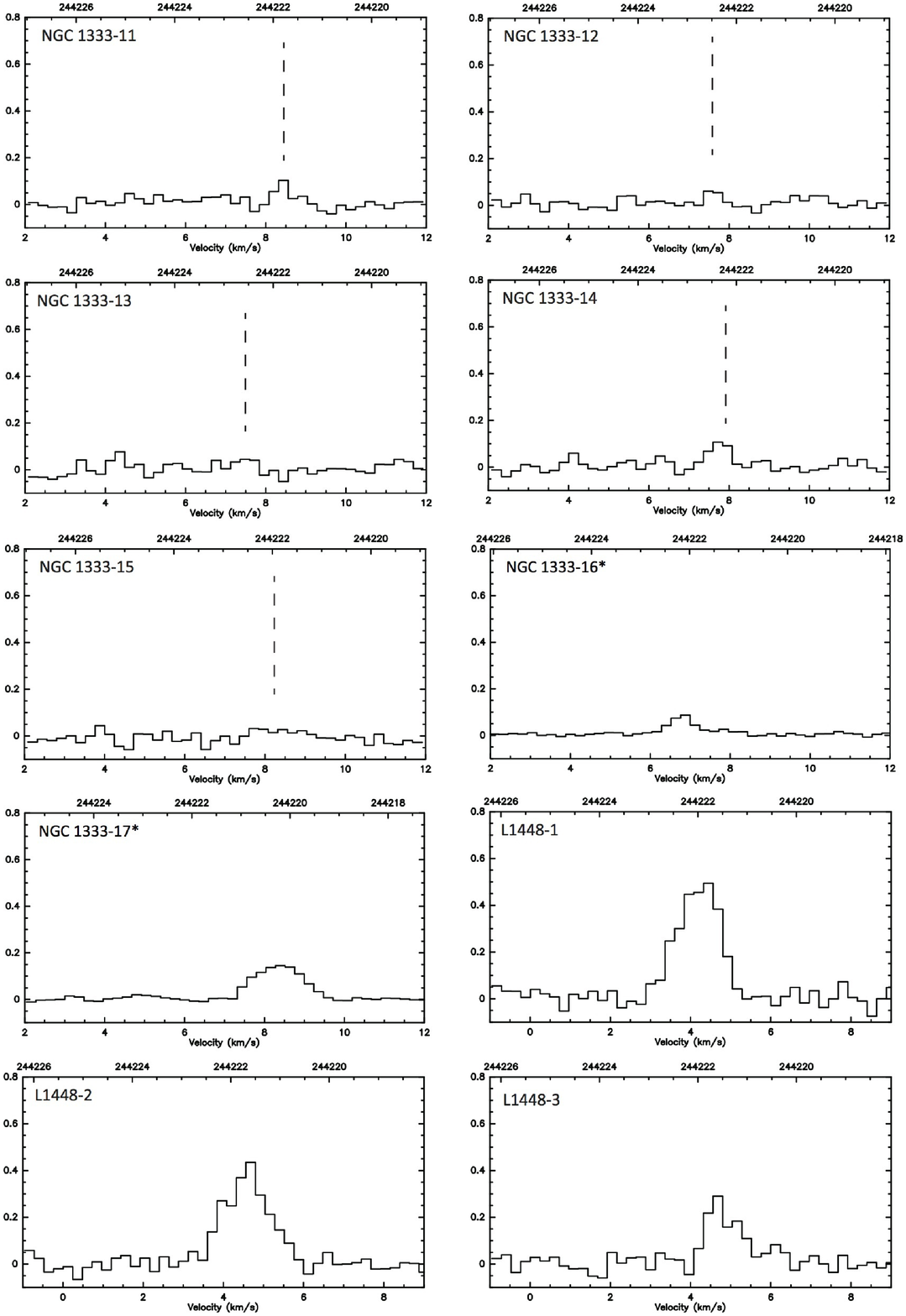}
\caption{Continued.}
\label{fg11}
\end{figure}
\addtocounter{figure}{-1}

\begin{figure}
\epsscale{1}
\plotone{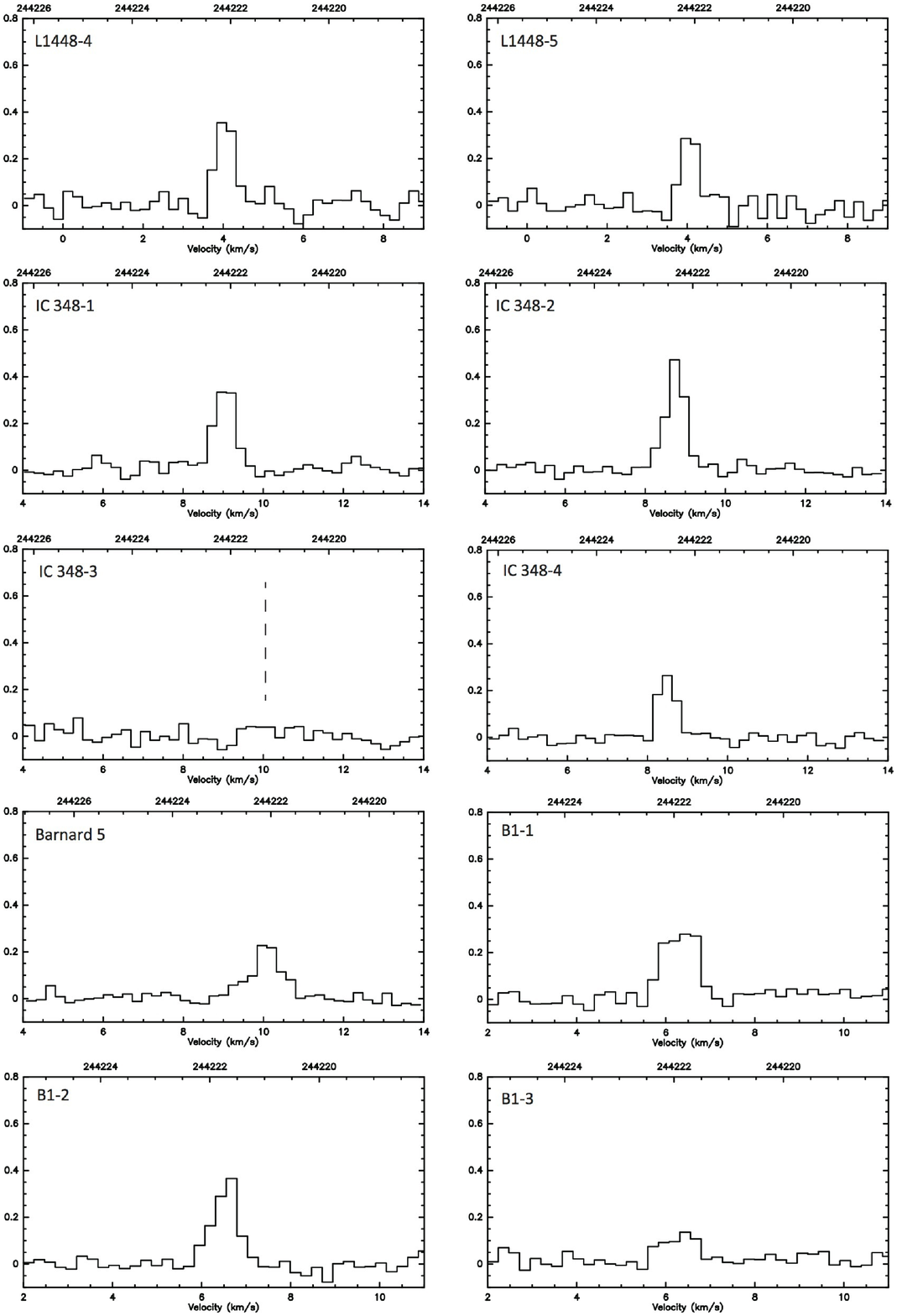}
\caption{Continued.}
\label{fg11}
\end{figure}
\addtocounter{figure}{-1}

\begin{figure}
\epsscale{1}
\plotone{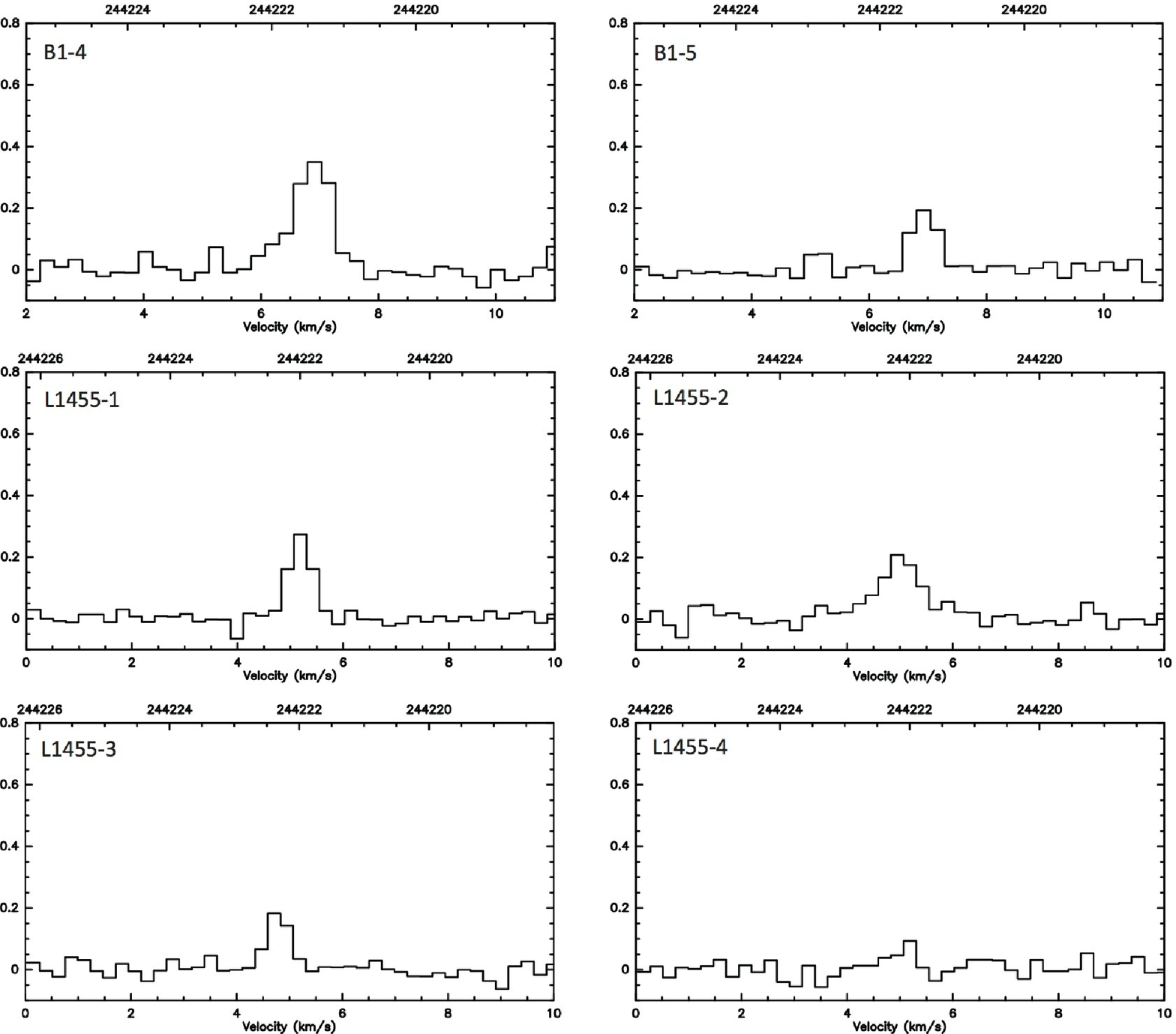}
\caption{Continued.}
\label{fg11}
\end{figure}

\begin{figure}
\rotate
\epsscale{1}
\plotone{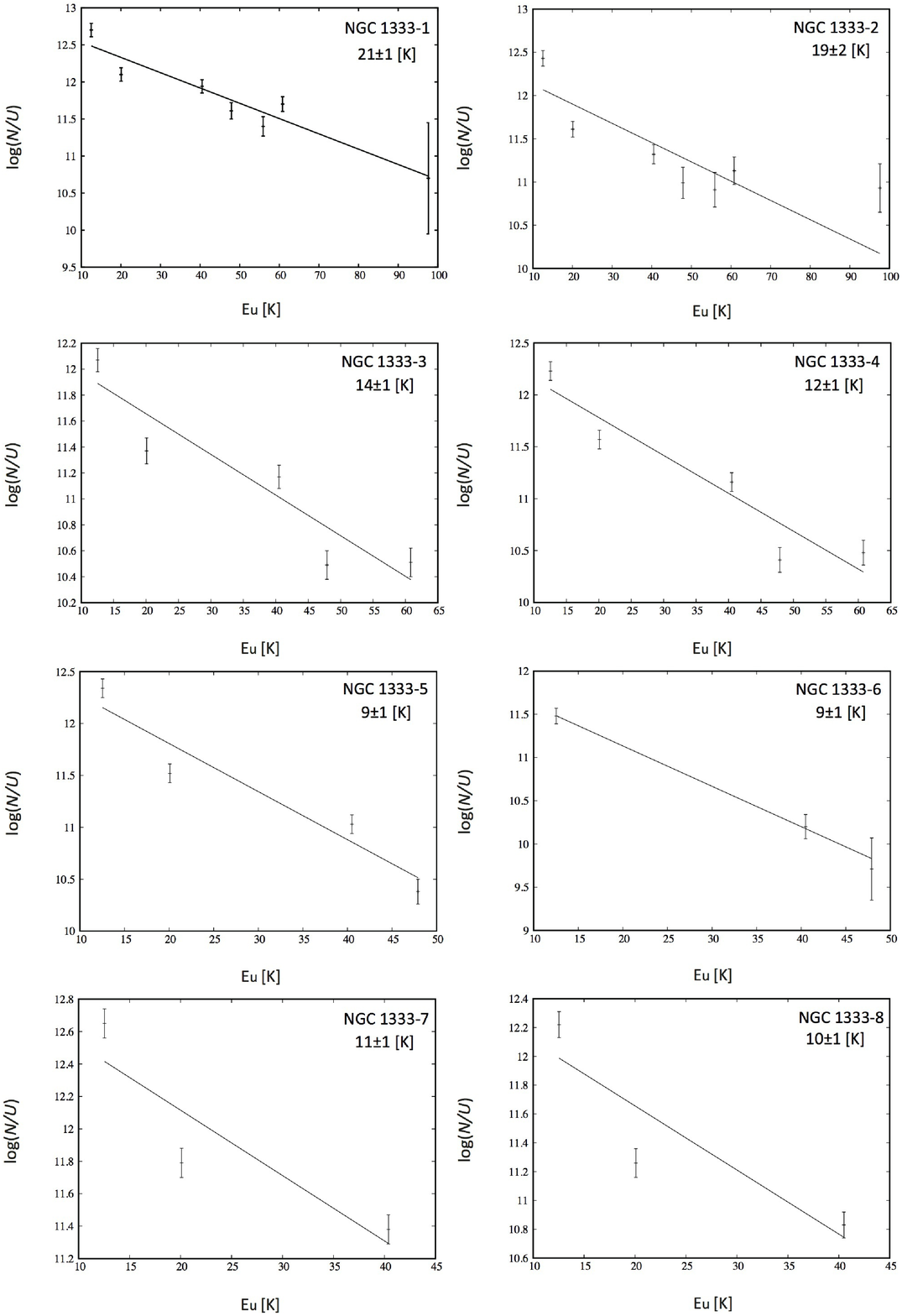}
\caption{Rotation diagrams of CH$_{3}$OH ($E$ state).  
Detected lines are marked by points with error bars.
A solid line indicates a single temperature fit.}
\label{fg12}
\end{figure}
\addtocounter{figure}{-1}

\begin{figure}
\rotate
\epsscale{1}
\plotone{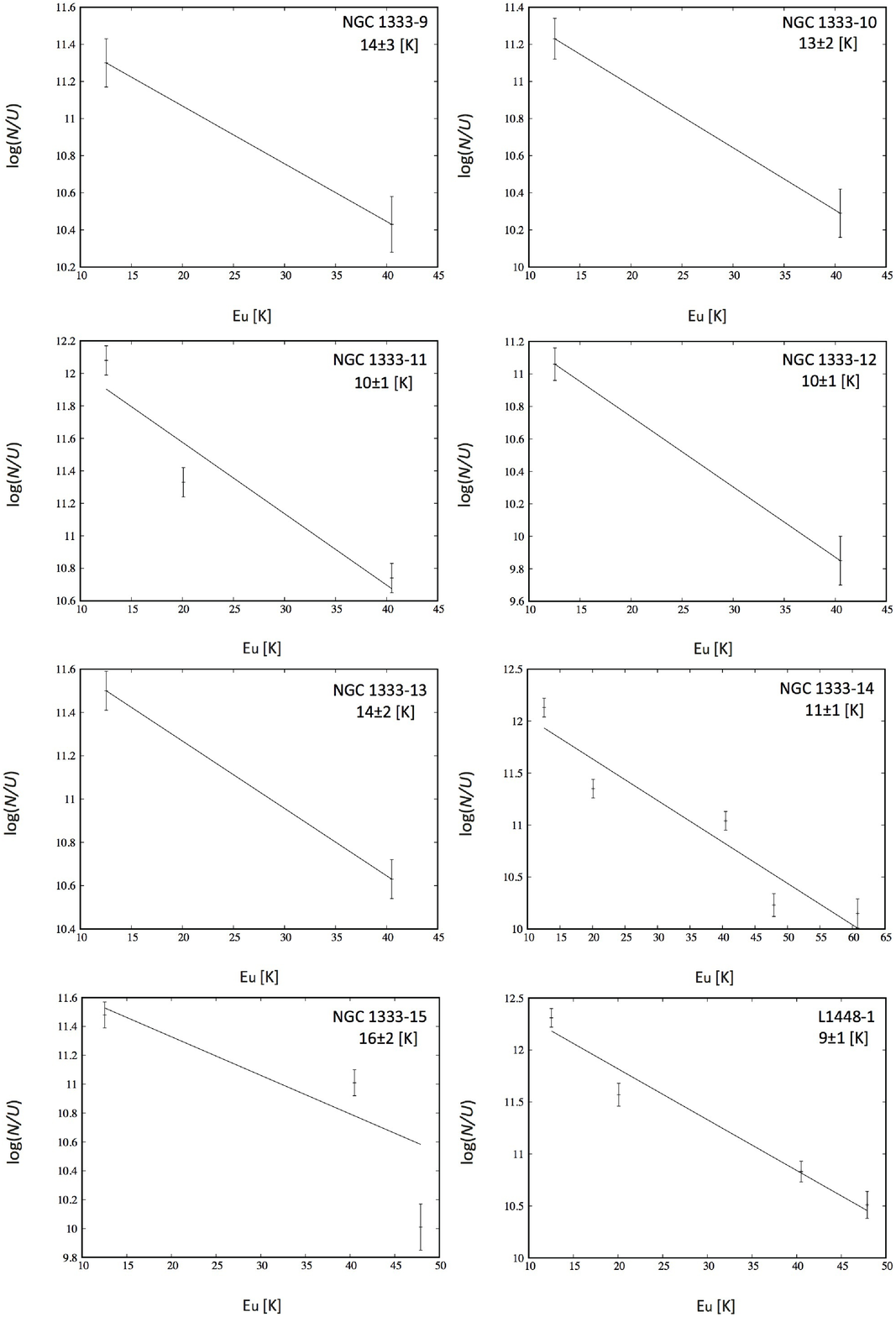}
\caption{Continued.}
\label{fg12}
\end{figure}
\addtocounter{figure}{-1}

\begin{figure}
\rotate
\epsscale{1}
\plotone{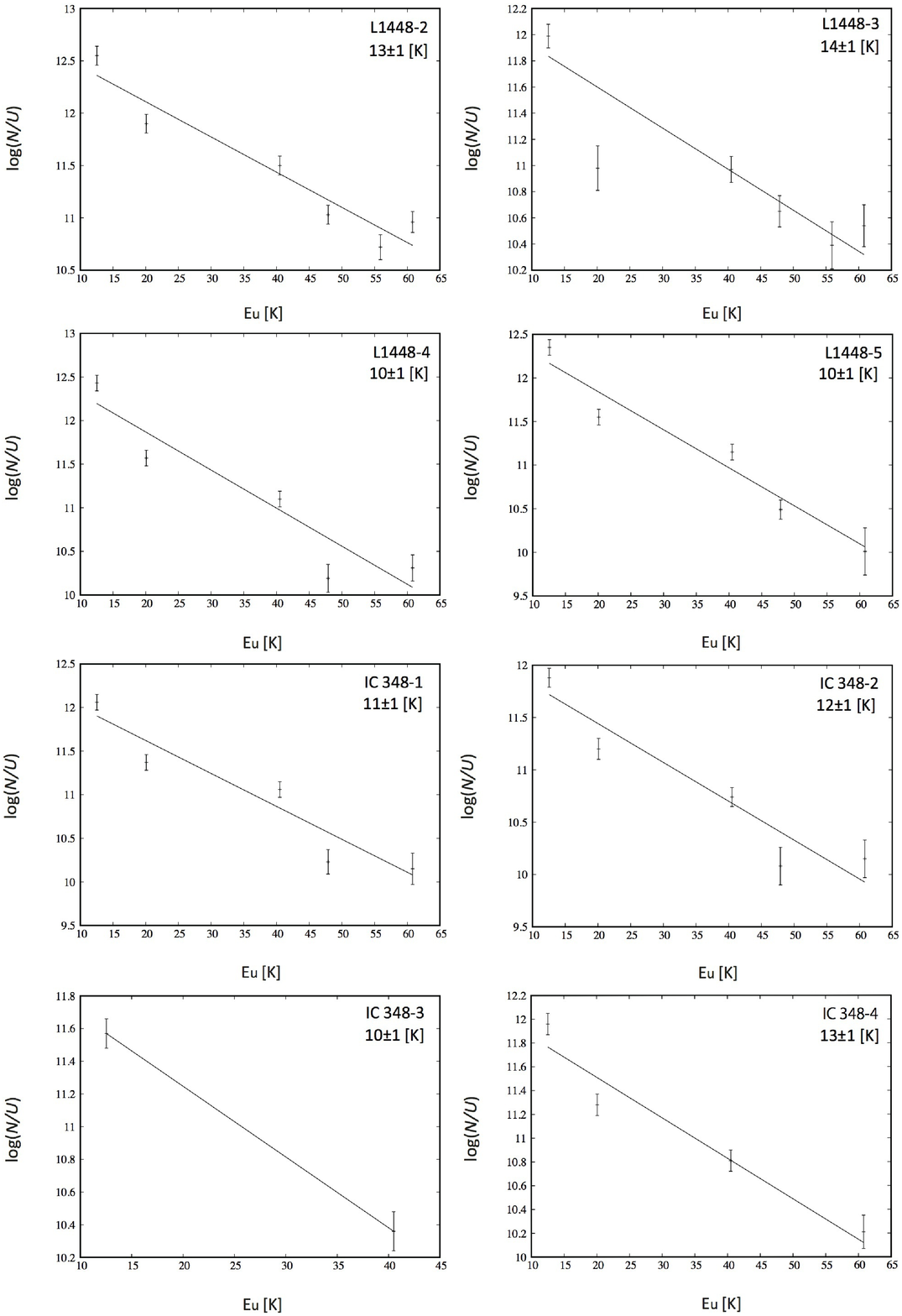}
\caption{Continued.}
\label{fg12}
\end{figure}
\addtocounter{figure}{-1}

\begin{figure}
\rotate
\epsscale{1}
\plotone{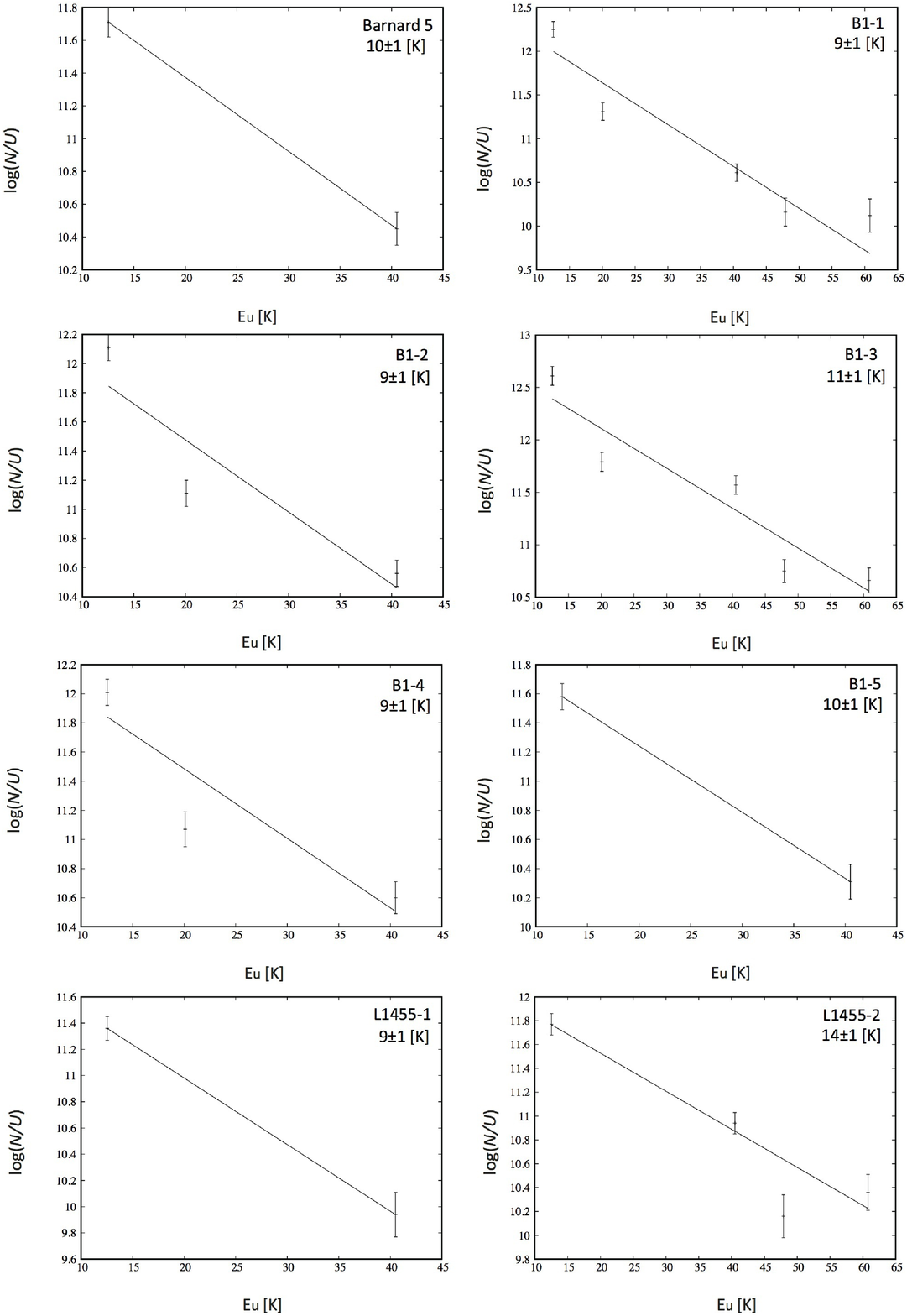}
\caption{Continued.}
\label{fg12}
\end{figure}
\addtocounter{figure}{-1}

\begin{figure}
\rotate
\epsscale{1}
\plotone{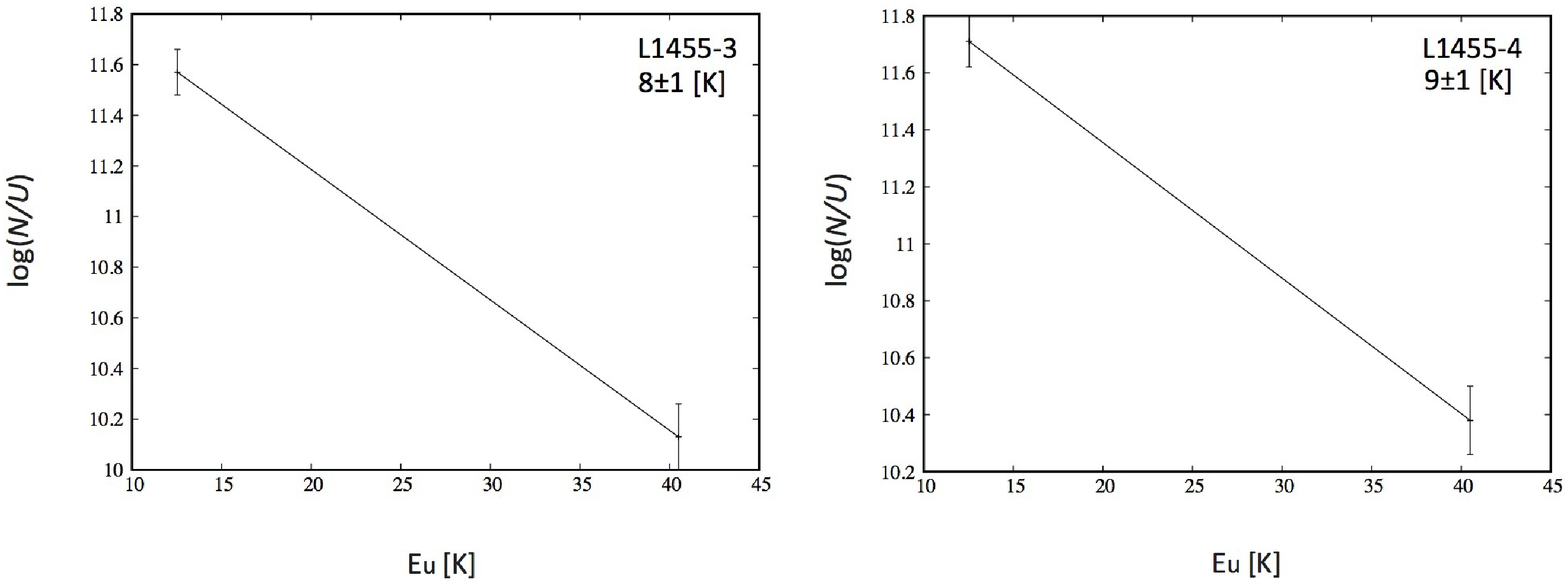}
\caption{Continued.}
\label{fg12}
\end{figure}
\addtocounter{figure}{-1}




\clearpage

\begin{thebibliography}{}

\bibitem[Bachiller et al.(1998)]{1998A&A...335..266B} Bachiller, R., Codella, C., Colomer, F., Liechti, S., \& Walmsley, C.~M.\ 1998, \aap, 335, 266 

\bibitem[Bachiller \& P{\'e}rez Guti{\'e}rrez(1997)]{1997ApJ...487L..93B} Bachiller, R., \& P{\'e}rez Guti{\'e}rrez, M.\ 1997, \apjl, 487, L93 

\bibitem[Bergin et al.(2016)]{2016ApJ...831..101B} Bergin, E.~A., Du, F., Cleeves, L.~I., et al.\ 2016, \apj, 831, 101 

\bibitem[Bizzocchi et al.(2014)]{2014A&A...569A..27B} Bizzocchi, L., Caselli, P., Spezzano, S., \& Leonardo, E.\ 2014, \aap, 569, A27 

\bibitem[Bottinelli et al.(2004)]{2004ApJ...617L..69B} Bottinelli, S., Ceccarelli, C., Neri, R., et al.\ 2004, \apjl, 617, L69 

\bibitem[Buckle \& Fuller(2002)]{2002A&A...381...77B} Buckle, J.~V., \& Fuller, G.~A.\ 2002, \aap, 381, 77 

\bibitem[Burke \& Hollenbach(1983)]{1983ApJ...265..223B} Burke, J.~R., \& Hollenbach, D.~J.\ 1983, \apj, 265, 223 

\bibitem[Cazaux et al.(2003)]{2003ApJ...593L..51C} Cazaux, S., Tielens, A.~G.~G.~M., Ceccarelli, C., et al.\ 2003, \apjl, 593, L51

\bibitem[Chandler \& Richer(2000)]{2000ApJ...530..851C} Chandler, C.~J., \& Richer, J.~S.\ 2000, \apj, 530, 851 

\bibitem[Chen et al.(2016)]{2016ApJ...826...95C} Chen, M.~C.-Y., Di Francesco, J., Johnstone, D., et al.\ 2016, \apj, 826, 95 

\bibitem[Codella et al.(2016)]{2016A&A...586L...3C} Codella, C., Ceccarelli, C., Cabrit, S., et al.\ 2016, \aap, 586, L3 

\bibitem[Cuadrado et al.(2015)]{2015A&A...575A..82C} Cuadrado, S., Goicoechea, J.~R., Pilleri, P., et al.\ 2015, \aap, 575, A82 

\bibitem[Fontani et al.(2012)]{2012MNRAS.423.1691F} Fontani, F., Palau, A., Busquet, G., et al.\ 2012, \mnras, 423, 1691 

\bibitem[Fukui et al.(2015)]{2015ApJ...807L...4F} Fukui, Y., Harada, R., Tokuda, K., et al.\ 2015, \apjl, 807, L4 

\bibitem[Garrod \& Herbst(2006)]{2006A&A...457..927G} Garrod, R.~T., \& Herbst, E.\ 2006, \aap, 457, 927 

\bibitem[Gerin et al.(2011)]{2011A&A...525A.116G} Gerin, M., Ka{\'z}mierczak, M., Jastrzebska, M., et al.\ 2011, \aap, 525, A116 

\bibitem[Goldsmith \& Langer(1999)]{1999ApJ...517..209G} Goldsmith, P.~F., \& Langer, W.~D.\ 1999, \apj, 517, 209 

\bibitem[Graninger et al.(2016)]{2016ApJ...819..140G} Graninger, D.~M., Wilkins, O.~H., \& {\"O}berg, K.~I.\ 2016, \apj, 819, 140 

\bibitem[Guzm{\'a}n et al.(2015)]{2015HiA....16..593G} Guzm{\'a}n, V., Pety, J., Gratier, P., et al.\ 2015, Highlights of Astronomy, 16, 593 

\bibitem[Hacar et al.(2017)]{2017A&A...606A.123H} Hacar, A., Tafalla, M., \& Alves, J.\ 2017, \aap, 606, A123 

\bibitem[Hatchell et al.(2005)]{2005A&A...440..151H} Hatchell, J., Richer, J.~S., Fuller, G.~A., et al.\ 2005, \aap, 440, 151 

\bibitem[Hatchell et al.(2007)]{2007A&A...468.1009H} Hatchell, J., Fuller, G.~A., Richer, J.~S., Harries, T.~J., \& Ladd, E.~F.\ 2007, \aap, 468, 1009 

\bibitem[Higuchi et al.(2014)]{2014AJ....147..141H} Higuchi, A.~E., Chibueze, J.~O., Habe, A., Takahira, K., \& Takano, S.\ 2014, \aj, 147, 141 

\bibitem[Higuchi et al.(2010)]{2010ApJ...719.1813H} Higuchi, A.~E., Kurono, Y., Saito, M., \& Kawabe, R.\ 2010, \apj, 719, 1813 

\bibitem[Hirota et al.(2008)]{2008PASJ...60...37H} Hirota, T., Bushimata, T., Choi, Y.~K., et al.\ 2008, \pasj, 60, 37 

\bibitem[Hirota et al.(2011)]{2011PASJ...63....1H} Hirota, T., Honma, M., Imai, H., et al.\ 2011, \pasj, 63, 1 

\bibitem[Imai et al.(2016)]{2016ApJ...830L..37I} Imai, M., Sakai, N., Oya, Y., et al.\ 2016, \apjl, 830, L37 

\bibitem[Kirk et al.(2006)]{2006ApJ...646.1009K} Kirk, H., Johnstone, D., \& Di Francesco, J.\ 2006, \apj, 646, 1009 

\bibitem[Koumpia et al.(2016)]{2016A&A...595A..51K} Koumpia, E., van der Tak, F.~F.~S., Kwon, W., et al.\ 2016, \aap, 595, A51 

\bibitem[Koumpia et al.(2017)]{2017A&A...603A..88K} Koumpia, E., Semenov, D.~A., van der Tak, F.~F.~S., Boogert, A.~C.~A., \& Caux, E.\ 2017, \aap, 603, A88 

\bibitem[Kristensen et al.(2010)]{2010A&A...516A..57K} Kristensen, L.~E., van Dishoeck, E.~F., van Kempen, T.~A., et al.\ 2010, \aap, 516, A57 

\bibitem[Lindberg \& J{\o}rgensen(2012)]{2012A&A...548A..24L} Lindberg, J.~E., \& J{\o}rgensen, J.~K.\ 2012, \aap, 548, A24 

\bibitem[Lindberg et al.(2015)]{2015A&A...584A..28L} Lindberg, J.~E., J{\o}rgensen, J.~K., Watanabe, Y., et al.\ 2015, \aap, 584, A28 

\bibitem[Lindberg et al.(2016)]{2016ApJ...833L..14L} Lindberg, J.~E., Charnley, S.~B., \& Cordiner, M.~A.\ 2016, \apjl, 833, L14 

\bibitem[Maret et al.(2005)]{2005A&A...442..527M} Maret, S., Ceccarelli, C., Tielens, A.~G.~G.~M., et al.\ 2005, \aap, 442, 527 

\bibitem[Matthews \& Wilson(2002)]{2002ApJ...574..822M} Matthews, B.~C., \& Wilson, C.~D.\ 2002, \apj, 574, 822 

\bibitem[Meixner \& Tielens(1993)]{1993ApJ...405..216M} Meixner, M., \& Tielens, A.~G.~G.~M.\ 1993, \apj, 405, 216 

\bibitem[{\"O}berg et al.(2011)]{2011ApJ...740...14O} {\"O}berg, K.~I., van der Marel, N., Kristensen, L.~E., \& van Dishoeck, E.~F.\ 2011, \apj, 740, 14 

\bibitem[Oya et al.(2014)]{2014ApJ...795..152O} Oya, Y., Sakai, N., Sakai, T., et al.\ 2014, \apj, 795, 152 

\bibitem[Pety et al.(2007)]{2007A&A...464L..41P} Pety, J., Goicoechea, J.~R., Hily-Blant, P., Gerin, M., \& Teyssier, D.\ 2007, \aap, 464, L41 

\bibitem[Sandell \& Knee(2001)]{2001ApJ...546L..49S} Sandell, G., \& Knee, L.~B.~G.\ 2001, \apjl, 546, L49 

\bibitem[Sandell et al.(1991)]{1991ApJ...376L..17S} Sandell, G., Aspin, C., Duncan, W.~D., Russell, A.~P.~G., \& Robson, E.~I.\ 1991, \apjl, 376, L17 

\bibitem[Sakai \& Yamamoto(2013)]{2013ChRv..113.8981S} Sakai, N., \& Yamamoto, S.\ 2013, Chemical Reviews, 113, 8981 

\bibitem[Sakai et al.(2014)]{2014Natur.507...78S} Sakai, N., Sakai, T., Hirota, T., et al.\ 2014, \nat, 507, 78 

\bibitem[Sakai et al.(2014)]{2014ApJ...791L..38S} Sakai, N., Oya, Y., Sakai, T., et al.\ 2014, \apjl, 791, L38 

\bibitem[Sakai et al.(2012)]{2012ApJ...754...70S} Sakai, N., Ceccarelli, C., Bottinelli, S., Sakai, T., \& Yamamoto, S.\ 2012, \apj, 754, 70 

\bibitem[Sakai et al.(2010b)]{2010ApJ...722.1633S} Sakai, N., Sakai, T., Hirota, T., \& Yamamoto, S.\ 2010, \apj, 722, 1633 

\bibitem[Sakai et al.(2010a)]{2010ApJ...718L..49S} Sakai, N., Shiino, T., Hirota, T., Sakai, T., \& Yamamoto, S.\ 2010, \apjl, 718, L49 

\bibitem[Sakai et al.(2009)]{2009ApJ...697..769S} Sakai, N., Sakai, T., Hirota, T., Burton, M., \& Yamamoto, S.\ 2009, \apj, 697, 769 

\bibitem[Sakai et al.(2008)]{2008ApJ...672..371S} Sakai, N., Sakai, T., Hirota, T., \& Yamamoto, S.\ 2008, \apj, 672, 371-381 

\bibitem[Saruwatari et al.(2011)]{2011ApJ...729..147S} Saruwatari, O., Sakai, N., Liu, S.-Y., et al.\ 2011, \apj, 729, 147 

\bibitem[Soma et al.(2015)]{2015ApJ...802...74S} Soma, T., Sakai, N., Watanabe, Y., \& Yamamoto, S.\ 2015, \apj, 802, 74 

\bibitem[Spezzano et al.(2016)]{2016A&A...586A.110S} Spezzano, S., Gupta, H., Br{\"u}nken, S., et al.\ 2016, \aap, 586, A110 

\bibitem[Stutzki et al.(1988)]{1988ApJ...332..379S} Stutzki, J., Stacey, G.~J., Genzel, R., et al.\ 1988, \apj, 332, 379 

\bibitem[Taquet et al.(2012)]{2012A&A...538A..42T} Taquet, V., Ceccarelli, C., \& Kahane, C.\ 2012, \aap, 538, A42 

\bibitem[Takakuwa et al.(2000)]{2000ApJ...542..367T} Takakuwa, S., Mikami, H., Saito, M., \& Hirano, N.\ 2000, \apj, 542, 367 

\bibitem[Tielens \& Hagen(1982)]{1982A&A...114..245T} Tielens, A.~G.~G.~M., \& Hagen, W.\ 1982, \aap, 114, 245 

\bibitem[Podio et al.(2014)]{2014A&A...565A..64P} Podio, L., Lefloch, B., Ceccarelli, C., Codella, C., \& Bachiller, R.\ 2014, \aap, 565, A64 

\bibitem[Watanabe \& Kouchi(2002)]{2002ApJ...571L.173W} Watanabe, N., \& Kouchi, A.\ 2002, \apjl, 571, L173 

\bibitem[Watanabe et al.(2015)]{2015ApJ...809..162W} Watanabe, Y., Sakai, N., L{\'o}pez-Sepulcre, A., et al.\ 2015, \apj, 809, 162 

\bibitem[Watanabe et al.(2014)]{2014ApJ...788....4W} Watanabe, Y., Sakai, N., Sorai, K., \& Yamamoto, S.\ 2014, \apj, 788, 4 

\bibitem[Watanabe et al.(2012)]{2012ApJ...745..126W} Watanabe, Y., Sakai, N., Lindberg, J.~E., et al.\ 2012, \apj, 745, 126 

\bibitem[Yamaguchi et al.(2012)]{2012PASJ...64..105Y} Yamaguchi, T., Takano, S., Watanabe, Y., et al.\ 2012, \pasj, 64, 105 

\bibitem[Yoshida et al.(2015)]{2015ApJ...807...66Y} Yoshida, K., Sakai, N., Tokudome, T., et al.\ 2015, \apj, 807, 66 





\end{thebibliography}
\end{document}